\documentclass[12pt]{article}
\usepackage[utf8]{inputenc} 
\DeclareUnicodeCharacter{2212}{-}

\usepackage{draft}
\usepackage{cite} 
\usepackage{tabularx}
\usepackage{tikz}

\usepackage{mathtools}
\usepackage[T1]{fontenc}
\usepackage{esint}

\usepackage{bbold}

\usepackage{soul}

\usepackage{multirow}

\usepackage[margin=1in]{geometry}
\usepackage{setspace}
\usepackage{color}
\usepackage{braket}
\usepackage{titling}
\usepackage{graphicx}
\usepackage[bf,margin=20pt,font={small}]{caption}
\usepackage{subcaption}
\usepackage{youngtab}
\usepackage{epigraph}
\usepackage{tikz}

\tikzset{
    Witten diagram/.style={
        execute at begin picture={
            \draw[blue, line width=1.5pt] circle[radius=\pgfkeysvalueof{/tikz/Witten/radius}];
            \path node (X){\phantom{X}};
        },
        baseline={(X.base)}
    },
    vertex/.style={circle,fill,inner sep=1.5pt,node contents={}},
    Witten/.cd,
    radius/.initial=3cm
}
\usetikzlibrary{patterns}
\usepackage{lipsum}
\usepackage{framed}
\usepackage{adjustbox}
\usepackage{mathtools}

\usetikzlibrary{math}
\newcommand{\eps}{\epsilon}
\usepackage{amsmath,bm}
\usepackage{amssymb,dsfont,amsfonts}
\usepackage{upgreek}
\allowdisplaybreaks
\renewcommand{\tr}{\operatorname{tr}}

\definecolor{royalblue}{rgb}{0.00000,0.44700,0.74100}
\definecolor{royalorange}{rgb}{0.85000,0.32500,0.09800}
\definecolor{royalyellow}{rgb}{0.92900,0.69400,0.12500}
\definecolor{purple}{rgb}{0.5804, 0.0, 0.82745098}
\definecolor{applegreen}{rgb}{0.55, 0.71, 0.0}
\definecolor{bittersweet}{rgb}{1.0, 0.44, 0.37}

\DeclareMathAlphabet{\mathpzc}{OT1}{pzc}{m}{it}

\usepackage{putex}

\usepackage{siunitx}

\usepackage{pgfplots}
\pgfplotsset{compat=1.10}
\usepgfplotslibrary{fillbetween}
\usetikzlibrary{patterns}

\def\XXint#1#2#3{{\setbox0=\hbox{$#1{#2#3}{\int}$ }
		\vcenter{\hbox{$#2#3$ }}\kern-.6\wd0}}

  \usepackage{adjustbox}
\usepackage{multirow}
\usepackage{tikz-cd}
\usepackage[setpagesize=false,pagebackref=false, linktocpage, bookmarksopen=true, colorlinks=true, linkcolor=blue,citecolor=blue,urlcolor=blue]{hyperref}
\usepackage{hyperref}

\numberwithin{equation}{section}

\newcommand{\cI}{{\cal I}}

\def\<{\langle}
\def\>{\rangle}

\def\ep{\epsilon}

\def\id{\mathds{1}}

\usetikzlibrary{decorations.pathmorphing}

\tikzset{snake it/.style={decorate, decoration=snake}}

\newcommand{\leftrarrows}{\mathrel{\raise.75ex\hbox{\oalign{%
				$\scriptstyle\leftarrow$\cr
				\vrule width0pt height.5ex$\hfil\scriptstyle\relbar$\cr}}}}
\newcommand{\lrightarrows}{\mathrel{\raise.75ex\hbox{\oalign{%
				$\scriptstyle\relbar$\hfil\cr
				$\scriptstyle\vrule width0pt height.5ex\smash\rightarrow$\cr}}}}
\newcommand{\Rrelbar}{\mathrel{\raise.75ex\hbox{\oalign{%
				$\scriptstyle\relbar$\cr
				\vrule width0pt height.5ex$\scriptstyle\relbar$}}}}

\makeatletter
\def\leftrightarrowsfill@{\arrowfill@\leftrarrows\Rrelbar\lrightarrows}
\newcommand{\xleftrightarrows}[2][]{\ext@arrow 3399\leftrightarrowsfill@{#1}{#2}}
\makeatother

\institution{NYU}{Center for Cosmology and Particle Physics, New York University, New York, NY 10003, USA}

\title{
Defect Fusion and Casimir Energy \\in Higher Dimensions
}

\authors{Oleksandr Diatlyk, Himanshu Khanchandani, Fedor K.~Popov,\\ and Yifan Wang
}

\abstract{
We study the operator algebra of extended conformal defects in more than two spacetime dimensions. Such algebra structure encodes the combined effect of multiple impurities on physical observables at long distances as well as the interactions among the impurities. These features are formalized by a fusion product which we define for a pair of defects, after isolating divergences that capture the effective potential between the defects, which generalizes the usual Casimir energy. We discuss general properties of the corresponding fusion algebra and contrast with the more familiar cases that involve topological defects. 
We also describe the relation to a different defect setup in the shape of a wedge.
We provide explicit examples to illustrate these properties using line defects and interfaces in the Wilson-Fisher CFT and the Gross-Neveu(-Yukawa) CFT and determine the defect fusion data thereof.

}

\date{\today}

\begin{document}
\maketitle 

\tableofcontents

\section{Introduction and Summary}

Impurities in many-body systems embody a vast array of rich physical phenomena, such as phase transitions, symmetry breaking and universality under renormalization group (RG) flows. Interestingly this is the case even when the bulk system is weakly coupled or free. 
One famous example is the Kondo effect \cite{Kondo:1964nea}, that describes a strongly coupled spin impurity in a sea of essentially free electrons (see \cite{Affleck:1995ge} for a review). Quantum Field Theory (QFT) provides a general and powerful framework to explain the dynamics of impurities in the continuum limit. For the prototypical Kondo model, this is achieved by modeling the impurity, in the so-called s-wave limit, by a boundary condition for the $d=2$ theory of free fermions. Under RG, it flows to a certain strongly coupled conformal boundary condition in the infra-red (IR) which can be identified by Conformal Field Theory (CFT) methods \cite{Affleck:1995ge,affleck1990current,affleck1991kondo,affleck1991critical}. More generally, under mild assumptions such as locality and reflection positivity, universality classes of isolated impurities that extend in $p\geq 1$ spacetime directions are modeled at low energies by conformal defects that preserve the $\mf{so}(p+1,1)$ conformal symmetry. We will focus on the case where the bulk system is gapless and described by a CFT. If the bulk is gapped instead, a $p$-dimensional conformal defect therein is not much different from a conventional $p$-dimensional local CFT coupled to certain topological backgrounds (in dimensions $d > p$).

In recent years, a lot of progress has been made in the study of conformal defects in CFT, especially thanks to the bootstrap philosophy. The bootstrap approach has proven to be extremely successful in constraining and solving the local operator data of CFT based on very general principles such as unitarity, locality and conformal symmetry (see \cite{Poland:2018epd,Hartman:2022zik,Poland:2022qrs} for recent reviews).  The challenge is to bring that success to the more general CFT observables that incorporate defects. Thus far most of the works on conformal defects have focused on the setup with a single defect insertion. This has already produced a large body of interesting results that capture the physics of isolated impurities, in particular from the  operator-product-expansion (OPE) between a single defect and one or more local (point) operators  \cite{Billo:2016cpy}. However impurities also interact with one another in a nontrivial way, especially when immersed in an interacting bulk system. In particular, one expects a generalization of the OPE between the corresponding conformal defects, which capture the physical signature of a collection of interacting impurities observed from far away. Such an OPE for conformal defects would also be important for formulating bootstrap-type constraints with multiple defects in CFT \cite{Gadde:2016fbj}.    

Here we consider the simplest nontrivial setting that consists of a pair of parallel $p$-dimensional conformal defects $\cD_1$ and $\cD_2$ extending along $\mR^p\subset \mR^d$ and separated in a transverse direction. Each conformal defect preserves the subalgebra $\mf{so}(p+1,1)\times \mf{so}(d-p)\subset \mf{so}(d+1,1)$ where $\mf{so}(p+1,1)$ is the conformal symmetry along the defect worldvolume and $\mf{so}(d-p)$ is the transverse rotation symmetry.\footnote{In this paper we focus on defects that are local with respect to other local operators. For example, the correlation functions of local operators $\la \cO_1(x_1)\dots \cO_n(x_n)\ra_\cD$ in the presence of the defect $\cD$ is single-valued. This excludes defects that are attached to topological defects of one-dimension higher (such as monodromy defects), and more generally defects that are not invariant under the transverse rotation $\mf{so}(d-p)$. \label{footnote:localdefect}
} 
In the special situations where one or both of the defects are topological, this OPE is non-singular (in fact topological) and equivalently defines the fusion product of the defects (see \cite{Gaiotto:2014kfa} for general aspects of topological defects). The topological defects that close under fusion give rise to generalized global symmetries in the theory, while the more general (non-topological) conformal defects are organized into modules with respect to these generalized symmetries. See recent reviews   \cite{McGreevy:2022oyu,Cordova:2022ruw,Schafer-Nameki:2023jdn,Brennan:2023mmt,Bhardwaj:2023kri,Shao:2023gho}  on topological defects which discuss these fusion products that contain important information about the generalized symmetries and their representations.
In particular, the generalized Kondo models with multiple impurities studied in \cite{jones1989critical,Affleck:1991yq,gan1995mapping,affleck1995conformal,gan1995solution,georges1995solution,mitchell2012universal,OBannon:2015cqy,lopes2020anyons,gabay2022multi} are solved by the topological OPE (fusion) of the relevant topological defects 
\cite{Bachas:2004sy} in the $d=2$ chiral (unfolded) free fermion theory.\footnote{These defects are chiral (thus topological) in the theory of left-moving fermions and they correspond to the conformal boundary condition in the non-chiral CFT by the folding trick \cite{affleck1995conformal}.}
On the other hand, the OPE (and fusion) of non-topological defects is more subtle due to divergences at small separation and less explored, and thus  will be the focus here. Previous works on such non-topological OPE between parallel defects can be found in \cite{Bachas:2007td,Bachas:2013ora,Konechny:2015qla} for non-topological defects in the $d=2$ free fermion and free boson theories. Specifically, the fusion product $\cD_1\circ\cD_2$ of a pair of parallel conformal defects $\cD_1,\cD_2$ are defined by a limit of the vanishing transverse separation between the defects, after subtracting a divergence that originates from the nontrivial Casimir energy due to the defect insertions \cite{Bachas:2007td}. This fusion product, together with the direct sum operation, defines an algebra of conformal defects in the $d=2$ CFT, which includes the well-studied topological defects as a subalgebra and is much richer. 

In this work, we generalize this fusion product to $p$-dimensional conformal defects in higher dimensions $d>2$. Previous related works that study correlation functions of two defects in specific CFTs can be found in \cite{Soderberg:2021kne,Rodriguez-Gomez:2022gbz,SoderbergRousu:2023zyj}. 
Here we describe the general structure of the defect fusion in Section~\ref{sec:genfusion} and comment on properties thereof in comparison to those of the more familiar fusion product that involves topological defects. In particular, we will see in Section~\ref{sec:fusionalgebra} that the fusion product is in general not associative, namely $\cD_1\circ (\cD_2 \circ \cD_3)\neq (\cD_1 \circ \cD_2) \circ \cD_3$ and furthermore a conformal defect $\cD$ in general may not have a dual $\cD'$ defined such that their fusion product contains the trivial defect $\id$ as a direct summand (i.e. $\cD\circ \cD'\ni \id$ and $\cD'\circ \cD\ni \id$). 
Another key difference between the topological fusion product and the more general fusion product introduced here is the divergence in the limit of vanishing separation as mentioned above. The divergence structure for the fusion of general $p$-dimensional conformal defects is constrained by conformal symmetry up to a few constant coefficients which we refer to as the Casimir coefficients, and the leading divergence captures the Casimir energy density $\cE_{\cD_1\cD_2}$ for the CFT vacuum in the presence of the parallel defects $\cD_1$ and $\cD_2$. These Casimir coefficients encode nontrivial interactions between the defect whose properties are to be further explored. In setups where the defects can be treated as probe particles, strings or branes in the CFT, these Casimir coefficients determine the effective potential between the defects (potentially of complicated shapes).
Here we focus on the fusion of flat (straight) defects in flat space (up to conformal transformations), in which case, only the leading Casimir coefficient (namely the Casimir energy density) contributes. For line defects modeled by a pair of probe particle and antiparticle separated by transverse distance $r$ in a conformal gauge theory, such as two conformal Wilson lines $W,\overline{W}$ of conjugate representations in the $d=4$ $\cN=4$ super-Yang-Mills theory, $\cE_{W\overline{W}}\over r$ is the familiar particle-antiparticle potential that is consistent with scale invariance and the coefficient $\cE_{W\overline{W}}$ depends quadratically on the charges of particles.

To illustrate the general properties of fusion we describe in Section~\ref{sec:genfusion}, we compute the fusion algebra of conformal defects explicitly in the bosonic $O(N)$ Wilson-Fisher CFT and the fermionic Gross-Neveu(-Yukawa) CFT. For concreteness, we consider the fusion of line defects and interface defects in these theories. In particular, we find that the pinning field defects (also known as magnetic line defects) in the $O(N)$ model  generate a non-associative algebra with interesting features such as symmetry enhancement under fusion. It contains an $S^{N-1}$ family of one dimensional associative subalgebras $\cA_{\rm id}=\{\cD(\hat n)\}$ generated by a pinning field defect $\cD(\hat n)$ pointing in the direction $\hat n \in S^{N-1}$ and each $\cD(\hat n)$ corresponds to an idempotent (i.e. $\cD(\hat n)\circ \cD(\hat n)=\cD(\hat n)$) in the full fusion algebra. The same algebra $\cA_{\rm id}$ also governs the fusion of the scalar Wilson line in the Gross-Neveu-Yukawa CFT. 
For interface fusion, we consider factorized interfaces $\cI_{\cB_1\cB_2}\equiv |\cB_1\ra \la \cB_2|$ constructed from a pair of conformal boundary conditions $\cB_1,\cB_2$ in the $O(N)$ Wilson-Fischer CFT and in the Gross-Neveu CFT. The fusion product  for these factorized interfaces is clearly given by $\cI_{\cB_1\cB_2}\circ \cI_{\cB_3\cB_4}=\cI_{\cB_1\cB_4}$, which is associative but not commutative. 

We also determine the Casimir energy density $\cE_{\cD_1\cD_2}$ that controls the leading divergence for defect fusion in these examples. Here we note that, for line defect $\cD$ and its orientation reversal $\overline{\cD}$,  
it is well-known that the Casimir energy (i.e. particle-antiparticle potential in the probe picture) is closely related to another type of defect observable, the cusp anomalous dimension $\Gamma_\cD(\tau)$ of the line defect $\cD$  with a cusp of size $\tau$ (i.e. a deflection angle of $ \theta = \pi-\tau$), in the limit of $\theta\to 0$ (see Section~\ref{sec:Casimirandwedge} for details). Here we generalize this relation to a pair of $p$-dimensional conformal defects $\cD_1,\cD_2$. The cusped line defect is then replaced by a wedge consisting of planar defects $\cD_1$ and $\cD_2$ meeting along $\mR^{p-1}$ at an opening angle $\theta$ in the flat space. By a conformal transformation, this is mapped to a configuration on $S^{d-p}\times \mH^p$, where both defects wrap $\mH^p$ and are located at two points on $S^{d-p}$ with angular separation $\theta$.\footnote{The information of the $(p-1)$-dimensional defect at the corner of the wedge is mapped to the boundary condition at the asymptotic infinity of $\mH^{p}$.} We will find  this last description to be an efficient approach to compute the Casimir energy density for conformal boundary conditions, such as in the $O(N)$ model, and correspondingly the leading divergences in the fusion of factorized interfaces. Our results both extend and provide nontrivial consistency checks for previous calculations of the Casimir energy between such boundary conditions obtained from other methods \cite{PhysRevA.46.1886, Diehl:2006mz, Gruneberg:2007av, Diehl:2011sy, Diehl:2014bpa}.

The rest of the paper is organized as follows. In Section~\ref{sec:genfusion}, we discuss general properties that arise in the fusion of conformal defects, emphasizing both similarities and  differences with the better-understood cases that involve topological defects. This general discussion is then followed by a number of explicit examples. In Section~\ref{sec:ONline}, we first determine the fusion algebra and the Casimir energy of the magnetic line defects in the $O(N)$ Wilson-Fisher CFT using $d=4-\ep$ expansion. Then in Section~\ref{sec:GNline},  we study the fusion of the scalar Wilson line in the Gross-Neveu-Yukawa CFT which we also study in an $\epsilon$ expansion in $d=4-\epsilon$ dimensions. We move onto the fusion of factorized interfaces in the $O(N)$ and Gross-Neveu CFTs in Section~\ref{sec:ONinterface}, where we extract the Casimir energy for a pair of boundary conditions from the free energy of the wedge using hyperbolic space. In particular, for the cases that involve extraordinary boundary conditions for the $O(N)$ model, we identify a simple dual classical mechanical problem in Section~\ref{sec:ExtraOrd}. We also discuss numerical methods to determine the Casimir energy directly at $d=3$ in the large $N$ limit in Section~\ref{sec:NumericalLargeN}. 
We end with a discussion of future directions in Section~\ref{sec:discuss}.

\section{General Properties of Defect Fusion}
\label{sec:genfusion}

In this section, we define the fusion product between general conformal defects in CFTs. As explained in the introduction, this captures the combined physical signature of a pair of parallel defects viewed from far away. 

\subsection{Fusion Algebra of Conformal Defects}
\label{sec:fusionalgebra}

We generalize the fusion product of conformal defect lines in $d=2$ CFTs \cite{Bachas:2007td} to that of $p$-dimensional conformal defects in higher dimensions as follows,\footnote{More generally, the conformal defects that participate in the fusion do not need to have the same dimensionality $p$ \cite{hybridfusion}.}
\ie 
(\cD_1 \circ \cD_2) (\Sigma) \equiv \lim_{r\to 0} e^{\sum_{n=0}^{\lfloor p/2 \rfloor}\int_\Sigma r^{2n-p} C^{(n)}_{\cD_1\cD_2} \cR^n} \cD_1(\Sigma_r)  \cD_2(\Sigma)\,,
\label{defectfusionprod}
\fe
where the worldvolumes of the parallel conformal defects $\cD_1$ and $\cD_2$ are denoted as $\Sigma_r$ and $\Sigma$ respectively, which are separated in the transverse $z$ direction by $r$. The possible divergences in the $r\to 0$ limit are constrained by conformal symmetry such that the counterterms are precisely those in \eqref{defectfusionprod} given by degree $n$ curvature invariants on $\Sigma$ (both intrinsic Riemann curvature and extrinsic curvature from the embedding) of the schematic form $\cR^n$ with constant coefficients $C^{(n)}_{\cD_1\cD_2}$ that render the limit (thus the fusion product $\cD_1\circ \cD_2$) finite.\footnote{For even $p$, the $n=p/2$ term in the exponent of \eqref{defectfusionprod} should be interpreted as a logarithmic divergence with coefficient $C^{({p/2})}_{\cD_1\cD_2}$.} We will refer to $C^{(n)}_{\cD_1\cD_2}$ as the Casimir coefficients for the fusion of the defects $\cD_1$ and $\cD_2$. We emphasize that all these Casimir coefficients $C^{(n)}_{\cD_1\cD_2}$ are physical (independent of local counterterms on the defect) and capture dynamical information for the pair of defects. In particular for $d=2$ and $p=1$, the only relevant Casimir 
coefficient $C^{(0)}_{\cD_1\cD_2}$ is proportional to the smallest scaling dimension $h^{\cD_1\cD_2}_{\rm gap}$ of defect changing operators between $\cD_1$ and $\cD_2$ \cite{CARDY1988377 , Affleck_1994, Affleck:1996mm},
\ie 
C^{(0)}_{\cD_1\cD_2}=h^{\cD_1\cD_2}_{\rm gap}-{c  \pi \over 24}\,,
\fe
where $c$ is the conformal central charge of the 2d CFT. This is also known as the Casimir energy of the vacuum with the defects $\cD_1,\cD_2$ inserted along the time direction. When either $\cD_1$ or $\cD_2$ is topological, all Casimir coefficients vanish identically since the fusion \eqref{defectfusionprod} is topological.

As mentioned before, here we focus on the simplest case where $\Sigma$ and $\Sigma_r$ are flat parallel $\mR^p$ hypersurfaces in the flat spacetime $\mR^d$. 
We will split the $\mR^d$ coordinates as $x^\m=(\vec y,z,\vec x)$ where $\vec y$ label the longitudinal $\mR^p$ directions to the defects and $\vec x$ are the common $\mR^{d-p-1}$ directions transverse to the defects.
Consequently all Casimir coefficients disappear in \eqref{defectfusionprod} since the curvatures vanish, except for the leading divergence with $n=0$ which captures the Casimir energy between the defects. To ease the notation, in the rest of the paper we will denote this coefficient as,
\ie 
\cE_{\cD_1\cD_2}\equiv C^{(0)}_{\cD_1\cD_2}\,,
\fe
and refer to it as the Casimir energy (coefficient) associated with the defects $\cD_1$ and $\cD_2$.

Let us comment on some general features of the fusion product defined in \eqref{defectfusionprod}. Firstly, this fusion product is commutative for $p\leq d-2$ as a consequence of the transverse rotation symmetry $\mf{so}(d-p)$ and locality (see footnote~\ref{footnote:localdefect}) and in general noncommutative for codimension-one defects (i.e. $p=d-1$). Secondly, the resulting defect $\cD_1\circ \cD_2$ is in general decomposable, and can be expressed as a direct sum of irreducible defects
\ie 
\cD_1\circ \cD_2 (\Sigma) =\bigoplus_{i} \cC_i(\Sigma)\cD_i (\Sigma)\,.
\fe
Each summand above is dressed by a decoupled $p$-dimensional TQFT living on the defect volume $\Sigma$ which we denote as  $\cC_i(\Sigma)$. This generalizes the fusion coefficients (valued in $\mZ_+$) in the same way as for the fusion of topological defects discussed in 
\cite{Kapustin:2007wm,Choi:2022zal,
Bhardwaj:2022yxj,Bartsch:2022ytj,Copetti:2023mcq}.
In contrast to the topological case, the fusion product of conformal defects as defined in \eqref{defectfusionprod} is in general not associative. This is not surprising since we are dealing with a truncated OPE for the defects and we will provide explicit examples in Section~\ref{sec:ONline}. 

Furthermore, given a conformal defect $\cD$, we define its dual $\overline\cD$ by its orientation reversal, namely
\ie 
\overline\cD(\Sigma)=\cD(\overline\Sigma)\,,
\fe
where the worldvolume $\overline\Sigma$ is related to $\Sigma$ by flipping the orientation. For topological defects that generate a (higher) fusion category, the fusion product of a defect and its dual, namely $\cD\circ \overline{\cD}$ (also $\overline{\cD} \circ \cD$), always contains the trivial topological defect $\id$, which is a property known as dualizability.\footnote{In TQFTs, it is possible to have topological defects that are not dualizable \cite{Kapustin:2010if}. A simple example is a factorized interface $|\cB_1\ra \la \cB_2|$ where $|\cB_1\ra$ and $|\cB_2\ra$ are two topological boundary conditions of the TQFT \cite{Roumpedakis:2022aik}.} However general conformal defects are not dualizable as we will illustrate in examples. 

\subsection{Casimir Energy, Wedge Free Energy and Hyperbolic Space}
\label{sec:Casimirandwedge}
\begin{figure}
    \centering
    \begin{subfigure}{0.37\textwidth}
        \centering
        \includegraphics[width=\textwidth]{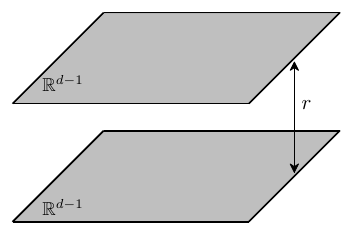}
        \caption{}
        \label{fig:subfig1}
    \end{subfigure}
    \hfill
    \begin{subfigure}{0.5\textwidth}
        \centering
        \includegraphics[width=\textwidth]{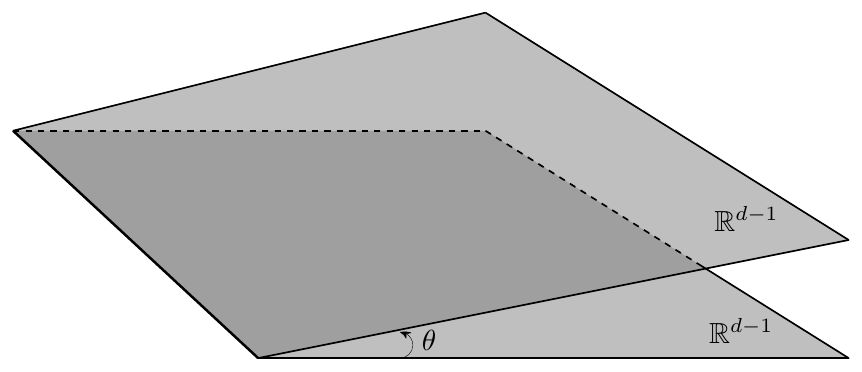}
        \caption{}
        \label{fig:subfig2SlabtoWedge}
    \end{subfigure}
    \caption{Slab and wedge geometry. In the limit of small angle $\theta$, the wedge imitates a parallel slab.}
    \label{fig:SlabtoWedge}
\end{figure}

As pointed out in the introduction, the Casimir energy for line defects modeled by a probe particle-antiparticle pair is related to the cusp anomalous dimension of the Wilson loop that represents the trajectory of the particle \cite{Drukker:2011za,Correa:2012at,Drukker:2012de,Correa:2012hh}. More concretely, consider a Wilson line $W$ in a conformal gauge theory with a cusp of size $\tau$ (the setup is the same as the one we depict in Figure \ref{fig:subfig2SlabtoWedge} except that here we are talking about lines instead of a surface). Then by the usual conformal transformation that maps plane to cylinder centered at the corner, this cusped Wilson line can be mapped to a pair of quark and antiquark lines running along the cylinder separated by a relative angle $\theta = \pi - \tau$ on $S^{d-1}$. In the limit of small $\theta$, the cusp anomalous dimension of the Wilson loop is related to the Casimir energy by \cite{Drukker:2011za}
\ie 
    \Gamma_{W}(\tau) = \frac{\cE_{W\overline W}}{\theta}\,, \quad \theta \to 0\,,
    \label{fromcuspadtocasE}
\fe 
where $\cE_{W\overline W}$ is the Casimir energy associated with the parallel defect lines, and in this case, the potential energy between a probe quark and a probe antiquark.

Here we generalize this relation \eqref{fromcuspadtocasE} to general conformal defects. As we will see in later sections, this gives us a useful tool to compute the Casimir energy coefficient in specific examples by evaluating partition functions on certain geometries. 

\begin{figure}
  \centering
  \includegraphics[width=1\textwidth]{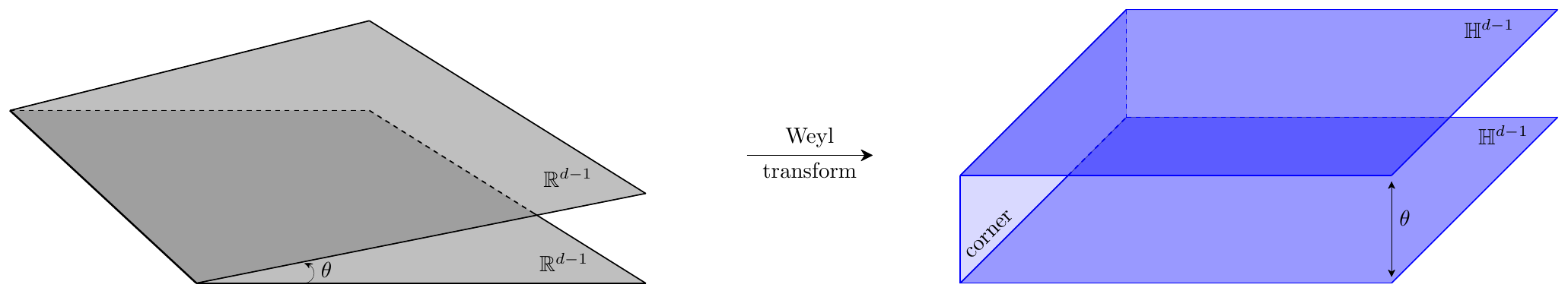}
 \caption{Weyl transformation of the wedge in flat space to hyperbolic space.}
  \label{fig: WeyltransfWedgetoSlab}
\end{figure}

To start, let us discuss the case of conformal boundaries $\cB_1$ and $\cB_2$. In particular, to extract the Casimir energy for a pair of conformal boundaries, one can calculate the free energy of the theory in a slab of width $r$ and the slab free energy in the limit of small $r$ (compared to the size of the slab) will give us the
Casimir energy (see Figure~\ref{fig:SlabtoWedge}). Equivalently, we can calculate the free energy of the CFT in a wedge, and in the limit of small opening angle $\theta$,  we expect to recover the Casimir energy (generalizing the case of a cusped line). Note that in general, CFT inside a slab and  CFT inside a wedge give rise to different observables, but in the limit of small opening angle, the wedge imitates the slab (see Figure~\ref{fig:SlabtoWedge}).\footnote{In the coordinates below: the wedge imitates the slab in the limit of small $\theta$ and in the region with large $\rho$ with fixed $\rho \theta$.} The geometry we use on the wedge may be described using cylindrical coordinates as below,
\begin{equation}
\label{cylindricalcoords}
    ds^2 =  d \vec{y}^2 + d \rho^2 + \rho^2 d \varphi^2 = \rho^2 \left( \frac{d \vec{y}^2 + d \rho^2}{\rho^2} + d \varphi^2 \right) = \rho^2 ds^2_{S^1_{\theta} \times \mH^{d-1}} \,,
\end{equation}
where $0 < \varphi < \theta$, $\vec{y}$ are the coordinates along the $d-2$-dimensional corner of the wedge, and $\rho$ is the distance from the corner. As indicated in the above equation, we can conformally map the setup to $S^1_\theta \times \mH^{d - 1}$ where $S^1_\theta$ represents a circular arc of size $\theta$ , $\mH^{d - 1}$ is the $d-1$-dimensional hyperbolic space and the corner of the wedge now locates at the asymptotic boundary of the $\mH^{d - 1}$ (see Figure~\ref{fig: WeyltransfWedgetoSlab}). The free energy on $S^1_\theta \times \mH^{d-1}$ in the limit of small $\theta$ takes the following form 
\begin{equation}
    F_{\cB_1\cB_2} = \frac{\textrm{Vol} (\mH^{d-1}) \ \cE_{\cB_1\cB_2}}{\theta^{d-1}}\,, \quad \theta \to 0\,,
\end{equation}
with $\cE_{\cB_1\cB_2}$ being the Casimir energy for the two conformal boundaries, generalizing \eqref{fromcuspadtocasE} for line defects.
In Section~\ref{sec:ONinterface}, we verify the above formula explicitly by comparing with the Casimir energy $\cE$ computed from the slab setup \cite{PhysRevA.46.1886, Diehl:2006mz, Gruneberg:2007av, Diehl:2011sy, Diehl:2014bpa}.

Note that the idea of using hyperbolic space to describe conformal defects in a wedge like configuration can be applied to defects of any dimension. Consider flat $p$-dimensional conformal defects $\cD_1$ and $\cD_2$ joined by a cusp which has an opening angle $\theta$. The geometry in this case with the origin located at the cusp can be written as
\begin{equation}
    ds^2 = d \vec{y}^2 + d \rho^2 + \rho^2 d s^2_{S^{d - p}} = \rho^2 \left( \frac{d \vec{y}^2 + d \rho^2}{\rho^2} + d s^2_{S^{d - p}} \right) = \rho^2 ds^2_{S^{d - p} \times \mH^{p}} \, ,
\end{equation}
where $\vec{y}$ are the coordinates on the $p-1$ dimensional corner of the wedge and the defects are separated by a relative angle $\theta$ along the $d - p$ dimensional sphere $S^{d-p}$. As is evident by using a conformal transformation, we can map the configuration to $S^{d-p} \times \mH^{p}$ where corner of the wedge sweeps an arc of size $\theta$ on $S^{d-p}$. As before, this allows us to relate the free energy on $S^{d-p} \times \mH^{p}$ to the Casimir energy for the parallel defects in the limit of small angle, 
\ie 
 F_{\cD_1\cD_2} = \frac{\textrm{Vol} (\mH^{p}) \ \cE_{\cD_1\cD_2}}{\theta^{p}}\,, \quad \theta \to 0\,.
 \label{hyperandcasgenp}
\fe
The relation \eqref{fromcuspadtocasE} for the cusp anomalous dimension for line defects ($p = 1$) is a special case of this formalism for which we get a usual cylinder after the Weyl transformation.

We emphasize that for general $p$, the $S^{d-p} \times \mH^{p}$ description makes manifest the residual conformal symmetry of the problem (as isometries), which leads to simplifications in the evaluation of the free energy as we will see in later sections.\footnote{The idea of using hyperbolic space to describe conformal defects was introduced in \cite{Giombi:2021uae} to study monodromy defects. The wedge configuration for $p=2$ was also studied in \cite{Drukker:2022beq} using hyperbolic space.} We also note that the wedge configuration is only fully specified if we also identify the $p-1$-dimensional defect operator at the corner (junction). As long as such a junction defect exists (which we assume to be the case), its specific choice does not matter in the relation \eqref{hyperandcasgenp} as it contributes sub-extensively compared to the volume factor ${\rm Vol}(\mH^{p})$. Therefore, we will often make convenient choices for the junction defect to facilitate the computation as we will explain in Section~\ref{sec:ONinterface}. Of course the free energies $\cF_{\cD_1\cD_2}$ (and related observables) we obtain that are exact functions of $\theta$ (before the small angle limit) contain information of the junction defect and will be a useful starting point to analyze bootstrap equations on conformal wedges. We will briefly comment on this in the conclusion.

\section{Fusing Line Defects at $d>2$}
\label{sec:ONline}

In this section we illustrate the general properties of defect fusion discussed in Section~\ref{sec:genfusion} using concrete examples. We start with the study of localized magnetic field lines in the critical $O(N)$ model introduced in \cite{Allais:2014fqa,Cuomo:2021kfm}. We denote these defects as $\cD(\hat n)$ where the unit vector $\hat n\in S^{N-1}$ specifies the orientation of the defect in the $O(N)$ directions. We will show that the fusion product among the defect lines labeled by $\hat n,\hat m \in S^{N-1}$ takes the following intuitive form\footnote{In the arXiv version of \cite{SoderbergRousu:2023zyj}, it was incorrectly concluded that the fused defect has a new fixed point, different from the ones that exist for individual defects. Here we will clarify that it is the same fixed point but the defect couples to a rotated field. We communicated this issue to the author of \cite{SoderbergRousu:2023zyj} shortly after their paper appeared. This was later corrected in the published version of \cite{SoderbergRousu:2023zyj}.}
\ie 
\cD(\hat n) \circ \cD(\hat m)=\cD\left({\hat n +\hat m\over \sqrt{2(1+\hat n \cdot \hat m)}} \right)\,.
\label{magnetfusion}
\fe 
We will also determine the associated Casimir energy using the $\epsilon$ expansion. Then in Section~\ref{sec:GNline}, we study a similar line defect in fermionic CFTs which was introduced in \cite{Giombi:2022vnz}. 

\subsection{Line defects in Wilson-Fisher theory}
Here we study the magnetic line defects in the $O(N)$ invariant  Wilson-Fisher theory described by the $\phi^4$ interaction in $d = 4 - \epsilon$ dimensions. Fusion of such magnetic line defects in this model was discussed earlier in \cite{SoderbergRousu:2023zyj}.\footnote{In \cite{SoderbergRousu:2023zyj}, they also studied a surface defect in the $O(N)$ model in $d= 6- \eps$ which is defined in a similar way.} Here we present the analysis again for clarity and also give an alternative way to study this fusion.  
We consider two parallel line defects and use coordinates $x = (y, z, \vec{x})$ such that the defects are extended along the $y$ direction and separated along the $z$ direction by distance $r$.  We will first consider the case where both the defects are coupled to the field $\phi^1$ so a common residual $O(N - 1)$ symmetry is preserved. The action for the coupled system with the two line defects is 
\ie 
    S= & \int d^d x \left( \frac{1}{2} (\partial_{\mu} \phi^I)^2 + \frac{\lambda_b \Lambda_T^{4-d}}{4} (\phi^I \phi^I)^2 \right)  \\
   & + h_{1,b}a_D^{\frac{d-4}{2}} \int d y \phi^1 (y, z = 0, \vec{x} = 0) + h_{2,b}a_D^{\frac{d-4}{2}} \int d y \phi^1 (y, z = r, \vec{x} = 0)\,.
    \label{WFcoupledaction}
\fe
Before proceeding to analyze this system, let us briefly discuss what scales are at play. First we have a UV cutoff of the bulk theory $\Lambda_T$ (roughly speaking the lattice size if one uses a lattice regularization), then there is a UV cutoff of the individual defects $\Lambda_D \sim a_D^{-1}$ (where $a_D$ is the width of the defect) and finally we have the distance between the defects $r$. We will assume the following hierarchy\footnote{Naively, we could have set $\Lambda_T=\Lambda_D$ as is commonly adopted in the literature \cite{
Goldberger:2001tn,Fredenhagen:2006dn,Michel:2014lva,Cuomo:2021kfm}, but in the limit $\Lambda_T \to \infty$ such a configuration could lead to a trivial defect. To make the statements more general and more robust we will treat these scales $\Lambda_T$ and $\Lambda_D$ separately.}  $\Lambda_T \gg \Lambda_D \gg \frac{1}{r}$, so that the bulk theory and the defects are already described by consistent CFT and DCFTs therein as $\Lambda_T,\Lambda_D \to \infty$. Thus, the renormalized couplings $\lambda$, $h_1$ and $h_2$ will flow to their fixed points (in a fixed scheme) in the coupled action \eqref{WFcoupledaction}, which take the following values  \cite{Cuomo:2021kfm}, 
\begin{equation}
\label{fixedpointvaluespinning}
    \lambda_* = \frac{8 \pi^2}{ (N + 8)} \epsilon\,, \quad h_{\pm} = \pm h_*\,,\quad  h_* = \left[\sqrt{N + 8} + \frac{4 N ^2 + 45 N + 170}{4 (N + 8)^{\frac{3}{2}}} \epsilon \right] + O(\epsilon^2)\,.
\end{equation}
There are two distinct conformal magnetic line defects in this theory depending on whether we tune the defect coupling to $h_{+}$ or $h_{-}$, which are obviously related by the bulk $\mZ_2$ global symmetry that reflects $\phi^1$. They correspond to the conformal lines $\cD(\pm \hat n)$ for $\hat n=(1,0,\dots,0)\in S^{N-1}$ respectively. In the following, we will consider all possible fusion products of two such defects $\cD_1,\cD_2=\cD(\pm \hat n)$ in the limit $r \rightarrow 0$ as defined in \eqref{defectfusionprod}.

\begin{figure}
    \centering
    \begin{subfigure}{0.2\textwidth}   \includegraphics[width=\linewidth]{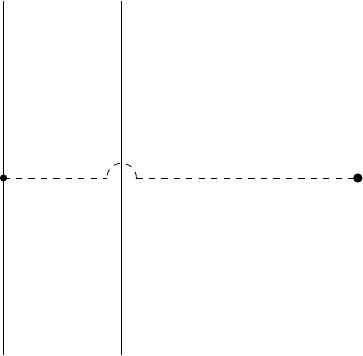}
        \label{fig:diagram1}
    \end{subfigure}
     \hspace{1.5em}\begin{subfigure}{0.2\textwidth}
        \includegraphics[width=\linewidth]{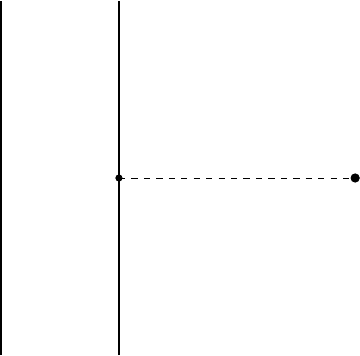}
        \label{fig:diagram2}
    \end{subfigure}
    \hspace{1.5em} \begin{subfigure}{0.2\textwidth}
        \includegraphics[width=\linewidth]{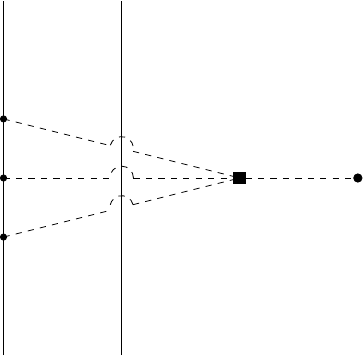}
        \label{fig:diagram3}
    \end{subfigure}\\
  \begin{subfigure}{0.2\textwidth}
        \includegraphics[width=\linewidth]{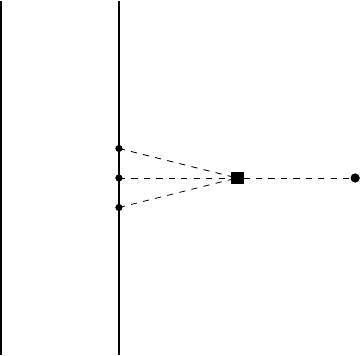}
        \label{fig:diagram4}
    \end{subfigure}
    \hspace{1.5em} \begin{subfigure}{0.2\textwidth}
        \includegraphics[width=\linewidth]{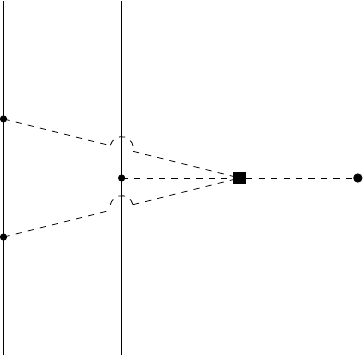}
        \label{fig:diagram5}
    \end{subfigure}
   \hspace{1.5em}  \begin{subfigure}{0.2\textwidth}
        \includegraphics[width=\linewidth]{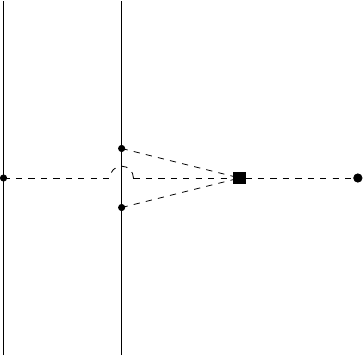}
        \label{fig:diagram6}
    \end{subfigure}
    \caption{Feynman diagrams that contribute to the one-point function $\langle \cD_1 \cD_2 \phi^1 (x) \rangle$. The filled circle represents defect coupling while the filled box represents the bulk coupling.}
    \label{1ptMagnONdiagrams}
\end{figure}

To determine the fusion product, we study the one-point function of the field $\phi^1$ when it is located far from both the defects.\footnote{In general, to fully identify a conformal defect, one needs to specify the one-point functions with all bulk primary operators. For the specific setup here (described by \eqref{WFcoupledaction}) it suffices to study the one-point function of $\phi^1$.} To leading order in $\lambda$, the diagrams that contribute are shown in Figure~\ref{1ptMagnONdiagrams}. We will start by writing down expressions for finite defect regulator $a_D$ and bare couplings $h_{1,b},h_{2,b}$, and then describe how they change when we tune the defects to criticality. After normalizing by the expectation value of the product of two defects, we obtain the following one-point function of $\phi^1$ with the defect insertions,
\ie
{}&\langle \cD_1 \cD_2 \phi^1 (x) \rangle 
\\
&=- h_{1,b} a_D^{\frac{d-4}{2}} \int d y \langle \phi^1 (x) \phi^1 (y, z = 0, \vec{x} = 0) \rangle  - h_{2,b} a_D^{\frac{d-4}{2}} \int d y \langle \phi^1 (x) \phi^1 (y, z = r, \vec{x} = 0) \rangle  \\
& + \frac{\lambda_b}{24} \Lambda_T^{4-d} a_D^{\frac{3(d-4)}{2}} \int d^d x' \left( \prod_{k = 1}^3 d y_k \right) \bigg\langle \phi^1 (x)  (\phi^I \phi^I)^2 (x') \bigg[ h_{1,b}^3 \left( \prod_{k = 1}^3 \phi^1 (y_k, z = 0, \vec{x} = 0)\right)  \\
&+ h_{2,b}^3 \left( \prod_{k = 1}^3 \phi^1 (y_k, z = r, \vec{x} = 0)\right) + 3 h_{1,b}^2 h_{2,b} \phi^1 (y_3, z = r, \vec{x} = 0)\prod_{k = 1}^2 \phi^1 (y_k, z = 0, \vec{x} = 0)  \\
& + 3 h_{1,b} h_{2,b}^2 \phi^1 (y_3, z = 0, \vec{x} = 0)\prod_{k = 1}^2 \phi^1 (y_k, z = r, \vec{x} = 0) \bigg] \bigg\rangle \,.
\label{D1D2phiset}
\fe 
We are assuming that the bulk is critical, so the corrections that induce a bulk mass term, for instance the tadpole contributions have been tuned away. Note that here and below, the expressions are only valid to first order in $\lambda$ unless otherwise specified. In other words, we will work to the leading order in the $\epsilon$ expansion where the correlation function $\langle \cD_1 \cD_2 \phi^1 (x) \rangle$ will be sufficient to pin down the fusion product of the defect lines. As we will see later, to correctly account for this correlator when the defects $\cD_1,\cD_2$ are tuned to their fixed points, one must include contributions that are higher order in $\lambda$ (and enhanced by divergences in the $\ep \to 0$ limit). Resummations of such contributions can be done via the Callan-Symanzik equation, which we implement explicitly in \eqref{CSonepointfunc} to
resum diagrams that are attached to a single defect (such as the first four diagrams in Figure~\ref{1ptMagnONdiagrams}). For diagrams that involve both defects (such as the last two diagrams in Figure~\ref{1ptMagnONdiagrams}), we present an alternative method to perform the resummation using the Schwinger-Dyson equation \eqref{resumbySD}.

Let us first analyze the contributions in \eqref{D1D2phiset}. Performing the integrals along the $y$ direction longitudinal to the defects in \eqref{D1D2phiset}, we obtain
\begin{equation}
\begin{split}
&\langle \cD_1 \cD_2 \phi^1 (x) \rangle = - \frac{h_{1,b} a_D^{\frac{d-4}{2}} \Gamma\left( \frac{d - 3}{2} \right)}{4 \pi^{\frac{d - 1}{2}} (\vec{x}^2 +z^2)^\frac{d - 3}{2}}  - \frac{h_{2,b} a_D^{\frac{d-4}{2}} \Gamma\left( \frac{d - 3}{2} \right)}{4 \pi^{\frac{d - 1}{2}} (\vec{x}^2 + (z - r)^2)^\frac{d - 3}{2}}  + \\
& +\frac{\lambda_b \Lambda_T^{4-d}  a_D^{\frac{3(d-4)}{2}} \Gamma\left( \frac{d - 3}{2} \right)^3 }{2 (3 d - 11) (4 - d) (4 \pi^{\frac{d - 1}{2}})^3 } \left( \frac{h_{1,b}^3}{(\vec{x}^2+z^2)^{\frac{ 3 d - 11}{2}}} + \frac{h_{2,b}^3}{(\vec{x}^2 + (z - r)^2)^{\frac{ 3 d - 11}{2}}} \right) \\
&+ 3 \lambda_b \Lambda^{4-d}_T a_D^{\frac{3(d-4)}{2}} (h_{1,b}^2 h_{2,b} I_1 + h_{1,b} h_{2,b}^2 I_2) \, ,
\end{split}
\label{oneptfperturbative}
\end{equation}
where the coefficients $I_{1,2}$ are  given by following integrals
\begin{equation}
\begin{split}
I_1 &=  \frac{\Gamma\left( \frac{d - 3}{2} \right)^4}{ (4 \pi^{\frac{d - 1}{2}})^4 }  \int \frac{d^{d - 2} \vec{x}' dz'}{ ((\vec{x} - \vec{x}')^2 + (z - z')^2)^{\frac{d - 3}{2} } ((\vec{x}')^2 + z'^2)^{d - 3} ((\vec{x}')^2 + (z' - r)^2 )^{\frac{d - 3}{2} }  } \, , \\
I_2 &= \frac{\Gamma\left( \frac{d - 3}{2} \right)^4}{ (4 \pi^{\frac{d - 1}{2}})^4 }  \int \frac{d^{d - 2} \vec{x}' dz'}{ ((\vec{x} - \vec{x}')^2 + (z - z')^2)^{\frac{d - 3}{2} } ((\vec{x}')^2 + z'^2)^{\frac{d - 3}{2}} ((\vec{x}')^2 + (z' - r)^2 )^{d - 3}} \, .
\end{split}
\end{equation}
It is easy to see that the above integrals are finite as long as $r > 0$, and only diverge when $r \rightarrow 0$. So the leading divergences in \eqref{oneptfperturbative} (at $r>0$) come from diagrams that only involve one of the defects rather than both. From these divergences we obtain the following relations between   the renormalized and the bare couplings at the leading order in $\epsilon$,
\begin{gather}
   \left\{
\begin{matrix}
 \lambda_b=\lambda +\mathcal{O}(\lambda^2)\,, \\
h_{1,b}=h_1+\frac{1}{\epsilon}\frac{\lambda h^{3}_1}{32\pi^2} (\Lambda_T a_D )^{\epsilon}+ \mathcal{O}(\lambda^2)\,,  \\
h_{2,b}=h_2+\frac{1}{\epsilon}\frac{\lambda h^{3}_2}{32\pi^2}(\Lambda_T a_D )^{\epsilon}+ \mathcal{O}(\lambda^2)\,. 
\end{matrix}
\right. 
\label{barecouplings}
\end{gather}
Note that here we are using a scheme where $\Lambda_T$ and $a_D$ are independent scales.

 We now resum diagrams that involve a single defect using the familiar Callan-Symanzik equation for the one-point function
 \ie 
 \braket{\mathcal{D}_1 \phi^1(x)} \equiv x_\perp^\frac{2-d}{2}\Phi\left(\Lambda_D x_\perp,h \right)\,,
 \fe 
 that we consider as a function of the renormalized coupling constant $h$, perpendicular distance $x_\perp=\sqrt{\vec{x}^2+z^2}$ to the defect and a defect width $\Lambda_D=a^{-1}_D$. The Callan-Symanzik equation states that physical observables should not depend on defect width $\Lambda^{-1}_D$ and renormalized coupling constant $h$,
\ie 
    \left(\Lambda_D \frac{\partial}{\partial \Lambda_D} + \beta(h) \frac{\partial}{\partial h} \right) \Phi\left(\Lambda_D x_\perp ,h \right) = 0\,, 
   \label{CSonepointfunc}
\fe 
 as long as we change $h$ accordingly to its beta function,
\ie 
  \beta(h)\equiv \Lambda_D \frac{d h}{d \Lambda_D} = -\frac{4-d}{2} h + \frac{\lambda}{16 \pi^2} h^3\,.
  \label{hbeta}
\fe 
The general solution to \eqref{CSonepointfunc} is given by the  method of characteristics.
Namely we look for $\bar{h}(x_\perp)$ that satisfies the following equation and boundary condition,
\begin{gather}
     -\frac{d \bar{h}}{d \log x_\perp} = -\frac{\epsilon}{2} \bar{h} + \frac{\lambda}{16 \pi^2} \bar{h}^3\, , \ \ \ \bar{h}\left(x_\perp=a_D\right)=h\,,
     \label{betafunctcoupling_h}
\end{gather}
and then $\Phi(\Lambda_D x_\perp,h)=\bar\Phi(\bar h(x_\perp))$ for general $\bar\Phi$ produces the desired solution to \eqref{CSonepointfunc}.
Solving \eqref{betafunctcoupling_h} explicitly, we have then (assuming $h>0$)
\begin{gather}
\label{gensoltoCS}
    \braket{\mathcal{D}_1 \phi^1(x)} = \frac1{x_\perp^\frac{d-2}{2}}\bar \Phi\left(\dfrac{1}{\sqrt{\frac{\lambda}{8\pi^2(4-d)}+\left(\frac{1}{h^2}-\frac{\lambda}{8\pi^2(4-d)}\right)\left(\Lambda_D x_\perp\right)^{d-4}}} \right)\,.
\end{gather}
Now, requiring that to the leading order in the $h$ expansion the solution should match the perturbative calculation in \eqref{oneptfperturbative},
\begin{gather}
    \braket{\mathcal{D}_1 \phi^1(x)}=- \frac{h a_D^{\frac{d-4}{2}} \Gamma\left( \frac{d - 3}{2} \right)}{4 \pi^{\frac{d - 1}{2}} x_\perp^{d - 3}} +\mathcal{O}(h^3,\lambda)\,, \quad  \quad 
\end{gather}
we can fix the unknown function $\bar\Phi$ in \eqref{gensoltoCS} and the final result takes the following form,
\begin{gather}
   \braket{\mathcal{D}_1 \phi^1(x)}=-\dfrac{\Gamma\left(\frac{d-3}{2}\right)}{4\pi^{\frac{d-1}{2}}}\frac{1}{(\vec{x}^2+z^2)^{\frac{d-2}{4}}}\frac{1}{\sqrt{\frac{\lambda}{8\pi^2 (4-d)} + \left(\frac{1}{h^2}-\frac{\lambda}{8\pi^2 (4-d)}\right)
   (\vec{x}^2+z^2)^{\frac{d-4}{2}}  a_D^{4-d} }}\,.
\end{gather}
Taking the limit $a_D \rightarrow 0$, we obtain the one-point function for the conformal defect (see \eqref{fixedpointvaluespinning}),
\begin{gather}
     \braket{\mathcal{D}_1 \phi^1(x)}=-\dfrac{h_*}{4\pi x_\perp^\frac{d-2}{2}}\,.
     \label{onepf}
\end{gather}
Thus after resumming the diagrams that are attached to a single defect, we obtain
that the two defects would individually flow to a conformal defect,
\ie 
&\langle \cD_1 \cD_2 \phi^1 (x) \rangle =  -\frac{1}{4\pi}\frac{h_{1*}}{ (\vec{x}^2 + z^2)^\frac{d-2}4}  -\frac{1}{4\pi} \frac{h_{2*}}{(\vec{x}^2 + (z - r)^2)^\frac{d-2}{4}} +\text{cross-terms}\,,
\label{toresumcrossterms}
\fe 
where $h_{1*},h_{2*}=\pm h_*$ (see \eqref{fixedpointvaluespinning}) depending on which of the two conformal lines $\cD(\pm \hat n)$ describes the fixed point. The ``cross-terms'' above refers to contributions from all the diagrams that involve both of the defects (such as the last two diagrams in Figure~\ref{1ptMagnONdiagrams}). In the following we will resum such contributions to fully determine $\langle \cD_1 \cD_2 \phi^1 (x) \rangle$ for $x_\perp\gg r$ to the leading order in $\ep$.

Naively,  if we omit these diagrams, the one-point function measured far away (i.e. $x_\perp^2=\vec{x}^2 + z^2 \gg r^2$) would simply be the sum of the one-point functions from a single defect, already at zeroth order in $\epsilon$,
\begin{equation} 
    \langle \cD_1 \cD_2 \phi^1 (x) \rangle =  -\frac{(h_{1*} + h_{2*})}{4 \pi x_\perp^{\frac{d-2}{2}}}\,. \label{eq:naivefusion}
\end{equation}
In particular, if both of the defects are described by the same conformal defect $\cD(\hat n)$ at large distance, the resulting effective coupling for $\phi^1$ on the defect would be doubled. This is in contradiction with the RG equation which admits one unique conformal line defect (up to a sign). This mismatch happens because  we have to take into account the infinite contributions coming from diagrams that involve both defects to obtain the correct result even at the leading order in $\eps$. To show this, let us take a closer look at the cross-terms in \eqref{toresumcrossterms}, coming from the diagrams involving both defects (for example the last two diagrams in Figure~\ref{1ptMagnONdiagrams}). Since we have already resummed all diagrams attached to a single defect, the integrals $I_{1,2}$ should be modified accordingly. Consequently, the leading pieces in the cross-terms take the following form 
\begin{equation}
\begin{split}
\hat{I}_1 &= \frac{1}{(4\pi)^3}\frac{\Gamma\left( \frac{d - 3}{2} \right)}{4 \pi^{\frac{d - 1}{2}} }  \int \frac{d^{d - 2} \vec{x}' dz'}{ ((\vec{x} - \vec{x}')^2 + (z - z')^2)^{\frac{d - 3}{2} } ((\vec{x}')^2 + z'^2)^\frac{d-2}{2} ((\vec{x}')^2 + (z' - r)^2 )^{\frac{d-2}{4} }  }\,, \\
\hat{I}_2 &=  \frac{1}{(4\pi)^3}\frac{\Gamma\left( \frac{d - 3}{2} \right)}{4 \pi^{\frac{d - 1}{2}} }  \int \frac{d^{d - 2} \vec{x}' dz'}{ ((\vec{x} - \vec{x}')^2 + (z - z')^2)^{\frac{d - 3}{2} } ((\vec{x}')^2 + z'^2)^{\frac{d-2}{4}} ((\vec{x}')^2 + (z' - r)^2 )^\frac{d-2}{2}}\,. \notag
\end{split}
\end{equation}
Let us estimate their contributions in the limit $x_\perp=\sqrt{x^2+z^2} \gg r$,
\begin{equation}
\begin{split}
    3\lambda  (h_{1*}^2 h_{2*} \hat{I}_1 + h_{1*} h_{2*}^2 \hat{I}_2)  \sim  \frac{3 \lambda \left(h_{1*}^2 h_{2*} + h_{2*}h_{1*}^2\right)}{4-d} \frac{1}{x_\perp^\frac{3d-10}{2}} \gg \frac{2h_*}{x_\perp^{\frac{d-2}{2}}}\,,
\end{split}
\end{equation}
which dominates over the naive solution \eqref{eq:naivefusion} at large distances and modifies the effective coupling to $\phi^1$ on the defect. To nail down the right value of this effective coupling, we have to resum the contributions that come from diagrams involving both of the defects. In particular, we have to sum up all the diagrams that  
behave as $(\lambda/ \epsilon)^n$ when the separation $r \rightarrow 0$ but are finite otherwise. Once we plug in the fixed point value of $\lambda$ in \eqref{fixedpointvaluespinning}, all of these contribution will add up to cancel the naive factor of two we encounter in \eqref{eq:naivefusion} when $h_{1*}=h_{2*}=h_*$. We could perform this resummation explicitly using again the Callan-Symanzik equation, but instead we would like to present another method for resummation of these contributions in the spirit of \cite{Polyakov:1972ay} using the Schwinger-Dyson equation. 

In the following, to ease the notation we will simply use $\phi^1(x)$ to denote the one-point function $\la \cD_1\cD_2 \phi^1(x)\ra$.
We notice that to leading order in $\lambda$ this resummation is equivalent to solving the classical equation for the field $\phi^1$, 
\ie 
\label{resumbySD}
\Box \phi^1 - \frac{\lambda_*}{x_{\perp}^{4-d}}  (\phi^1)^3 = 0\,,
\fe 
where $\Box$ is the scalar Laplacian,
with specific boundary conditions determined by the insertion of defects,
\ie 
\phi^1(x) =\begin{cases}
    -\frac{h_{1*}}{4\pi \left(\vec{x}^2+z^2\right)^\frac{d-2}{4}}\,, \quad \vec{x},z \to 0\,, 
    \\
    -\frac{h_{2*}}{4\pi \left(\vec{x}^2+(z-r)^2\right)^\frac{d-2}{4}}\,,\quad  \vec{x},(z-r) \to 0 \,.
\end{cases} 
\fe  
 Note that in \eqref{resumbySD} we have taken into account that at the fixed point of the bulk theory the coupling constant must depend on the distance.
 
We first consider the case when both defects are aligned (e.g. $\cD_{1,2}$ are both described $\cD(\hat n)$ in the conformal limit) so that both defect coupling constants are tuned to the same value $h_{1*}=h_{2*}=h_*$. Then one can check that near both of the defects, the corrections are small and we can trust the approximation
\begin{gather}
\phi^1(x) \approx -\frac{h_*}{2\pi x_\perp^\frac{d-2}{2}}\, , \quad |\vec x|,|z| \lesssim r \, 
\label{neardefectbehavior}
\end{gather}
On the other hand, in the limit $x_\perp \gg r$, we expect that the field $\phi(x)$ should depend only on $x_\perp$ and satisfy the following equation
\begin{gather}
     (\phi^1)''(x_\perp) + \frac{d-2}{x_\perp} (\phi^1)'(x_\perp)- \frac{\lambda_*}{x_\perp^{4-d}} (\phi^1)^3(x_\perp) = 0\,, 
\end{gather}
subject to the behavior \eqref{neardefectbehavior} for small $x_\perp$.

 To solve this equation we perform the following change of variables
\begin{gather}
    \phi^1(x_\perp)=  -\frac{g(\eps \log(x_\perp))}{4\pi x_\perp^\frac{d-2}{2}}\,,
\end{gather}
that leads to the following equation for the function $g(\eps \log(x_\perp))$,
\begin{gather}
 \dot{g} - \eps \ddot{g} =  \frac{1}{4}(d-2) g - \frac{\lambda_*}{16\eps \pi^2} g^3 \, . 
\end{gather}
Taking into account that we work in the limit $d \to 4$ and $\lambda_* \propto (4-d)$, we can neglect the term containing $\ddot{g}$ in the above equation and obtain the following
\begin{gather}
    \eps \dot{g} = \frac{\eps}{2} g - \frac{\lambda_*}{16\pi^2} g^3 \, .
\end{gather}
Notice that this coincides with the RG equation for the renormalized coupling constant $h$ \eqref{hbeta} and consequently we conclude that in the limit $x_\perp \to \infty$,
\begin{gather}
    \phi^1(x_\perp) \to -\frac{h_*}{4\pi x_\perp^\frac{d-2}{2}}\,.
\end{gather}

Next, we consider the case when the two defects are anti-aligned and the coupling constants have the opposite fixed point values $h_{1} = - h_{2} = h_{*}$. Close to the defects,  we have the following approximation
\begin{gather}
\phi^1(\vec{x}, z)=\phi_{(0)}(\vec{x},z) =\frac{z h_* r}{ 4 \pi \left(\vec{x}^2+z^2\right)^\frac{d+2}{4}}\,,\quad  |\vec{x}|,|z| \lesssim r  \, .  \label{eq:ApproxFusion2}
\end{gather}
Let us compute the first correction to this behaviour for large $x_\perp$ treating the interaction term $\lambda_*$ in \eqref{resumbySD} perturbatively as in 
\ie 
\phi^1 = \phi_{(0)} + \lambda_* \phi_{(1)} +\ldots\,.
\fe 
The solution is 
\ie  
\phi_{(1)} = \frac{ h_*^3 r^3}{7 (4\pi)^3  (4-d) x_\perp^{\frac{3}{2}d+1}} \left(z^3 + \frac{6}{5(d-6)} z x_\perp^2 \right) \ll \phi_{(0)}\,, \quad x_\perp \to \infty \, ,
\fe 
therefore we can conclude that the profile of $\phi^1$ does not receive large corrections in $x_\perp \to \infty$, and we can trust the solution $\phi^1(\vec{x}, z)=\phi_{(0)}(\vec{x},z)$ even far away from the pair of defects. Since $\phi_{(0)}$ vanishes in the limit $r \rightarrow 0$, we conclude that the fused defect is trivial.

Notice that vanishing of the one-point function of $\phi^1$ does not rule out the possibility that the fused defect $\cD_1\circ \cD_2$ could be a direct sum $\cD(\hat{n})\oplus \cD (- \hat{n})$ for $\hat n=(1,0,\dots,0)$.\footnote{We thank Zohar Komargodski for a question and discussions on this point.} To exclude this possibility, we need to show that the one-point function of the $\mZ_2$ even operator $(\phi^1)^2$ also vanishes (and $\la \cD(\hat n) (\phi^1)^2(x)\ra=\la \cD(-\hat n) (\phi^1)^2(x)\ra$ is nonzero). However, to the order we are working at, the one-point function of $(\phi^1)^2$ is proportional to the square of that for $\phi^1$ and in particular $\la \cD_1 (\phi^1)^2(x)\ra$ is given by the square of \eqref{onepf} and $\la \cD_1\cD_2 (\phi^1)^2(x)\ra$ is given by the square of \eqref{eq:ApproxFusion2} in the limit $r \rightarrow 0$ (the corrections to this start at order $\lambda^2$). This implies  the fused defect is indeed trivial.

To sum up, if $h_{1*}+h_{2*} = 0$, then the one-point function of $\phi^1$ and $(\phi^1)^2$ (similarly for the $O(N)$ singlet $(\phi^I)^2$) fall off faster than $1/ x_\perp^{d-2\over 2}$ and $1/x_\perp^{d-2}$ respectively. This implies that no defect is detectable in the limit of long distances, resulting in a trivial defect. On the other hand, if we start with $h_{1*}+h_{2*} \neq 0$ then  $\phi^1(x_{\perp})$ would flow back to the conformal solution described by $h= h_* \operatorname{sgn}(h_{1*}+h_{2*})$.
Thus we have established the following fusion rules 
\ie 
        \cD(\hat n) \circ \cD(\hat n) = \cD(\hat n) \, , \quad  \cD(-\hat n)\circ \cD(-\hat n) =\cD(-\hat n)\, , \quad \cD(\hat n) \circ \cD(-\hat n) = \id \,,
\fe 
confirming \eqref{magnetfusion} for the special case where the two defects are aligned or anti-aligned in the $O(N)$ directions.

\subsection{Symmetry Enhancement}
We can generalize the above computations to a situation when the two original defects are coupled to different components of $\phi^I$, for instance $\phi^2$ and $\phi^1$. We have the same action as in \eqref{WFcoupledaction}, but now $h_{2, b}$ couples to $\phi^2$ instead. In that situation, we start with having only an unbroken $O(N-2)$ symmetry. But as we show now, at long distances (or equivalently, in the conformal limit for the fused defect), it gets enhanced back to $O(N-1)$. The resulting one-point function of $\phi^1$ to leading order in the coupling constant $\lambda$ is computed by the following,
\begin{equation}
\begin{split}
&\langle \cD_1 \cD_2 \phi^1 (x) \rangle \\
& =- \frac{h_{1,b} a_D^{\frac{d-4}{2}}\Gamma\left( \frac{d - 3}{2} \right)}{4 \pi^{\frac{d - 1}{2}} (\vec{x}^2 + z^2 )^\frac{d - 3}{2}}+ \frac{\lambda_b \Gamma\left( \frac{d - 3}{2} \right)^3 h_{1,b}^3 \Lambda_T^{4-d} a_D^{\frac{3(d-4)}{2}}}{2 (3 d - 11) (4 - d) (4 \pi^{\frac{d - 1}{2}})^3 (\vec{x}^2 + z^2 )^{\frac{ 3 d - 11}{2}}}  + \lambda_b  h_{1,b} h_{2, b}^2 \Lambda_T^{4-d} a_D^{\frac{3(d-4)}{2}}I_2 \,. 
\end{split}
\end{equation}
As discussed in the previous subsection, when we study these defects at finite transverse distance $r \gg a_D$, the coupling $h_{1,b}$ gets resummed to $h_{\pm}$ (when we sum over all leading diagrams stemming from the same defect). This yields again naively
\begin{equation} \label{onepfSymmEnhW}
    \langle \cD_1 \cD_2 \phi^1 (x) \rangle = - \frac{h_{1*} }{4 \pi \sqrt{\vec{x}^2 + z^2 }} = \mp \frac{\sqrt{N + 8}}{4 \pi \sqrt{\vec{x}^2 + z^2 }} \, ,
\end{equation}
and the same holds for the other component 
\begin{equation} \label{onepfSymmEnhW2}
    \langle \cD_1 \cD_2 \phi^2 (x) \rangle = - \frac{h_{2*} }{4 \pi \sqrt{\vec{x}^2 + z^2 }} = \mp \frac{\sqrt{N + 8}}{4 \pi \sqrt{\vec{x}^2 + (z - r)^2 }} \,,
\end{equation}
which are valid close to either one of the defects but corrections are important for more general $x$ including when $x_\perp\gg r$.
 To see this, note that we have the following RG equations for $h_1$ and $h_2$ 
\begin{equation}
    \begin{split}
         \mu \frac{d h_1}{d \mu} &= - \frac{\epsilon}{2} h_1 + \frac{\lambda}{16 \pi^2} h_1 (h_1^2 + h_2^2) \, , \\
         \mu \frac{d h_2}{d \mu} &= - \frac{\epsilon}{2} h_2 + \frac{\lambda}{16 \pi^2} h_2 (h_1^2 + h_2^2) \,.
    \end{split}
\end{equation}
This implies that at the fixed point for the coupled two-defect system $h_1^2 + h_2^2 = h_*^2$ to leading order in $\eps$ and the naive solution from $h_1^2 + h_2^2 = 2 h_*^2$ is not allowed. Thus we have to take into account diagrams that involve both of the defects. 

As in the previous subsection, we will resum these diagrams using the Schwinger-Dyson equation with boundary conditions specified by the defects. In fact, we can tackle the more general case when $h_1$ couples to $\hat n$ direction of the field $\vec\phi$ while $h_2$ couples to the component along $\hat m$. In this case, the equation reads
\ie  
    \Box \vec{\phi} - \frac{\lambda_*}{x_\perp^{4-d}}(\phi^I \phi^I) \vec{\phi} = 0\,, 
    \label{genSDphi}
\fe 
with the following boundary behavior near the defects,
\ie
    \vec{\phi}(x) =\begin{cases}
        -\frac{h_* \hat{n}}{4\pi \left(\vec{x}^2+z^2\right)^\frac{d-2}{4}}\,, 
~ \vec{x},z \to 0\,, 
\\
 -\frac{h_* \hat{m}}{4\pi \left(\vec{x}^2+(z-r)^2\right)^\frac{d-2}{4}}\,,~ \vec{x},(z-r) \to 0 \, .
    \end{cases} 
\fe 
We can solve the above equation following the similar steps as the previous section. Far away from the defects, equation \eqref{genSDphi} becomes,
\begin{gather}
    (\vec{\phi})''(x_\perp) + \frac{d-2}{x_\perp} (\vec{\phi})'(x_\perp) -  \frac{\lambda_*}{x_\perp^{4-d}} (\phi^I \phi^I) \vec{\phi} = 0\,, 
\end{gather}
with the following boundary condition in the near defects region,
\begin{gather}
    \vec{\phi}(x) = -\frac{h_*\left(\hat{n} + \hat{m}\right)}{4\pi x_\perp^\frac{d-2}{2}}\,, \quad x_\perp \to 0\,.
\end{gather}
We can then argue as we have done in the previous subsection that at large distances,
\begin{gather}
    \vec{\phi}(x) \to -\frac{h_*(\hat{n}+\hat{m})}{4\pi x_\perp^\frac{d-2}{2} \sqrt{2 (1+\hat{n} \cdot \hat{m})}}\,.
\end{gather}
This establishes the fusion rules as in \eqref{magnetfusion}
\begin{gather}
    \mathcal{D}(\hat{n}) \circ \mathcal{D}(\hat{m}) = \mathcal{D}\left(\frac{\hat{n} + \hat{m}}{\sqrt{2 (1+ \hat{n} \cdot \hat{m}})} \right) \, .
\end{gather}
Note that even though we started from a defect configuration which only preserves $O(N - 2)$ symmetry, at distances larger than the separation of the two defects, there is a symmetry enhancement and we recovers $O(N - 1)$ symmetry.

\subsection{Casimir Energy}\label{CasimirenergyO(N)magnetic defects}
Now we are in position to compute the Casimir energy of the two defects to the leading order in $\eps$. We start with the configuration of defects which are coupled to $\hat{n}\cdot \vec{\phi}$ and $\hat{m}\cdot \vec{\phi}$ respectively where $\hat n,\hat m \in S^{N-1}$. To isolate this scheme independent observable, we need to calculate the expectation value of the product of defects and normalize this quantity by the product of the expectation values of the individual defects:
\begin{equation}
    \langle \langle \cD_1(\hat n) \cD_2(\hat m) \rangle \rangle  =  \frac{\langle \cD_1 \cD_2 \rangle }{\langle \cD_1 \rangle \langle \cD_2 \rangle} =\exp\left(- \frac{\cE(\hat n ,\hat m) L}{r}\right)\,.
    \label{linedefect2pf}
\end{equation}
In the above equation, $L$ is the length of the defect, while $r$ is the separation between the two defects. We always work in the limit $L \gg r$. To leading order in $\lambda$, we have contributions from the diagrams shown in Figure \ref{1ptMagnONdiagrams1}:
\begin{figure}[t]
    \centering
    \begin{subfigure}{0.15\textwidth}   \includegraphics[width=\linewidth]{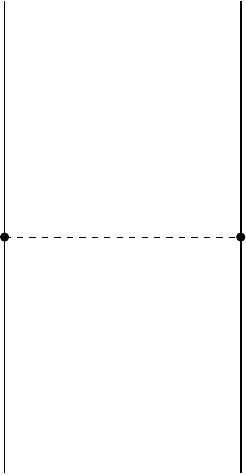}
        \caption{}
        \label{casMagn1}
    \end{subfigure}
     \hspace{1.5em}\begin{subfigure}{0.15\textwidth}
        \includegraphics[width=\linewidth]{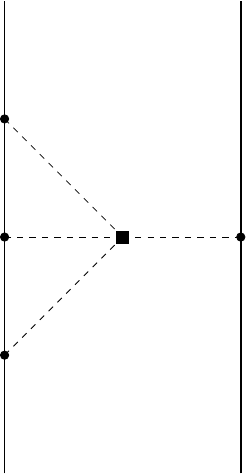}
        \caption{}
        \label{casMagn2}
    \end{subfigure}
    \hspace{1.5em} \begin{subfigure}{0.15\textwidth}
        \includegraphics[width=\linewidth]{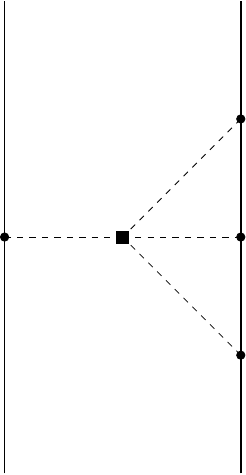}
        \caption{}
        \label{casMagn3}
    \end{subfigure}
    \hspace{1.5em} \begin{subfigure}{0.15\textwidth}
        \includegraphics[width=\linewidth]{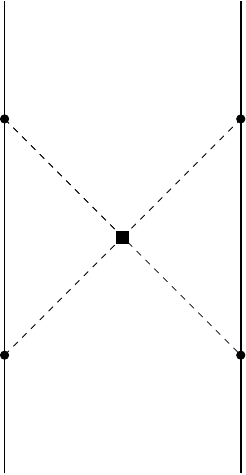}
        \caption{}
        \label{casMagn4}
    \end{subfigure}
 
    \caption{Feynman diagrams that contribute to the defect two-point function $  \langle \langle \cD_1(\hat n) \cD_2(\hat m) \rangle \rangle$ in the critical  $O(N)$ vector model.}
    \label{1ptMagnONdiagrams1}
\end{figure}
\begin{gather}
    \frac{\cE(\hat n ,\hat m) L}{r} 
    = -(\hat n\cdot\hat m)\frac{h_{1,b} h_{2,b}a^{d-4}_D L \Gamma\left( \frac{d - 3}{2} \right) }{4 \pi^{\frac{d - 1}{2}} (r^2)^{\frac{d - 3}{2}}} + \frac{\lambda_b \Lambda_T^{4-d}  a^{2(d-4)}_D L}{(r^2)^{\frac{3 d - 11}{2}}} \left( \frac{ \Gamma\left( \frac{d - 3}{2} \right)}{4 \pi^{\frac{d - 1}{2}}} \right)^4 \times  \\ \nonumber
    \bigg[\frac{2  (\hat n\cdot\hat m) h_{1,b} h_{2,b} (h_{1,b}^2 + h_{2,b}^2) \pi^{\frac{d - 1}{2}} }{\Gamma\left( \frac{d - 3}{2} \right) (4 - d) (3 d - 11) } + \frac{\left(1+2(\hat n\cdot\hat m)^2\right) h_{1,b}^2 h_{2,b}^2 \pi^{\frac{d - 1}{2} } \Gamma\left( \frac{- d + 5}{2} \right)^2 \Gamma\left( \frac{3 d - 11}{2} \right) }{2 \Gamma\left( d - 3\right)^2 \Gamma\left( 5 -  d \right)}\bigg] \, .   
\end{gather}
Notice that this calculation is being done for finite separation $r$. Consequently the diagrams that involve both defects do not diverge (as $\ep\to0$)  and there is no need to sum up an infinite series as when we were calculating the one-point functions $\la \cD_1\cD_2\phi^1(x)\ra$ in the limit $r\to 0$. As usual, when we substitute bare couplings in terms of renormalized couplings \eqref{barecouplings}, the divergences cancel and we get to leading order in $\epsilon$
\begin{equation}
\begin{split}
    &  \frac{\cE(\hat n ,\hat m) L}{r} =  -(\hat n\cdot\hat m)\frac{h_1 h_2 L }{4 \pi r } - (\hat n\cdot\hat m)\frac{\left(\gamma + \log 4 \pi + 2 \log \left(\frac{r}{a_D}\right) \right) \epsilon h_1 h_2 L}{8 \pi r} \\
    &+\frac{L}{r} \bigg[ (\hat n\cdot\hat m)\frac{\lambda  h_1 h_2 (h_1^2 + h_2^2) \left( 3 + \gamma + \log 4 \pi + 2 \log \left(\frac{r}{a_D}\right) \right) }{128 \pi^3 }
    + \frac{(1+2(\hat n\cdot\hat m)^2) \lambda h_1^2 h_2^2}{512 \pi}\bigg] \, . 
\end{split}
 \label{Casimirenergywithout diverg}
\end{equation}
After plugging in the fixed point values of the couplings \eqref{fixedpointvaluespinning}, the Casimir energy is given by
\ie 
     \cE(\hat n ,\hat m) = &   -(\hat n\cdot\hat m)\frac{N + 8}{4 \pi} - \frac{\epsilon}{4 \pi} \left( (\hat n\cdot\hat m)\frac{N^2 - 3 N  - 22 }{2( N + 8)} - \frac{\left(1+2(\hat n\cdot \hat m)^2\right) \pi^2 (N + 8)}{16}  \right) \,.
\fe 
For example, depending on if we are fusing aligned or anti-aligned defects, Casimir energies are given by
\begin{align}
    \cE_{++} &=\cE(\hat n ,\hat n) = -\frac{N + 8}{4 \pi} - \frac{\epsilon}{4 \pi} \left( \frac{N^2 - 3 N  - 22 }{2( N + 8)} - \frac{3 \pi^2 (N + 8)}{16}  \right), \\
    \cE_{+-} &=\cE(\hat n ,-\hat n) = \frac{N + 8}{4 \pi} + \frac{\epsilon}{4 \pi} \left( \frac{N^2 - 3 N  - 22 }{2( N + 8)} + \frac{3 \pi^2 (N + 8)}{16}  \right) .
\end{align}
This is consistent with the results in \cite{Rodriguez-Gomez:2022gbz}.\footnote{Our results match those of \cite{Rodriguez-Gomez:2022gbz} at the level of integrands, however there was an algebra error in the evaluations of the integrals there (see equations (46)-(50) in \cite{Rodriguez-Gomez:2022gbz}).} On the other hand, when we fuse defects that couple to different components $\phi^1$ and $\phi^2$ respectively, which corresponds to $\hat n= \hat e_1=(1,0,\dots,0)$ and $\hat m= \hat e_2=(0,1,\dots,0)$, the leading contribution to the Casimir energy comes from the order $\lambda$ term:
\begin{gather}
   \cE(\hat e_1 ,\hat e_2)=  \frac{ (N + 8) \epsilon \pi}{64}\,.
\end{gather}

\subsection{Fusing Scalar Wilson Lines in Fermionic CFTs}
\label{sec:GNline}

In this section, we extend our analysis to fermionic theories. In particular, we focus on the Gross-Neveu-Yukawa model.
\subsubsection{Gross-Neveu-Yukawa model}
\label{sec:lineinGNY}
Similar to the magnetic line defect in $O(N)$ model, there is also a scalar Wilson line in the Gross-Neveu-Yukawa model which was studied in \cite{Giombi:2022vnz}. In this section we study the  fusion of two such lines. The action for the coupled system with two insertions of these lines is given by 
\begin{align}
    S &= \int d^d x \left( \frac{(\partial_{\mu} s)^2}{2} - \left( \bar{\Psi}_i \gamma\cdot \partial \Psi^i + g_{1,b} s \bar{\Psi}_i \Psi^i\right) + \frac{g_{2,b}}{24} s^4   \right) \notag  \\
    &+ h_{1,b} \int d y s (y, z = 0, \vec{x} = 0) + h_{2,b} \int d y s (y, z = r, \vec{x} = 0)\,,
    \label{GNYlinecoupleaction}
\end{align}
where $i=1,\ldots,N_f$ and we will use $N = N_f c_d$ with $c_d$ being the number of components of a Dirac fermion in $d$ dimensions, i.e. $c_d=2^{[\frac{d}{2}]}$, where $[\frac{d}{2}
]$ is the greatest integer less than or equal to $\frac{d}{2}$. As was shown in \cite{Giombi:2022vnz}, in $d = 4 - \epsilon$ expansion, the Wilson line defect for the scalar field $s(x)$ has a fixed point. The couplings (both bulk and defect) at the fixed point  are given below,
\begin{align}
        g_{1*}^2 &= \frac{(4\pi)^2}{ N + 6 } \epsilon\, , \hspace{0.5cm} g_{2*} = \frac{(4\pi)^2 \left( -N + 6  + \sqrt{N ^2 + 132 N + 36} \right)}{6(N + 6)} \epsilon\, , \notag \\
        h_*^2 &=  \frac{108}{6 -  N + \sqrt{N ^2 + 132 N + 36}}+{H \epsilon}\,,
    \label{GNYfp}
\end{align}
{where the correction $H$ to the line defect fixed point coupling $H$ is given in \eqref{HNextorder}}. Note that the defect preserves the $U(N_f)$ global symmetry in the bulk but breaks the $\mZ_2$ parity symmetry since $s(x)$ is parity odd. Correspondingly, there are two inequivalent conformal scalar Wilson lines related by parity. We refer to them as $\cD^\pm$ which are defined by fixed point couplings $h=\pm h_*$ respectively.

Following the same reasoning  as in the scalar $O(N)$ model case, it can be shown that for $r \neq 0$, both of the lines in \eqref{GNYlinecoupleaction} must flow individually to one of the two conformal defects $\cD^\pm$. By inspecting the diagrams, it is easy to see that their fusion products are identical to the scalar theory when both the defects are aligned (or anti-aligned) in the same $O(N)$ direction. However the defect Casimir energy will be different, due to one additional diagram  (last diagram in Figure~\ref{1ptMagnGNdiagrams1}), compared to the $O(N)$ model case. For simplicity, we will choose the defect and bulk regulators to satisfy $a_D \Lambda_T=1$ below.

\subsubsection{Casimir Energy}
\begin{figure}[t]
    \centering
    \begin{subfigure}{0.13\textwidth}   \includegraphics[width=\linewidth]{FusionCasimirFreeMagnON1.pdf}
        \caption{}
        \label{casMagnGN1}
    \end{subfigure}
     \hspace{1.5em}\begin{subfigure}{0.13\textwidth}
        \includegraphics[width=\linewidth]{FusionCasimirFreeMagnON2.pdf}
        \caption{}
        \label{casMagnGN2}
    \end{subfigure}
    \hspace{1.5em} \begin{subfigure}{0.13\textwidth}
        \includegraphics[width=\linewidth]{FusionCasimirFreeMagnON3.pdf}
        \caption{}
        \label{casMagnGN3}
    \end{subfigure}
    \hspace{1.5em} \begin{subfigure}{0.13\textwidth}
        \includegraphics[width=\linewidth]{FusionCasimirFreeMagnON4.pdf}
        \caption{}
        \label{casMagnGN4}
    \end{subfigure}
    \hspace{1.5em} \begin{subfigure}{0.13\textwidth}
        \includegraphics[width=\linewidth]{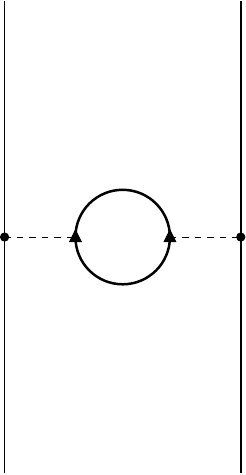}
        \caption{}
        \label{casMagnGN5}
    \end{subfigure}
 
    \caption{ Feynman diagrams that contribute to the defect two-point function $\langle \langle \cD_1 \cD_2 \rangle \rangle$ in the Gross-Neveu-Yukawa model. Here solid lines represent fermion propagators.}
    \label{1ptMagnGNdiagrams1}
\end{figure}

Similarly to the scalar case in Section~\ref{CasimirenergyO(N)magnetic defects} the Casimir energy is defined as:
\begin{equation}
    \langle \langle \cD_1 \cD_2 \rangle \rangle  =  \frac{\langle \cD_1 \cD_2 \rangle }{\langle \cD_1 \rangle \langle \cD_2 \rangle} =\exp\left(- \frac{\cE L}{r}\right)\, ,
    \label{linedefect2pfF}
\end{equation}
and as before, $L$ is the length of the defect while $r$ is the separation between the defects. To leading order in $\epsilon$, we have contributions from the diagrams shown in Figure~\ref{1ptMagnGNdiagrams1}, 
\begin{align}
\label{GNcme}
   - \frac{\cE L}{r} &= 
     h_{1,b} h_{2,b} \int dy_1 dy_2 \langle s_0 s_r \rangle - \frac{g_{2,b} h^3_{1,b} h_{2,b}}{24} \int dy_1 dy_2  d^d x \langle (s_0)^3 s_r s^4(x)  \rangle \notag \\     &- \frac{g_{2,b} h_{1,b} h^3_{2,b}}{24} \int dy_1 dy_2  d^d x \langle s_0 (s_r)^3 s^4(x)  \rangle - \frac{g_{2,b} h^2_{1,b} h^2_{2,b}}{16} \int dy_1 dy_2  d^d x \langle (s_0)^2 (s_r)^2 s^4(x)  \rangle 
     \notag \\
    &+ \frac{g_{1, b}^2 h_{1,b} h_{2,b}}{2} \int dy_1 d y_2  d^d x d^d y\langle s_0 s_r \bar{\Psi}_i {\Psi^i} s (x)   \bar{\Psi}_i {\Psi^i} s (y)  \rangle    \,,
\end{align}
where we have introduced the shorthand $s_0 = s(y,z=0,\vec{x}=0)$ and $s_r=s(y,z=r,\vec{x}=0)$ for the field $s$ inserted at two different line defects. 
Performing the integrations, we get 
\begin{equation}
\begin{split}
   - \frac{\cE L}{r} &= \frac{h_{1,b} h_{2,b} L \Gamma\left( \frac{d - 3}{2} \right) }{4 \pi^{\frac{d - 1}{2}} (r^2)^{\frac{d - 3}{2}}} - \frac{g_{2,b} L}{ 6 (r^2)^{\frac{3 d - 11}{2}}} \left( \frac{ \Gamma\left( \frac{d - 3}{2} \right)}{4 \pi^{\frac{d - 1}{2}}} \right)^4 \bigg[ \frac{2 h_{1,b} h_{2,b} (h_{1,b}^2 + h_{2,b}^2) \pi^{\frac{d - 1}{2}} }{\Gamma\left( \frac{d - 3}{2} \right) (4 - d) (3 d - 11) }  \\
    &+ \frac{3 h_{1,b}^2 h_{2,b}^2 \pi^{\frac{d - 1}{2} } \Gamma\left( \frac{- d + 5}{2} \right)^2 \Gamma\left( \frac{3 d - 11}{2} \right) }{2 \Gamma\left( d - 3\right)^2 \Gamma\left( 5 -  d \right)}\bigg]   - \frac{g_{1, b}^2 h_{1,b} h_{2,b} L N \Gamma \left( \frac{d}{2} - 1 \right)^2 \Gamma \left(d - \frac{7}{2} \right) }{64 \Gamma(d - 2) (4 - d) \pi^{d - \frac{1}{2}} (r^2)^{d - \frac{7}{2}} }\,.  
\end{split}
\end{equation}
Plugging in the bare coupling constants in terms of the renormalized one  \cite{Giombi:2022vnz},
\begin{equation}
    h_b = \mu^{\frac{\epsilon}{2}} \left( h + \frac{g_2 h^3}{192 \pi^2 \epsilon} + \frac{g_1^2 h N}{32 \pi^2 \epsilon} \right), \hspace{0.4cm} g_{1, b} = \mu^{\frac{\epsilon}{2}} \left( g_1 + O(g_1^3, g_2^2) \right) , \hspace{0.54cm} g_{2, b} = \mu^{\epsilon} \left( g_2 + O(g_1^3, g_2^2) \right) \,, 
\end{equation}
and expanding in $\epsilon$, we obtain the following 
\ie 
\label{eq:CasimirGNYcoupling}
 &   - \frac{\cE L}{r} = \frac{L}{r} \bigg[\frac{h_1 h_2 }{4 \pi } + \frac{\left(\gamma + \log 4 \pi + 2 \log (\mu r) \right) \epsilon h_1 h_2 }{8 \pi } - \frac{g_2  h_1^2 h_2^2}{1024 \pi } +    \\
    & \frac{g_1^2 h_1 h_2 N  \left(\psi ^{(0)}\left(\frac{1}{2}\right) -2 \log (\mu r)-2-\log (\pi ) \right) }{128 \pi^3 } - \frac{g_2 h_1 h_2 (h_1^2 + h_2^2) \left( 3 + \gamma + \log 4 \pi + 2 \log (\mu r) \right) }{768 \pi^3 }  \bigg] \,.
\fe 
Note that for simplicity, here we are using the RG scheme such that $\Lambda_T = 1/a_{D} = \mu$ so we only have a single RG scale in the problem. 

In order to compute the Casimir energy from \eqref{eq:CasimirGNYcoupling} to order $\epsilon$, we need the fixed point coupling $h_\star$ for the GNY line defect to order $\epsilon$ which was previously studied in \cite{Pannell:2023pwz,Barrat:2023ivo}.\footnote{We thank Simone Giombi and Anurag Pendse for pointing out a mistake in the previous version of the paper and for pointing out these references.}  For completeness, we provide an alternative derivation in Appendix \ref{FixedPointNextOrder} with additional consistency checks and record the result here,\footnote{Note that the previous reference \cite{Barrat:2023ivo} contains a mistake in their (B.11) for the beta function of the defect coupling $h$. We thank Simone Giombi and Anurag Pendse for discussions on this point.}
\begin{gather}
    h^2_{\star}=\frac{108}{6 -  N + \sqrt{N ^2 + 132 N + 36}}+H \epsilon
\end{gather}
where
\begin{gather}
    H=\frac{31 N^3+8739 N^2+180954 N+31536}{2 (N+6) \left(-N^3-126 N^2+756 N+216+\left(N^2+60 N+36\right) \sqrt{N^2+132 N+36}\right)}\nonumber\\
    +\frac{\sqrt{N^2+132 N+36} \left(-31 N^2+\left(648 \pi ^2-3453\right) N+5256\right)}{2 (N+6) \left(-N^3-126 N^2+756 N+216+\left(N^2+60 N+36\right) \sqrt{N^2+132 N+36}\right)}\,.
    \label{HNextorder}
\end{gather}
Plugging in the fixed point values of couplings \eqref{GNYfp}, we obtain the following results to leading order in $\epsilon$, depending on the type of the line defect $\cD^\pm$ in the Gross-Neveu-Yukawa model,  
\begin{align}
    &\cE_{++} =  -\frac{27}{ \pi(6 -  N + \sqrt{N ^2 + 132 N + 36})}-\epsilon\frac{H}{4\pi} - \frac{27 \epsilon \left(8 N+9 \left(8+\pi ^2\right)\right)}{8 \pi  (N+6) (N-6 - \sqrt{N^2 + 132 N + 36})}\, , \notag\\
    &\cE_{+-} = \frac{ 27}{ \pi(6 -  N + \sqrt{N ^2 + 132 N + 36} ) }+\epsilon\frac{H}{4\pi} - \frac{27 \epsilon \left(- 8 N+9 \left(-8+\pi ^2\right)\right)}{8 \pi  (N+6) (N-6 - \sqrt{N^2 + 132 N + 36})}\,.
\end{align}  
Note that the $\mZ_2$ symmetry flips $\cD^+$ to $\cD^-$ and hence flips the sign of the fixed point value of the defect coupling constant. However as is clear from the \eqref{eq:CasimirGNYcoupling}, the Casimir energy is invariant under this flip and hence we have $\cE_{--}=\cE_{++}$ and $\cE_{-+}=\cE_{+-}$.

\section{Fusing Interfaces and Boundaries in $d>2$}
\label{sec:ONinterface}
In this section, we study fusion of codimension-one defects. 
In particular, we focus on factorized interfaces $\cI_{\cB_1\cB_2}\equiv |\cB_1\ra \la \cB_2|$ constructed from a pair of conformal boundaries $|\cB_1\ra$ and $|\cB_2\ra$.  The nontrivial information in the fusion product of such interfaces is contained in the fusion of corresponding conformal boundaries, with nontrivial Casimir energy. To determine the boundary Casimir energy, we will make extensive use of wedge geometry and study CFT inside a wedge. In Section~\ref{eq:on4eps} we review various boundary conditions of the scalar $O(N)$ model. We then discuss the simple case of free scalars in a wedge and then move on to describe the critical $O(N)$ model in the same configuration. Along the way, we will also clarify some confusing claims and issues discussed in the literature about the critical $O(N)$ model in a wedge. In Section~\ref{sec:ExtraOrd} we consider the case with extraordinary boundary condition which only exists for the critical $O(N)$ model, and study its fusion with other boundary conditions by recasting it into a particular classical mechanical problem. Then in Section~\ref{sec:GNBoundary} we generalize our calculations to fermionic theories and study the fusion of conformal boundaries in free fermions and the Gross-Neveu model. Finally in Section~\ref{sec:NumericalLargeN}, we make progress towards studying defect fusion directly in $d=3$ by numerically solving the large $N$ $O(N)$ model in a three dimensional wedge.

\subsection{$O(N)$ Model in $d = 4 - \epsilon$}
\label{eq:on4eps}
Here we will consider fusion of boundary conditions for the critical $O(N)$ model. There are two well-known $O(N)$ symmetric conformally invariant boundary conditions, which are Dirichlet and Neumann. In the free theory, they are characterized respectively by either vanishing of the field or its normal derivative at the boundary 
\begin{equation}
    \phi|_{\rm bdy} = 0 \quad  \textrm{or} \quad \partial_{\perp} \phi|_{\rm bdy} = 0 \, .
    \label{freeDN}
\end{equation}
The interacting theory defined by the familiar $O(N)$ invariant action,
\begin{equation}
    S= \int d^d x \left( \frac{1}{2} (\partial_{\mu} \phi^I)^2 + \frac{\lambda}{4} (\phi^I \phi^I)^2 \right) \, .
    \label{actionON}
\end{equation}
has a perturbative fixed point in $d = 4 - \epsilon$ dimensions as we have reviewed around \eqref{fixedpointvaluespinning}. At  the interacting bulk fixed point, the Dirichlet and Neumann boundary conditions in \eqref{freeDN} continue perturbatively into what are known as ordinary (O) and special (sp) boundary fixed points respectively. In addition to these two, there is an additional boundary condition in the interacting theory called the extraordinary (E) boundary condition.\footnote{We are studying the theory in $d = 4 - \epsilon$ so the distinction between extraordinary and extraordinary-log boundary conditions \cite{Metlitski:2020cqy} will not be important for us. We will also not study the extraordinary-log interface because it continues to a dimension two interface and not a codimension one interface in $d = 4 - \epsilon$  \cite{Krishnan:2023cff, Giombi:2023dqs, Raviv-Moshe:2023yvq, Trepanier:2023tvb }.} This is characterized by a non-zero vacuum expectation value for the field $\phi^I$ and spontaneously breaks the $O(N)$ symmetry to $O(N-1)$. This non-zero vacuum expectation value is given by the following solution in flat half-space to the classical equation of motion
\begin{equation}
    \langle \phi^N (x) \rangle = \pm\sqrt{\frac{2}{\lambda}} \frac{1}{z} \, , 
\end{equation}
where the boundary is located at $z= 0$ and we picked $\phi^N$ to be the field that gets non-zero vacuum expectation value. In the absence of extra interactions localized on the boundary, these are all the known conformal boundary conditions in this model. For more details on boundary conditions in the $O(N)$ model see \cite{Die86a, Die97, cardy_1996, Diehl:1981zz, McAvity:1993ue, McAvity:1995zd, PhysRevLett.38.1046, 10.1143/PTP.70.1226, McAvity:1995zd, Liendo:2012hy, Gliozzi:2015qsa, Bissi:2018mcq, Kaviraj:2018tfd, Carmi:2018qzm, Giombi:2020rmc} and references therein. 

The Casimir energy for fusion of conformal boundaries in the critical $O(N)$ model for ordinary and special boundary conditions was considered before by studying the model in a slab geometry in \cite{ PhysRevA.46.1886, Diehl:2006mz, Gruneberg:2007av, Diehl:2011sy, Diehl:2014bpa}. In particular, for a slab of width $r$, the free energy per unit area inside the slab for small $r$ should go like 
\begin{equation}
    F/A = \frac{\cE}{r^{d-1}}\, .
\end{equation}
The coefficient $\cE$, called the Casimir energy, for the case of $O(N)$ model with different pairs of boundary conditions (special or ordinary)  was found to be \cite{Diehl:2011sy}
\begin{align} 
    \cE^{\left(\textrm{sp}, \textrm{sp}\right)} &= - \frac{N \pi^2}{1440}  + \frac{N \epsilon \pi^2}{2880} \left( 1 - \gamma - \log (4 \pi) + \frac{2 \zeta'(4)}{\zeta(4)} + \frac{5(N + 2)}{2 (N + 8) } \right) - \frac{N \pi^2 \epsilon^{\frac{3}{2}}}{72 \sqrt{6}} \left(  \frac{N + 2}{N + 8}\right)^{\frac{3}{2}} \, , \notag \\
    \cE^{(\textrm{O}, \textrm{O})} &= - \frac{N \pi^2}{1440}  + \frac{ N \epsilon \pi^2}{2880} \left( 1 - \gamma - \log (4 \pi) + \frac{2 \zeta'(4)}{\zeta(4)} + \frac{5(N + 2)}{2 (N + 8) } \right) + O(\epsilon^2) \, ,  \label{CasimirEnONRes} \\
    \cE^{(\textrm{sp}, \textrm{O})} &= \frac{ 7 N \pi^2}{11520} + \frac{N \epsilon \pi^2}{23040} \left( \frac{5(N + 2)}{N + 8} + 6 \log 4 - 7 \left( 1 - \gamma - \log (\pi) + \frac{2 \zeta'(4)}{\zeta(4)} \right) \right) + O(\epsilon^2) \, . \notag
\end{align}
Notice that for the $(\textrm{sp}, \textrm{sp})$ case, the expansion involves a non-analytic term of order $\epsilon^{3/2}$, that correspond to the presence of a zero mode (see discussion in the later part of Section~\ref{sec:freeEonwedge}).  

Here we will reproduce the above and provide new results when one of the boundaries is an extraordinary boundary condition by studying the $O(N)$ model in a wedge geometry \cite{JLCardy_1983, Antunes:2021qpy, Bissi:2022bgu}. As explained in Section~\ref{sec:Casimirandwedge}, at small angles, the wedge imitates the slab and can be used to calculate the Casimir energy. Furthermore, to facilitate the computation, we will also map the wedge to hyperbolic space as described in Section~\ref{sec:Casimirandwedge}. The geometry we use may be described using cylindrical coordinates as below,
\begin{equation}
\label{wedgecoords}
    ds^2 =  d \vec{y}^2 + d \rho^2 + \rho^2 d \varphi^2 = \rho^2 \left( \frac{d \vec{y}^2 + d \rho^2}{\rho^2} + d \varphi^2 \right) = \rho^2 ds^2_{S^1_{\theta} \times \mH^{d-1}} \,,
\end{equation}
where $\vec{y}$ are the coordinates on the corner of the wedge, $\rho$ is the distance from the corner and $0 < \varphi < \theta$ is the angular coordinate around the corner.

\subsubsection{Free Scalar in a Wedge}
We first discuss a free massive scalar in the geometry of $S_\theta^1 \times \mH^{d - 1}$. Having in mind that we will soon move on to the interacting case, we will refer to the Dirichlet and Neumann boundary conditions in \eqref{freeDN} as ordinary and special respectively. With respect to these boundary conditions, we have the following mode decompositions of the scalar field $\phi$ on $S_\theta^1$:  
\begin{equation}
\begin{split}
    \phi(x) =  \sqrt{\frac{2}{\theta}} \sum_{n = 1}^{\infty} \phi_n (\rho, \vec{y}) \sin \left (\frac{n \pi \varphi}{\theta} \right) \hspace{1cm}  &(\textrm{O}, \textrm{O}) \, , \\ 
    \phi(x) = \frac{\phi_0 (\rho, \vec{y})}{\sqrt{\theta}} +   \sqrt{\frac{2}{\theta}} \sum_{n = 1}^{\infty} \phi_n (\rho, \vec{y}) \cos \left (\frac{n \pi \varphi}{\theta} \right) \hspace{1cm}  &(\textrm{sp}, \textrm{sp}) \, ,  \\
    \phi(x) =   \sqrt{\frac{2}{\theta}} \sum_{n = 0}^{\infty} \phi_n (\rho, \vec{y}) \sin \left (\frac{(n + \frac{1}{2}) \pi \varphi}{\theta} \right) \hspace{1cm}  &(\textrm{O}, \textrm{sp}) \, .
\label{KKreduction}
\end{split}
\end{equation}
Next, we calculate the two-point function of the field $\phi$ which will be used later to study the interacting theory. Let us first focus on the (O,O) case and the other cases will be similar. After performing a Kaluza-Klein reduction on the $S_\theta^1$, we obtain the following action for massive scalars $\phi_n$ on the $\mH^{d - 1}$,
\begin{equation}
\label{modesaction}
\begin{split}
     S &=  \frac{1}{2} \sum_{n = 1}^{\infty} \int \frac{d \rho d \vec{y}}{\rho^{d-1}} \left[(\partial_{i} \phi_n)^2 + \left(m^2 + \frac{n^2 \pi^2 }{\theta^2}  \right) \phi_n^2 \right]\,,
\end{split}
\end{equation}
where $m^2$ is the mass term for the mode $\phi_n$ on $S^1_\theta\times \mH^{d-1}$.

The two-point function of the field $\phi$ on the hyperbolic cylinder is given by a sum over bulk-to-bulk propagators of each mode $\phi_n$,
\begin{equation}
    \langle \phi (x_1) \phi(x_2)  \rangle = \frac{2}{\theta} \sum_{n = 1}^{\infty} \sin \left (\frac{n \pi \varphi_1 }{\theta} \right) \sin \left (\frac{n \pi \varphi_2 }{\theta} \right) G_{\Delta_n} (\xi), \quad \xi = \frac{\vec{y}_{12}^{2} + \rho_{12}^2}{4 \rho_1 \rho_2}\,,
    \label{phi2pf}
\end{equation}
where $\xi$ is a cross-ratio and $\Delta_n$ is the scaling dimension of the corresponding operator
on the boundary of the $\mH^{d-1}$ (i.e. corner of the wedge). The scaling dimension $\Delta_n$ is related to the mass of $\phi_n$ by the usual holographic dictionary on $\mH^{d-1}$,
\ie 
\Delta_n(\Delta_n - d+2) = m^2 + {n^2 \pi^2\over \theta^2}\,.
\label{dimmassrelation}
\fe

For our purpose here, the scalar $\phi$ is conformally coupled on $S^1_\theta\times \mH^{d-1}$ with scalar curvature $R=-(d-1)(d-2)$ and therefore the induced mass satisfies,
\ie 
m^2={(d-2)\over 4(d-1)}R = -{(d-2)^2\over 4}\,,
\label{confmass}
\fe
which fixes the corner scaling dimensions via \eqref{dimmassrelation}\footnote{These are scaling dimensions of the operators located at the boundary of $\mathbb{H}^{d-1}$, so in the original setup, at the intersection of two boundaries that make up the wedge.},
\ie 
 \Delta_n = \frac{d}{2} - 1 + \frac{n \pi}{\theta}\,.
 \fe 
Note that the other solution $\Delta_n = {d\over 2} - 1 - {n\pi\over \theta}$ to \eqref{dimmassrelation} for the conformal case is forbidden because it would violate the unitarity bound for $\theta < \pi$ (we will be interested in small $\theta$). Finally the bulk-bulk propagator for general dimensions $\Delta$ is given by 
\begin{equation}
    G_{\Delta} (\xi) = \frac{\Gamma \left(\Delta \right)}{2 \pi^{\frac{d}{2} - 1} \Gamma \left( \Delta + 2 - \frac{d}{2} \right) (4 \xi)^{\Delta} } {}_2 F_1 \left( \Delta, \Delta - \frac{d-3}{2}, 2 \Delta - d + 3; - \frac{1}{\xi} \right).
\end{equation}
From this two-point function \eqref{phi2pf} in the wedge we can extract the one-point function of the operator $\phi^2$. We take the short distance $(\xi \rightarrow 0, \varphi_1 \rightarrow \varphi_2)$ limit of \eqref{phi2pf} and use the MS scheme to obtain
\begin{equation} \label{OnePointScalarWedge}
    \langle \phi^2 (\varphi) \rangle =  \frac{ 2 \Gamma \left(\frac{3 - d}{2}\right) }{ (4 \pi)^{\frac{d - 1}{2}} \theta }  \sum_{n = 1}^{\infty} \sin^2 \left (\frac{n \pi \varphi }{\theta} \right)  \frac{ \Gamma \left(\Delta_n \right)}{ \Gamma \left(3 - d + \Delta_n\right)}\,. 
\end{equation}
As a consistency check, we want to evaluate this sum in $d = 4$ to compare with the results of \cite{Bissi:2022bgu}. 
To this end, we separate the divergent and finite pieces of the above sum at large $n$ 
\begin{equation}
\begin{split}
     \langle \phi^2 (\varphi) \rangle &=  \frac{ 2 \Gamma \left(\frac{3 - d}{2}\right) }{ (4 \pi)^{\frac{d - 1}{2}} \theta }  \sum_{n = 1}^{\infty} \sin^2 \left (\frac{n \pi \varphi }{\theta} \right) \bigg[ \left(\frac{n \pi }{\theta}\right)^{d - 3} -\frac{(d - 4)(d - 3)(d - 2)}{24} \left(\frac{n \pi }{\theta}\right)^{d - 5} \\
     & + \frac{ \Gamma \left(\frac{d}{2} - 1 + \frac{n \pi }{\theta} \right)}{ \Gamma \left(2-\frac{d}{2} +\frac{n \pi }{\theta}\right)} - \left(\frac{n \pi }{\theta}\right)^{d - 3} + \frac{(d - 4)(d - 3)(d - 2)}{24} \left(\frac{n \pi }{\theta}\right)^{d - 5}  \bigg]\,.
\end{split}    
\end{equation}
Now the second line of the above equation is finite, and may be numerically evaluated in any dimension (it actually vanishes in $d = 4$) and the first line may be evaluated by zeta function regularization. In $d = 4$, this gives 
\begin{gather}
    \langle \phi^2 (\varphi) \rangle = \frac{\pi^2 - \theta^2}{48 \pi^2 \theta^2} - \frac{1}{16 \theta^2 \sin^2 \left( \frac{\pi \varphi}{\theta} \right)}  \hspace{1cm}  (\textrm{O}, \textrm{O}) \, , \notag\\
     \langle \phi^2 (\varphi) \rangle = \frac{\pi^2 - \theta^2}{48 \pi^2 \theta^2} + \frac{1}{16 \theta^2 \cos^2 \left( \frac{\pi \varphi}{\theta} \right)}   \hspace{1cm}  (\textrm{sp}, \textrm{sp})  \,.\label{OnePoint4dOO}
\end{gather}
For the (\textrm{sp}, \textrm{O}) case, we need to sum over half integers instead, 
\begin{equation} \label{OnePoint4dOsp}
    \langle \phi^2 (\varphi) \rangle = - \frac{2 \theta^2 + \pi^2 \left[ 1  + 6 \cot \left( \frac{\pi \varphi}{\theta} \right) \csc \left( \frac{\pi \varphi}{\theta} \right)  \right]}{96 \pi^2 \theta^2}   \hspace{1cm}  (\textrm{O}, \textrm{sp})\,.   
\end{equation}
These results agree with what was found in \cite{Bissi:2022bgu}.\footnote{Note that the one-point function of scalar operators on $S^1 \times \mH^{d-1}$ is constant along $\mH^{d-1}$ and is related by a Weyl transformation to the wedge in flat space.}

\subsubsection{Interacting Wedge in $\epsilon$ Expansion}
The $\epsilon$ expansion of the $O(N)$ model inside a wedge in $4 - \epsilon$ dimensions was studied in \cite{JLCardy_1983, Bissi:2022bgu}. Here we will only emphasize one aspect of it which was missed in the previous studies. 

As we saw in the previous section, when we impose Neumann boundary conditions on both boundaries corresponding to the (sp, sp) case, the lowest mode at the corner of the wedge is the $n = 0$ mode with scaling dimension $\Delta_0={d\over 2} - 1$. It turns out that this lowest operator at the corner acquires an anomalous dimension of order $\sqrt{\epsilon}$. This behavior was missed in the previous studies of this setup. 

We start by considering the bulk-wedge two-point function of the bulk field $\phi^I$ and a boundary operator $O^J_n$ in the fundamental representation of $O(N)$ group. This two-point function is constrained by conformal symmetry and $O(N)$ symmetry to take the following form
\begin{equation}
    \langle \phi^I (\rho, \varphi, \vec{y}) O_n^J(\vec{y}_1)  \rangle = \frac{\delta^{IJ} f(\varphi)}{\rho^{\Delta_{\phi} - \Delta_n} \left( (\vec{y} - \vec{y}_1)^2 + \rho^2 \right) ^{\Delta_n}}\,. \label{eq:confonept}
\end{equation}
Here we are again considering a wedge  in the flat space. Since the bulk field satisfies an equation of motion, this implies the following relation in the leading order in $\lambda$
\begin{equation}
    \Box  \langle \phi^I (\rho, \varphi, \vec{y}) O_n^J(\vec{y}_1)  \rangle   = \lambda_* (N + 2) \langle \phi^2 \rangle  \langle \phi^I (\rho, \varphi, \vec{y}) O_n^J(\vec{y}_1)  \rangle  \,,
\end{equation}
which can be used to deduce the dimensions of the fields $O^J_n$.  Thus, acting with the Laplacian on the equation \eqref{eq:confonept} we get
\begin{gather}
   \Box  \langle \phi^I (\rho, \varphi, \vec{y}) O_n^J(\vec{y}_1)  \rangle = \notag\\
   \left( \frac{\left( \Delta_{\phi} - \Delta_n \right)^2 + \partial_{\varphi}^2}{\rho^2}  + \frac{2 \Delta_n \left( 2 \Delta_{\phi} - d + 2 \right)}{(\vec{y} - \vec{y}_1)^2 + \rho^2} \right) \frac{\delta^{IJ} f(\varphi)}{\rho^{\Delta_{\phi} - \Delta_n} \left( (\vec{y} - \vec{y}_1)^2 + \rho^2 \right) ^{\Delta_n}}\,, 
\end{gather}
where $\Delta_{\phi} = d - 2$ to first order in $\epsilon$. Then we can solve this equation order by order in $\epsilon$ by expanding 
\begin{equation}
    f = f^{(0)} +  f^{(1)}\,, \hspace{1cm} \Delta_n = \Delta_n^{(0)} +  \gamma_n\,.
\end{equation}
Focusing now on the case of $ n = 0$ mode for (sp, sp) boundary condition, to zeroth order in $\epsilon$, this corresponds to $\Delta_n^{(0)} = \Delta_{\phi}$ and $f^{(0)}= f_0^{(0)} = \textrm{const} $. The equation for $f^{(1)}$ is 
\begin{equation}
    \epsilon \partial_{\varphi}^2 f^{(1)} + \gamma_0^2 f_0^{(0)} - \lambda_* (N + 2) f_0^{(0)} \left( \frac{\pi^2 - \theta^2}{48 \pi^2 \theta^2} + \frac{\csc^2 \left( \frac{\pi \varphi}{\theta} \right)}{16 \theta^2} \right) = 0\,,
\end{equation}
with the following solution,
\ie 
    \epsilon f^{(1)}(\varphi)  = \mathcal{A} + \mathcal{B} \varphi + \frac{\varphi^2 f_0^{(0)}}{2 } \left( \frac{\lambda_{*} (N + 2) (\pi^2 - \theta^2)}{48 \pi^2 \theta^2} - \gamma_0^2\right) - \frac{\lambda_* (N + 2) f_0^{(0)}}{16 \pi^2} \log \sin \left( \frac{\pi \varphi}{\theta} \right)\,. 
\fe 
Imposing Neumann boundary conditions at both $\varphi = 0$ and $\theta$ requires setting the following derivatives to zero
\begin{equation}
   \begin{split}
       \epsilon \partial_{\varphi}f^{(1)}(\varphi)|_{\varphi = 0} &=   - \frac{\lambda_* (N + 2) f_0^{(0)}}{16 \pi^2 \varphi}  + \mathcal{B}\,, \\
       \epsilon \partial_{\varphi}f^{(1)}(\varphi)|_{\varphi = \theta} &=   - \frac{\lambda_* (N + 2) f_0^{(0)}}{16 \pi^2 (\varphi - \theta)}  + \mathcal{B} + \theta f_0^{(0)} \left( \frac{\lambda_{*} (N + 2) (\pi^2 - \theta^2)}{48 \pi^2 \theta^2} - \gamma_0^2\right)\,.
   \end{split} 
\end{equation}
The $1/\varphi$ and $1/ (\varphi - \theta)$ terms correspond to exchanged operators at the boundaries and should not be set to zero. Since we have special boundary conditions at both the boundaries, the exchanged operator is the field $\phi$ at the boundary, i.e. $\hat{\phi}$, which has dimensions $d/2 - 1= 1$, and this explains the powers of $\varphi$ and $(\varphi - \theta)$ in the exchange terms. Setting other terms to zero then requires $\mathcal{B} = 0$ and
\begin{equation}
\gamma_0 = \sqrt{\frac{\lambda_{*} (N + 2) (\pi^2 - \theta^2)}{48 \pi^2 \theta^2}} \implies \Delta_0 = \frac{d}{2} - 1 + \sqrt{\frac{(N + 2) (\pi^2 - \theta^2) \epsilon}{6 \theta^2 (N + 8)}}\,,
\end{equation}
where we used the fixed point value of the coupling constant $\lambda_*$ given in \eqref{fixedpointvaluespinning}. Note that this $\sqrt{\epsilon}$ piece in the anomalous dimension vanishes for $\theta = \pi$, which is expected since in this case, the wedge reduces to a plane and hence results in a BCFT and there is no term of order $\sqrt{\epsilon}$ for a BCFT \cite{Giombi:2020rmc}. This $\sqrt{\epsilon}$ behavior of the anomalous dimension was missed in \cite{Bissi:2022bgu} which led them to incorrectly conclude that the $n = 0$ mode is not conformal for $\theta \neq \pi$.

Another way to see this $\sqrt{\epsilon}$ behavior of anomalous dimension is to add explicitly a $c \phi^2$ interaction localized on the corner of the wedge. This is the lowest dimension $O(N)$ symmetric operator and is classically marginal on the corner. Such an interaction for a codimension two defect was studied in the context of Renyi twist defect in Wilson-Fisher theory \cite{Metlitski:2009iyg}. They found that for $c = 0$, the operator $\phi^2$ on the wedge is actually marginally relevant in $d = 4 - \eps$. They found that there is a non-trivial stable fixed point for the coupling $c$ with the value of the coupling being of order $\sqrt{\eps}$ at the fixed point. We suspect that calculating the anomalous dimension of the lowest mode in the presence of this coupling will reproduce the $\sqrt{\eps}$ anomalous dimension that we found above using equations of motion. 

\subsubsection{Free Energy on $S_\theta^1 \times \mH^{d - 1}$}
\label{sec:freeEonwedge}
We now move on to calculate the Casimir energy for a pair of boundaries. To do that, we just need to compute the free energy on $S_\theta^1 \times \mH^{d - 1}$. In the leading order in $\lambda$ this is given by  
\ie 
    F(\theta) =&\, F_{0}(\theta) + I(\theta)\\
    =&\,
    \frac{ N \textrm{Vol}(\mH^{d-1})}{ 2 (4 \pi)^{\frac{d - 1}{2}} \Gamma(\frac{d - 1}{2})}  \int_{-\infty}^{\infty} d \nu \frac{|\Gamma(i \nu + \frac{d-2}{2})|^2}{|\Gamma(i \nu)|^2}  \sum_{n}^{\infty} \log \left( \nu^2 + \frac{\omega_n^2 \pi^2 }{\theta^2} \right) + \frac{\lambda}{4} \int  d^d x \langle \left(\phi^I \phi^I \right)^2 \rangle\,, \label{eq:CasEnHd}
\fe 
where the frequency sums are defined such that $\omega_n=1, 2, \ldots,\infty$ for (O,O), $\omega_n=0, 1, \ldots,\infty$ for (sp,sp) and $\omega_n=1/2, 3/2, \ldots,\infty$ for (O,sp). To deduce the above result, we have used that the eigenvalues of the scalar Laplacian $-\Box_{\mH^{d-1}}$ on hyperbolic space  is $\n^2 +{(d-2)^2\over 4}$ (see also \eqref{modesaction} and \eqref{confmass}) and the spectral density can be found in  \cite{Camporesi:1994ga, Bytsenko:1996rr}. We are interested in the leading term in the free energy in the expansion at very small $\theta$. To simplify the expression in \eqref{eq:CasEnHd}, we first differentiate $F_0(\theta)$  with respect to $\theta$
\begin{equation}
\begin{split}
    \frac{\partial F_{0}(\theta)}{\partial \theta} &= -\frac{ N \textrm{Vol}(\mH^{d-1})}{  (4 \pi)^{\frac{d - 1}{2}} \Gamma(\frac{d - 1}{2})} \sum_{n = 1}^{\infty}  \int_{-\infty}^{\infty} d \nu \frac{|\Gamma(i \nu + \frac{d-2}{2})|^2 \omega_n^2 \pi^2 }{|\Gamma(i \nu)|^2 \left( \nu^2 + \frac{\omega_n^2 \pi^2 }{\theta^2} \right)  \theta^3} \\
    &= -\frac{ N \textrm{Vol}(\mH^{d-1}) \Gamma(\frac{3 - d}{2}) }{  (4 \pi)^{\frac{d - 1}{2}} } \sum_{n}^{\infty}   \frac{\Gamma \left( \frac{d-2}{2} + \frac{\omega_n \pi }{\theta} \right) \omega_n^2 \pi^2 }{\Gamma \left( 2 - \frac{d}{2} + \frac{\omega_n \pi }{\theta} \right)  \theta^3}\,,
\end{split}
\end{equation}
where the integral over $\nu$ is evaluated using Cauchy theorem. To proceed further, we expand the gamma functions at small $\theta$ \cite{pjm/1102613160} and then perform the sum over $n$ which gives 
\ie 
        &\frac{\partial F_{0}(\theta)}{\partial \theta} =  -\frac{ N \textrm{Vol}(\mH^{d-1}) \Gamma(\frac{3 - d}{2}) }{  (4 \pi)^{\frac{d - 1}{2}} } \sum_{n}^{\infty}   \frac{ (n \pi)^{d - 1} }{\theta^d}\,.
        \fe 
Integrating and dropping the integration constant, which is not important in the limit $\theta \rightarrow 0$, we obtain the following for the different pairs of boundary conditions,         
        \ie 
        & F_{0}^{(\textrm{sp},\textrm{sp})}(\theta) = F_{0}^{(\textrm{O},\textrm{O})}(\theta) 
        =\frac{ N \textrm{Vol}(\mH^{d-1}) \Gamma(\frac{3 - d}{2}) }{  (4 \pi)^{\frac{d - 1}{2}} (d-1) }  \frac{  \pi^{d - 1} \zeta(1 - d) }{\theta^{d-1}}\,,  \\
        & F_{0}^{(\textrm{O}, \textrm{sp})}(\theta) =  \frac{ N \textrm{Vol}(\mH^{d-1}) \Gamma(\frac{3 - d}{2}) }{  (4 \pi)^{\frac{d - 1}{2}} }  \frac{  \pi^{d - 1} (2^{1- d} - 1) \zeta(1 - d) }{\theta^{d-1}}\,.
\fe 
As discussed in Section~\ref{sec:Casimirandwedge}, the coefficient of the ${\textrm{Vol}(\mH^{d-1})\over \theta^{d-1}}$ dependence determines the Casimir energy. To take into account the contribution  $I(\theta)$ of the interaction term in \eqref{eq:CasEnHd}, at leading order in $\epsilon$, we may just use the one-point functions in the free theory from \eqref{OnePoint4dOO} and \eqref{OnePoint4dOsp}. For instance, for $(\textrm{O}, \textrm{O})$ and $(\textrm{sp},\textrm{sp})$, we have
\begin{equation}
    \frac{\lambda}{4} \int  d^d x \langle \left(\phi^I \phi^I \right)^2 \rangle \bigg|_{\theta \ll 1} = \frac{ \textrm{Vol}(\mH^{d-1}) N (N + 2) \epsilon \pi^2  }{1152 (N + 8) \theta^3 }\,. 
\end{equation}
Combining the free and the interaction pieces, and expanding in $d = 4 - \epsilon$, we indeed recover the results in \eqref{CasimirEnONRes}, up to the $\epsilon^{3/2}$ piece for the $(\textrm{sp},\textrm{sp})$ case which we will come to below.

In the case of (\textrm{sp},\textrm{sp}) the situation is more subtle because of the presence of a zero-mode giving rise to the non-analytical contributions.  To properly take it into account we should first find the spectrum of the following differential equation
\begin{gather}
    \left(-\partial^2_{\varphi}-\Box_{\mH^{d-1}}+\lambda_{*}\left(\phi^J \phi^J\right)\right)\phi^{I}=E \phi^{I}\,.
\end{gather}
The above equation implies that bulk-wedge two-point function should satisfy
\begin{gather}
    \left[-\partial^2_{\varphi}-\Box_{\mH^{d-1}}+\lambda_{*}(N+2)\langle \phi^{2}\left(\varphi\right)\rangle\right] \langle \phi^{I}\mathcal{O}(\vec{y}_{\parallel})\rangle=E\langle \phi^{I}\mathcal{O}(\vec{y}_{\parallel})\rangle\,,
\end{gather}
where $\mathcal{O}(\vec{y}_{\parallel})$ is an arbitrary boundary operator, and we have used Wick theorem to compute the interaction term to the leading expansion for small $\epsilon$. 
By implementing the reduction of the scalar field on $S^1$  as in \eqref{KKreduction},  we  obtain the following equation for the individual wedge modes,
\begin{gather}
    \left[-\Box_{\mH^{d-1}}+\frac{\pi^{2}n^2}{\theta^{2}}+\lambda_{*}(N+2)\langle \phi^{2}(\varphi)\rangle\right] \langle \phi_{n}^{I}\mathcal{O}(\vec{y}_{\parallel})\rangle=E_{n}\langle \phi_{n}^{I}\mathcal{O}(\vec{y}_{\parallel})\rangle\,.
\end{gather}
For non-zero $n$, we can safely neglect the interaction term in the equation since we are working to leading order in $\ep$, and deduce  that $E_{n \neq 0}=\nu^{2}+\frac{\pi^{2}n^2}{\theta^{2}}$. However, for $n=0$, we must include the first correction to the zero mode energy $E_{0}$ by  $\lambda_{*}$ given below,
\ie 
 E_{0}=\nu^{2}+E^{(1)}_{0}~~{\rm with}~~  E^{(1)}_{0}
   =\dfrac{\lambda_{*}(N+2)}{\theta}\int \limits_{0}^{\theta}d\varphi \langle \phi^{2}\left(\varphi\right)\rangle \,,
   \label{energycorrecONn=0}
\fe 
where the one-point function in $d=4$ is given in \eqref{OnePoint4dOO}. To regularize the divergent integral in \eqref{energycorrecONn=0} we use dimensional regularization,
\begin{gather}
    \int \limits_{0}^{\theta}d\varphi \dfrac{1}{\cos^{2}{\left(\frac{\pi \varphi}{\theta}\right)}} \Longrightarrow  \int \limits_{0}^{\theta}d\varphi \dfrac{1}{\left[\cos{\left(\frac{\pi \varphi}{\theta}\right)}\right]^{\alpha}}\,.
    \label{dimreg1ptfunct}
\end{gather}
After performing the integration, and setting $\alpha=2$ it can be shown that the integral in \eqref{dimreg1ptfunct} vanishes and thus only the $\varphi$ independent part of $\la \phi^2(\varphi)\ra$ in \eqref{OnePoint4dOO} contributes, producing
\begin{gather}
    E^{(1)}_{0}=\frac{\lambda_{*}(N+2)(\pi^{2}-\theta^{2})}{48 \pi^{2}\theta^{2}}=\frac{(\pi^{2}-\theta^{2})(N+2)}{6 \theta^{2}(N+8)}\epsilon\,.
\end{gather}
Now, we can calculate contribution of the zero mode to the free energy in the case of  $(\textrm{sp},\textrm{sp})$,
\begin{gather}
    \delta F=\frac{N \textrm{Vol}(\mH^{d-1})}{2\left(4\pi\right)^{\frac{d-1}{2}}\Gamma\left(\frac{d-1}{2}\right)}\int \limits_{-\infty}^{+\infty}d\nu\frac{\left|\Gamma\left(i \nu+\frac{d-2}{2}\right)\right|^{2}}{\left|\Gamma\left(i \nu\right)\right|^2}\log{\left(\nu^2+E^{(1)}_{0}\right)}\,.
\end{gather}
After first taking derivative over $E^{(1)}_{0}$, we can easily compute the integral over $\nu$ as before,
\begin{gather}
   \frac{\partial \delta F  }{\partial E^{(1)}_{0}}  = \frac{1}{2}\frac{N \textrm{Vol}(\mH^{d-1})}{\left(4\pi\right)^{\frac{d-1}{2}}}\Gamma\left(\frac{3-d}{2}\right)\left(E^{(1)}_{0}\right)^{\frac{d-3}{2}} +\mathcal{O}\left(\frac{1}{\theta^{d-1}}\right)\,.
\end{gather}
Integrating over $E^{(1)}_{0}$, and neglecting the constant of integration which is independent of $\theta$ and hence is not important for the leading $\theta \rightarrow 0$ behavior, we get
\begin{gather}
    \delta F=\frac{1}{d-1}\frac{N \textrm{Vol}(\mH^{d-1})}{\left(4\pi\right)^{\frac{d-1}{2}}}\Gamma\left(\frac{3-d}{2}\right)\left(E^{(1)}_{0}\right)^{\frac{d-1}{2}}\,.
\end{gather}
In the case of $d=4$ and small $\theta$, we have 
\begin{gather}
   \delta F=-\frac{N \textrm{Vol}(\mH^{3})}{72 \theta^3}\frac{\pi^2\epsilon^{\frac{3}{2}}}{\sqrt{6}}\left(\frac{N+2}{N+8}\right)^{\frac{3}{2}}\,,
\end{gather}
which produces the order $\epsilon^{3/2}$ contribution to the Casimir energy in \eqref{CasimirEnONRes} as promised.

\subsubsection{Classical Mechanical Analogy for Extraordinary Boundary Conditions}
\label{sec:ExtraOrd}
In this section, we discuss the situation when one of the boundaries of the wedge has extraordinary boundary conditions, i.e. one of the field components, $\phi^N$, acquires a non-zero vacuum expectation value. Working at leading order in $\ep$, this vacuum expectation value is given by the classical solution of the equation
\begin{gather}
    \left(\partial_\varphi^2  + \frac{(d-2)^2}{4}\right)\phi^N - \lambda (\phi^N)^3=0\,.
\end{gather}
Multiplying this equation by $\partial_\varphi \phi^N$ we arrive at the following conservation law
\begin{gather}
    \Tilde{E}=\frac{1}{2}\left(\partial_\varphi \phi^N\right)^2 + \frac{(d-2)^2}{8} (\phi^N)^2 - \lambda \frac{(\phi^N)^4}{4} = {\rm const~in}~\varphi\,.
\end{gather}
To simplify the analysis, we rescale the angle and the field variables as below,
\begin{gather}
\label{rescale}
    \varphi = \frac{2}{d-2}t\,, \quad  \phi^N = \frac{1}{\sqrt{\lambda}} \frac{d-2}{2} y\,, \quad  \Tilde{E}=\frac{1}{\lambda} \left(\frac{d-2}{2}\right)^4 E\,,
\end{gather}
which produces the following conserved quantity,
\begin{gather}
    E = \frac{1}{2} \left(\partial_t y\right)^2 + V(y)\,,\quad V(y) =  \frac{1}{2} y^2 - \frac14 y^4 \,.\label{eq:dimlesspot}
\end{gather}
that describes classical trajectories $y(t)$ subject to potential $V(y)$ and conserved energy $E$.

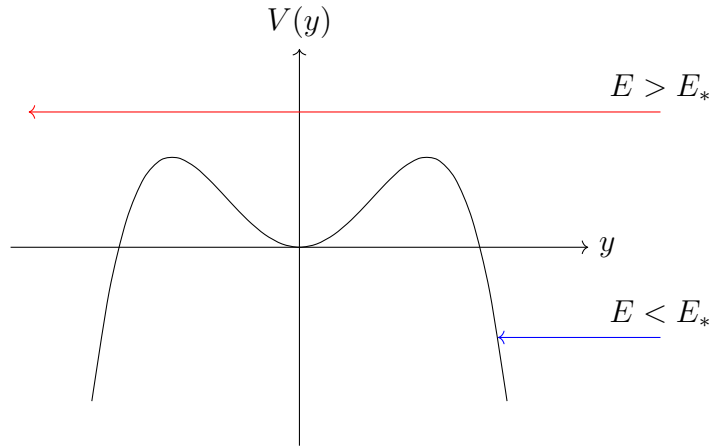
\begin{figure}[!htb]
    \centering
    \begin{tikzpicture}[domain=-2.3:2.3,scale=1.2]
  \draw[->] (-3.2,0) -- (3.2,0) node[right] {$y$};
  \draw[->] (0,-2.2) -- (0,2.2) node[above] {$V(y)$};
  \draw[red] [->] (4,1.5) node[left,above,black]{$E>E_*$} --(-3,1.5) ;
  \draw[blue] [->] (4,-1) node[left,above,black]{$E<E_*$} --(2.2,-1) ;
  \draw plot[smooth] (\x,{\x*\x-0.25*\x*\x*\x*\x}) node[right]{};
\end{tikzpicture}
    \caption{The classical potential $V(y)$ that governs the $O(N)$ CFT on a wedge with extraordinary boundary conditions.}
    \label{fig:classpot}
\end{figure}

To leading order in $\ep$, the free energy of the $O(N)$ CFT on the wedge corresponds to the action of this classical system,
\begin{gather}
    F=S = \int dt \left[\frac12 (\partial_t y)^2 - V(y)\right]\,,
\end{gather}
Thus, the study of the extraordinary boundary conditions and their fusion with various boundary conditions is reduced to the study of various trajectories in the potential \eqref{eq:dimlesspot}. Let us note that the extraordinary boundary condition corresponds to the situation when the trajectories start at spatial infinity $\phi^N\left(\varphi=0\right)=+\infty$ and after finite time (measured by $\varphi$ or $t$ related by rescaling \eqref{rescale}) reach one of the following values at $\varphi=\theta$:
\begin{enumerate}
\item it reaches $\phi^N\left(\varphi=\theta\right)=0$ (corresponding to extraordinary-ordinary (E-O)),
\item it reaches a turning point $\partial_\varphi \phi^N\left(\varphi=\theta\right)=0$ (corresponding to extraordinary-special (E-sp)),
\item it passes through the potential to $\phi^N\left(\varphi=\theta\right)=-\infty$ or returns back to $\phi^N\left(\varphi=\theta\right)=+\infty$ (extraordinary-extraordinary (E-E)).
\end{enumerate}

\begin{table}[!htb]
   \begin{center}
\begin{tabular}{ |c|c|c| } 
\hline
\multirow{3}{*}{\shortstack{$E>E_*$}} & $(\textrm{E-O})$ & $\begin{cases}
        \phi^N(\varphi=0)=+\infty \\  
        \phi^N(\varphi=\theta)=0
\end{cases}$ \\
\cline{2-3}
& $(\textrm{E-E}^{-})$ & $\begin{cases}
        \phi^N(\varphi=0)=+\infty \\  
        \phi^N(\varphi=\theta)=-\infty
\end{cases}$ \\
\hline
\end{tabular}
\begin{tabular}{|c|c|c|}
\hline
\multirow{3}{*}{\shortstack{$E<E_*$}} & $(\textrm{E-sp})$ & $\begin{cases}
        \phi^N(\varphi=0)=+\infty \\  
        (\phi^N)'(\varphi=\theta)=0
\end{cases}$ \\
\cline{2-3}
& $(\textrm{E-E}^{+})$ & $\begin{cases}
        \phi^N(\varphi=0)=+\infty \\  
        \phi^N(\varphi=\theta)=+\infty
\end{cases}$ \\
\hline
\end{tabular}
\end{center}
    \caption{Correspondence between  boundary conditions for the CFT on a wedge and boundary conditions for the classical trajectories. The critical energy is located at $E_*=\frac{1}{4}$.}
    \label{ExtrBCs}
\end{table}

Note that there are two possible extraordinary boundary conditions, because the field $\phi^N$ near the boundary can behave like $\pm 1/ z$ where $z$ is the perpendicular distance. We can either have same signs on both the boundaries (which we will call (E-E$^{+}$)) or opposite signs (which we will call (E-E$^{-}$)). We summarize all the possibilities in Table~\ref{ExtrBCs}. Let us also point out that for all the trajectories we discuss in the table above, there are analogous ones related by the transformation $\phi^N \rightarrow - \phi^N$. But the free energy is invariant under this transformation, so they will have the same free energy as the cases that we discuss here.

All of these cases correspond to various cases of fusion between boundary conditions in the $O(N)$ CFT. We start by identifying the exact trajectories in \eqref{ClassicalsolExtr} and \eqref{eq:exact_traj_S} and use these exact trajectories to calculate the free energy of the solution in \eqref{eq:exact_traj_free}. Then starting from \eqref{eq:extra_ord_small_th}, we explain an alternative strategy which does not require the knowledge of an exact trajectory to calculate the free energy at small opening angle $\theta$, which is all we need for determining the Casimir energy. We present the final results for the Casimir energy for all the combinations of boundary conditions in \eqref{eq:results_extra_Casimir}.

To start, we notice that $V(y)$ has a critical point at $y=\pm 1$ that gives the critical energy $E_*=\frac14$. Then we have two qualitatively different scenarios depending on $E$,
\begin{enumerate}
    \item $E>\frac14$: this corresponds to the red trajectory in Figure~\ref{fig:classpot}. The solution in this case is given by a Jacobi elliptic sine amplitude function (denoted by $sn$) \cite{Abramowitz}:
    \begin{gather}
        y(t) =-\sqrt{1-\sqrt{1-4 E}} \operatorname{sn}\left(\sqrt{\frac{1+\sqrt{1-4 E}}{2}} (t-T(E)), \frac{1-\sqrt{1-4E}-2E}{2 E}\right)\, ,
        \label{ClassicalsolExtr}
    \end{gather}
    where $T(E)$ is the half-period of this function,
    \begin{equation}
    T(E) = \dfrac{iK\left(\frac{-1+\sqrt{1-4E}+4E}{2E}\right)+2K\left(\frac{1-\sqrt{1-4E}-2E}{2 E}\right)}{\sqrt{\frac{1+\sqrt{1-4 E}}{2}}} \,,
    \end{equation}
    where 
    \ie 
    K(m)=\int \limits_{0}^{\pi/2}\frac{d \alpha}{\sqrt{1-m \sin^2{\alpha}}}
    \fe
    is a complete elliptic integral of the first kind. It is clear that in this case $y'(t)\neq 0$ thus Neumann (special) boundary condition is not admissible at the opposite side of the wedge. Instead the allowed trajectory can reach the point $y=0$, which corresponds to the Dirichlet boundary condition (the case of (E-O) in Table~\ref{ExtrBCs}). Also, the trajectory can go to $y=-\infty$ which corresponds to the extraordinary-extraordinary (E-E$^-$) case in Table~\ref{ExtrBCs}. To identify the range of $t$, we note that the solution \eqref{ClassicalsolExtr} blows up at $t = 0, 2 T(E)$ and vanishes at $t=T(E)$. We conclude from Table~\ref{ExtrBCs} that $t \in [0,T(E)]$ for (E-O) and $t\in [0,2T(E)]$ for (E-E$^-$).
    
    \item $E<\frac14$: this corresponds to the blue trajectory in the Figure~\ref{fig:classpot}.  The corresponding solution with the energy $E$ is given by a Jacobi elliptic delta amplitude function (denoted by $dn$) \cite{Abramowitz}  
    \begin{gather}
    \label{eq:exact_traj_S}
        y(t)  = \sqrt{\frac{2}{2-k}} \operatorname{dn}\left( \frac{i}{\sqrt{2-k}} (t-T(E)),k\right)\,, \quad
        E = \frac{1-k}{(k-2)^2}\,,
    \end{gather}
    with $k \in [0,2)$, and $T(E)$ is the half period given by 
    \begin{gather}
        T(E)=\sqrt{2-k}K(1-k)\,.
    \end{gather}
    In this situation, the trajectory can never reach $y=0$ but instead the turning point $y'=0$ which corresponds to the Neumann boundary condition (the (E-sp) case in Table~\ref{ExtrBCs}). Furthermore it can bounce back and go back to $y = +\infty$ which corresponds to the (E-E$^+$) case. As before, the ranges for $t$ are fixed to be $t \in [0,T(E)]$ for (E-sp) and $t\in [0,2T(E)]$ for (E-E$^+$).
\end{enumerate}

Let us first focus on the case of extraordinary-ordinary (E-O) boundary conditions for which $t \in [0,T(E)]$ and solution \eqref{ClassicalsolExtr} satisfies the following boundary conditions:
\begin{gather}
    y(t=0)=+\infty\,, \quad    \ \ \   y\left(t=T(E)\right)=0 \,.
\end{gather}
From \eqref{rescale} and \eqref{ClassicalsolExtr}, we deduce that the half period $T(E)$ is related to $\theta$ by 
\begin{gather}
    \theta=\frac{2}{d-2}T(E)\,.
    \label{periodEO}
\end{gather}
Since we are interested in the limit $\theta \rightarrow 0$ and thus $T(E)\to 0$, we take energy $E \rightarrow +\infty$. Expanding \eqref{ClassicalsolExtr} in this limit we obtain
\begin{gather}
    T(E)=\frac{e^{i \frac{\pi}{4}}K(2)}{E^{\frac{1}{4}}}+\mathcal{O}\left(\frac{1}{E^{\frac{3}{4}}}\right)\, \Longrightarrow E=\dfrac{1}{\theta^4}\left(\frac{1}{d-2}\dfrac{\Gamma\left(\frac{1}{4}\right)^2}{2\sqrt{\pi}}\right)^{4}+\mathcal{O}\left(\dfrac{1}{\theta^{2}}\right)\,,
    \label{periodExtrDiri}
\end{gather}
where we have used the following representation of $K(2)$ \cite{Abramowitz},
\begin{gather}
    e^{i \frac{\pi}{4}}K(2)=\frac{\pi}{2}e^{i \frac{\pi}{4}} {}_2 F_1\left(\frac{1}{2},\frac{1}{2},1,2\right)=\dfrac{\Gamma\left(\frac{1}{4}\right)^2}{4\sqrt{\pi}}\,.
\end{gather}
Now let us compute the corresponding free energy  density of this solution, which is simply given by evaluating the Lagrangian on this classical configuration,
\begin{gather}
    \mathcal{L}= \frac{1}{2}\left(\partial_\varphi \phi^N\right)^2 -\left[ \frac{(d-2)^2}{8} (\phi^N)^2 - \lambda \frac{(\phi^N)^4}{4} \right]=\dfrac{1}{\lambda}\left(\frac{d-2}{2}\right)^4\left( \frac{1}{2} \left(\partial_t y\right)^2 - V(y)\right)\,.
\end{gather}
However, the above expression is divergent near $t \sim 0$, corresponding to the contribution to the free energy coming from each individual BCFTs residing on the boundaries of the wedge. These divergences are universal ($E$ independent) and can be easily take into account. 
First we rewrite the action as
\begin{gather}
    S=\int \limits_{0}^{T(E)}dt\left(\frac{1}{2} \left(\partial_t y\right)^2 - V(y)\right)=\int \limits_{0}^{T(E)}dt  \left(\partial_t y\right)^2 -E \ T(E)  \, .
\end{gather}
The integral of the first term on the RHS can be recasted in the following way \cite{arnol2013mathematical},
\ie 
    A(E)=\int \limits_{0}^{T(E)}dt \left(\partial_t y\right)^2=-\int \limits_{0}^{+\infty}dy \ \partial_t y = \int \limits_{0}^{+\infty}dy \sqrt{2(E-V(y))}\,,
    \fe 
 which satisfies the following,
    \ie 
    \dfrac{d A(E)}{d E}=\int \limits_{0}^{+\infty}dy \dfrac{1}{\sqrt{2(E-V(y))}}= -\int \limits_{0}^{+\infty} \frac{dy}{\partial_t y}=T(E) \Rightarrow \notag
    A(E)=\int \limits_{0}^{E} T(E)dE+{\rm const}\,,
\fe 
where the constant term does not depend on energy $E$ and contains all divergences.
Since we are interested only in the $E$ dependence at large $E$, we use \eqref{periodExtrDiri} to evaluate the action,
\ie 
    S=&\,\int \limits_{0}^{E} T(E)dE-E \ T(E) 
    \\
    =&  \,
    \dfrac{\Gamma\left(\frac{1}{4}\right)^2}{12\sqrt{\pi}}E^{\frac{3}{4}}+\mathcal{O}\left(\frac{1}{E^{\frac{1}{4}}}\right)=\dfrac{1}{3 \theta^3}\left(\dfrac{\Gamma\left(\frac{1}{4}\right)^2}{4\sqrt{\pi}}\right)^{4}\left(\frac{2}{d-2}\right)^{3}+\mathcal{O}\left(\frac{1}{\theta^2}\right)\,,
\fe 
which determines the wedge free energy in the case of (E-O) boundary conditions,
\begin{gather}
   F^{(\textrm{E-O})}
    =\dfrac{1}{3\lambda \theta^3}\textrm{Vol}(\mH^{d-1})\left(\dfrac{\Gamma\left(\frac{1}{4}\right)^2}{4\sqrt{\pi}}\right)^{4}\,.
    \label{eq:exact_traj_free}
\end{gather}
Similarly, in the case of (E-E$^-$) boundary conditions, the action in the limit $E \rightarrow +\infty$ takes the following form,
\begin{gather}
    S=\int \limits_{0}^{2T(E)}dt\left(\frac{1}{2} \left(\partial_t y\right)^2 - V(y)\right)=2\int \limits_{0}^{T(E)}dt \left(\partial_t y\right)^2 -2E \ T(E)=\dfrac{\Gamma\left(\frac{1}{4}\right)^2}{6\sqrt{\pi}}E^{\frac{3}{4}}+\mathcal{O}\left(\frac{1}{E^{\frac{1}{4}}}\right)\,.
\end{gather}
Since the period is doubled compared with the case of (E-O) (see \eqref{periodEO}), we have the following identification between $E$ and $\theta$,
\begin{gather}
     \theta=\frac{4}{d-2}T(E)\, \Longrightarrow E=\dfrac{1}{\theta^4}\left(\frac{1}{d-2}\dfrac{\Gamma\left(\frac{1}{4}\right)^2}{\sqrt{\pi}}\right)^{4}+\mathcal{O}\left(\dfrac{1}{\theta^{2}}\right)\,.
\end{gather}
Consequently the free energy in this case evaluates to,
\begin{gather}
    F^{(\textrm{E-E$^-$})}
    =\dfrac{16}{3\lambda \theta^3}\textrm{Vol}(\mH^{d-1})\left(\dfrac{\Gamma\left(\frac{1}{4}\right)^2}{4\sqrt{\pi}}\right)^{4}=16 F^{(\textrm{E-O})}\,.
\end{gather}

The method presented above relies on the fact that we have found the exact form of the classical solution \eqref{ClassicalsolExtr}  and it gives us access to the wedge free energy as an exact function of $E$ (equivalently $\theta$). However since we are mainly interested in the $E\to \infty$ limit, we can also determine the Casimir energy without the explicit solution and by directly expanding around $E=\infty$. To illustrate how this works we again consider the case of  (E-O) boundary conditions. 

In the $E \rightarrow +\infty $ limit, the period of motion from $y=+\infty$ to $y=0$ can be determined as follows. We first note the following integral representation for the half-period,
\begin{gather}
\label{eq:extra_ord_small_th}
    T(E)=\int \limits_{0}^{+\infty}\dfrac{dy}{\sqrt{2(E-\frac{1}{2}y^2+\frac{1}{4}y^4 )}}
    =\dfrac{1}{\sqrt{2}E^{\frac{1}{4}}}\int \limits_{0}^{+\infty}dx \dfrac{1}{\sqrt{1+\frac{1}{4}x^4-\frac{1}{2\sqrt{E}}x^2 }}\,,
\end{gather}
where in the second step we have rescaled $y=E^{\frac{1}{4}}x$. Expanding around $E\to \infty$ gives,
\begin{gather}
    T(E)=\dfrac{1}{\sqrt{2}E^{\frac{1}{4}}}\int \limits_{0}^{+\infty}dx \dfrac{1}{\sqrt{1+\frac{1}{4}x^4 }}+\mathcal{O}\left(\frac{1}{E^\frac12}\right)=\dfrac{1}{E^{\frac{1}{4}}}\dfrac{\Gamma\left(\frac{1}{4}\right)^2}{4\sqrt{\pi}} + \mathcal{O}\left(\frac{1}{E^\frac12}\right)\,,
\end{gather}
which indeed coincides with the first term of period $T(E)$ in \eqref{periodExtrDiri}.

Analogous calculations for the (E-sp) and (E-E$^\pm$) cases are straightforward. The Casimir energy then follows from the relation \eqref{hyperandcasgenp} with the free energy after plugging in the fixed point coupling \eqref{fixedpointvaluespinning}. Below we summarize the final results to leading order in $\epsilon$ for the fusion of extraordinary boundary conditions with various boundary conditions,
\begin{align}
\label{eq:results_extra_Casimir}
   \cE^{\textrm{(E-O)}} &= \dfrac{1}{6144}\dfrac{N + 8}{ \pi^4 \epsilon}\Gamma\left(\frac{1}{4}\right)^8, \quad 
     \cE^{\textrm{(E-E$^-$)}} = \dfrac{1}{384}\dfrac{N + 8}{ \pi^4 \epsilon}\Gamma\left(\frac{1}{4}\right)^8 \, , \\
     \cE^{\textrm{(E-sp)}} &= \dfrac{1}{24576}\dfrac{N + 8}{ \pi^4 \epsilon}\Gamma\left(\frac{1}{4}\right)^8, \quad  
     \cE^{\textrm{(E-E$^+$)}} = \dfrac{1}{1536}\dfrac{N + 8}{ \pi^4 \epsilon}\Gamma\left(\frac{1}{4}\right)^8\,. \notag
\end{align}
It would be interesting to understand if the hierarchy among the Casimir energies above can be understood in terms of the RG flows connecting the different boundary conditions on the one side of the wedge. 

\subsection{Gross-Neveu Model in $d=2+\ep$} 
\label{sec:GNBoundary}

In this section, we study the Casimir energy of fusion between conformal boundaries in the Gross-Neveu (GN) CFT described by the following action
\begin{gather}
S=-\int d^{d}x\sqrt{g}\left(\bar{\Psi}_{i}\gamma \nabla \Psi^{i}+\frac{g}{2}\left(\bar{\Psi}_{i}\Psi^{i}\right)^{2}\right) \, ,
    \label{GNaction}
\end{gather}
where again $i  = 1, \dots, N_f$ and we introduce $N = N_f c_d$ with $c_d$ defined earlier in Section~\ref{sec:lineinGNY}. In $d=2+\epsilon$, the theory admits a fixed point with critical coupling \cite{PhysRevD.10.3235} (or see \cite{moshe2003quantum} for a review)
\ie 
g_{*}=\dfrac{\pi}{N_f-1}\epsilon\,,
\fe 
The conformal boundary conditions in this model have been recently studied in \cite{Giombi:2021cnr} and are defined as
\begin{gather}
\gamma_\perp \Psi^i = \pm \Psi^i, \quad \gamma_\perp = (\vec{n}\cdot \vec{\gamma})\,,
\end{gather}
where $\vec{n}$ is a normal vector to the boundary and the two possibilities are related by parity. These boundary conditions ensure that there is no particle current through the boundary \cite{Johnson:1975zp,DePaola:1999im,Milonni:1994xx}. As we have done for the scalars, we will study this fermionic model inside a wedge with an opening angle $\theta$.
We show below that the leading contribution to the free energy $F$ in the limit of small opening angle $\theta$ is of the expected form 
\begin{gather}
    \dfrac{F}{V}=\dfrac{\cE_{GN}}{\theta^{d-1}}\,,
\end{gather}
where the Casimir energy $\cE_{GN}$  for $d=2+\epsilon$ for different boundary conditions is,
 \begin{equation}
 \begin{split}
      \cE^{+-}_{GN}= \cE^{-+}_{GN} &=-\dfrac{N_f \pi}{24}+\dfrac{\epsilon \pi}{48}N_f\left(\gamma+\log{\pi}-2\frac{\zeta'(2)}{\zeta(2)}\right)+\mathcal{O}(\epsilon^{2})\,, \\
    \cE^{++}_{GN}= \cE^{--}_{GN} &=\dfrac{N_f \pi}{12}-\dfrac{\epsilon \pi}{24}N_f\left(\gamma+\log{4\pi}-2\frac{\zeta'(2)}{\zeta(2)}-\dfrac{3}{2}\dfrac{(2N_f-1)}{(N_f-1)}\right)+\mathcal{O}(\epsilon^{2})\,.
 \end{split}
 \end{equation}
In Section~\ref{sec:free_fermions_wedge}, we first study properties of free fermions inside a wedge. We then consider the interacting model in Section~\ref{sec:GNepsWedge} and determine the Casimir energy.

\subsubsection{Free Fermions in a Wedge}
\label{sec:free_fermions_wedge}
We start by analyzing the two-point function of free fermions in the wedge geometry. The main result of this subsection which will be useful in the following subsections is the one-point function of the fermion bilinear operator inside the wedge given in \eqref{psibarpsisum} and its expansion in $d = 2 + \eps$ dimensions given in \eqref{psibarpsid=2}. 

To simplify notation we consider a single fermion with a fixed flavor $\psi=\Psi^{1}$. The Dirac equation in the wedge with cylindrical coordinates \eqref{wedgecoords} is given by\footnote{For simplicity, we will often drop the argument of the propagator when
there is no room for confusion.} 
\begin{equation}
    \left( \gamma^{\rho} \partial_{\rho_{1}} + \frac{\gamma^{\varphi}}{\rho_{1}} \partial_{\varphi} + \frac{\gamma^{\rho}}{2 \rho_{1}} + \Vec{\gamma} \cdot \Vec{\partial}_{ 1}  \right) G_{\psi}(\varphi_1,\rho_1,\varphi_2,\rho_2,\Vec{y}_{12}) = -\frac{\delta^{(d)}(x_{1}-x_{2})}{\rho_{1}}\,,
    \label{DiraceqWedge}
\end{equation}
where $\Vec{y}_{12}\equiv \vec{y}_{1}-\vec{y}_{2}$. In addition, we have a choice of boundary conditions at the boundaries of the wedge, i.e. at $\varphi = 0$ and at $\varphi=\theta$,
\begin{equation}
    \gamma^{\varphi} G_{\psi} = +G_{\psi}  \hspace{1cm} \textrm{or} \hspace{1cm}  \gamma^{\varphi} G_{\psi} = -G_{\psi}\,.
    \label{BCfermionswedge}
\end{equation}
Consequently, we label the possible wedge configurations by $++, --$, $+-$  or $-+$, where the first sign corresponds to the boundary condition at $\varphi=0$, and the second at $\varphi=\theta$. This two-point function is related by a Weyl transformation to that on $S^1 \times \mH^{d-1}$. We find the following solutions depending on the boundary conditions,
\begin{equation}
\label{greenfunct}
    G_{\psi}=
    \left\{
    \begin{aligned}
         &\frac{\pi}{\theta}\sum_{n \in \mathbb{Z}_{\geq 0} } \ \alpha_n \frac{ \hat{P}_{\pm}\left(n,\varphi_{1}\right)  G^{\mH^{d-1}}_{\psi} (\rho_1, \rho_2, \Vec{y}_{12}) \hat{P}_{\mp}\left(-n,\varphi_{2}\right)}{ 2 \pi (\rho_1 \rho_2)^{\frac{d - 1}{2}}} \,, &++/ -- \\
         &\frac{\pi}{\theta}\sum_{n \in \mathbb{Z}_{\geq 0} + \frac{1}{2} }\alpha_n \frac{ \hat{P}_{\pm}\left(n,\varphi_{1}\right) G^{\mH^{d-1}}_{\psi} (\rho_1, \rho_2, \Vec{y}_{12})  \hat{P}_{\mp}\left(-n,\varphi_{2}\right)}{ 2 \pi (\rho_1 \rho_2)^{\frac{d - 1}{2}}}\,, &+-/-+
    \end{aligned}
    \right. \, ,
\end{equation}
where we have introduced the projector and constant $\alpha_n$ defined below,
\begin{gather}
  \hat{P}_{\pm}\left(n,\varphi\right)\equiv  e^{-\frac{i \pi \gamma^{\varphi} }{4} } \left( e^{i \frac{\pi n}{\theta} \varphi} \pm \gamma^{\varphi} e^{- i \frac{\pi n}{\theta} \varphi}  \right) \,, \qquad \alpha_n= \left\{
\begin{matrix}
1, \quad  n \neq 0 \\
\ \frac{1}{2}, \quad n=0 \, \
\end{matrix}
\right. .
\end{gather}
This projector ensures that the corresponding two-point function satisfies the correct boundary conditions. 

The Green's function $G_\psi^{\mH^{d-1}}$ on $\mH^{d-1}$ with a mass $\frac{n\pi}{\theta}$ satisfies the following equation,
\begin{equation}
    \left(\rho_1 \left(\gamma^{\rho} \partial_{\rho_1} + \Vec{\gamma} \cdot \Vec{\partial}_{1} \right) - \frac{d-2}{2} \gamma^{\rho} + \frac{n\pi}{\theta} \right) G^{\mH^{d-1}}_{\psi}(\rho_1, \rho_2, \vec{y}_{12}) = -\rho_{1}^{d-1}\delta(\rho_1-\rho_2)\delta^{(d-2)}\left(\vec{y}_{1}-\vec{y}_{2}\right) \, .
\label{equationmassiveGreenFuncHd-1}
\end{equation}
As was reviewed in \cite{Giombi:2021cnr}, there are two solutions to \eqref{equationmassiveGreenFuncHd-1}  corresponding to two possible boundary behaviors. The first solution is called the standard quantization and satisfies the following boundary condition,
\ie 
\gamma^{\rho}G^{\mH^{d-1}}_{\psi}(\rho_1, \rho_2, \vec{y}_{12})\big|_{\rho_1\rightarrow 0}=-\sgn(n)G^{\mH^{d-1}}_{\psi}(\rho_1, \rho_2, \vec{y}_{12})\big|_{\rho_1\rightarrow 0}\,,
\fe 
which is valid for any $n$. In this case the leading boundary spinor operator has dimension $\Delta_n=\frac{d-1}{2}+\frac{n \pi}{\theta}$.\footnote{As in the case of scalar, these operators live at the boundary of $\mathbb{H}^{d-1}$ so in the original setup, at the intersection of the two boundaries that make up the wedge.} The other solution is called alternative quantization and satisfies the following boundary condition, 
\ie 
\gamma^{\rho}G^{\mH^{d-1}}_{\psi}(\rho_1, \rho_2, \vec{y}_{12})\big|_{\rho_1\rightarrow 0}=\sgn(n)G^{\mH^{d-1}}_{\psi}(\rho_1, \rho_2, \vec{y}_{12})\big|_{\rho_1\rightarrow 0}
\fe 
which is unitary only for $\frac{n \pi}{\theta}<\frac{1}{2}$. In this case the leading boundary spinor operator has dimension $\Delta_n = \frac{d-1}{2}-\frac{n \pi}{\theta}$. 

Since we are interested in unitary theory the only possibility for all modes but $n=0$ modes is to choose standard quantization solution with the following explicit form \cite{Giombi:2021cnr},
\ie
\label{massgreenfunctionHd-1}
 G^{\mH^{d-1}}_{\psi}  
       = & \frac{-\Gamma\left( \frac{d - 1}{2} + \frac{n \pi }{\theta} \right)   }{\Gamma \left( \frac{1}{2} + \frac{n \pi }{\theta} \right) 4^{1 + \frac{n \pi }{\theta}} (4\pi)^{\frac{d}{2} - 1} \sqrt{\rho_1 \rho_2} } \\
      &\times \bigg[ \frac{ \left(\gamma^{\rho} \rho_{12} + \Vec{\gamma} \cdot \vec{y}_{12} \right)  }{(1 + \xi)^{\frac{d - 3}{2}} \xi^{1 + \frac{n \pi }{\theta}} } {}_2 F_1 \left( \frac{n \pi }{\theta} + \frac{3-d}{2} , 1 + \frac{n \pi }{\theta}, 1 + \frac{2 n \pi }{\theta}, - \frac{1}{\xi} \right)  \\  
& - \textrm{sgn}(n) \frac{\gamma^{\rho} \left( - \gamma^{\rho} (\rho_{1} + \rho_2 ) + \Vec{\gamma} \cdot \vec{y}_{12}\right)  }{ (1 + \xi)^{\frac{d - 1}{2} } \xi^{\frac{n \pi }{\theta}} } {}_2 F_1 \left(  \frac{n \pi }{\theta} + \frac{3 - d}{2} , \frac{n \pi }{\theta}, 1 + \frac{2 n \pi }{\theta}, - \frac{1}{\xi} \right)    \bigg]\,,
\fe 
where we have introduced the cross-ratio $\xi\equiv \frac{\rho^2_{12}+\vec{y}_{12}^2}{4\rho_1\rho_2}$. 

On the other hand, for $n=0$  we have the following two choices \cite{Giombi:2021cnr},
\ie 
 G^{\mH^{d-1}}_{\psi}=-\dfrac{\Gamma\left(\frac{d-1}{2}\right)}{2\left(4\pi\right)^{\frac{d-1}{2}}}\left[\frac{ \left(\gamma^{\rho} \rho_{12} + \Vec{\gamma} \cdot \vec{y}_{12} \right)  }{\sqrt{\rho_1 \rho_2} }\dfrac{1}{\xi^{\frac{d-1}{2}}}\pm\frac{\gamma^{\rho} \left( - \gamma^{\rho} (\rho_{1} + \rho_2 ) + \Vec{\gamma} \cdot \vec{y}_{12}\right)  }{ \sqrt{\rho_1 \rho_2} }\frac{1}{\left(1+\xi\right)^{\frac{d-1}{2}}} \right]\, ,
    \label{GreenfuncHd-1zeromodes}
\fe 
which satisfy $\gamma^{\rho}G^{\mH^{d-1}}_{\psi}(\rho_1, \rho_2, \vec{y}_{12})\big|_{\rho_1\rightarrow 0}=\pm G^{\mH^{d-1}}_{\psi}(\rho_1, \rho_2, \vec{y}_{12})\big|_{\rho_1\rightarrow 0}$ respectively. As we show below, the free energy does not depend on this choice in the leading and next to leading order in $\epsilon$.

Finally, as a consistency check, using \eqref{equationmassiveGreenFuncHd-1} we can verify that the solution  \eqref{greenfunct} indeed satisfies \eqref{DiraceqWedge}:
\ie 
   & \left( \gamma^{\rho} \partial_{\rho_{1}} + \frac{\gamma^{\varphi}}{\rho_1} \partial_{\varphi_{1}} + \frac{\gamma^{\rho}}{2 \rho_1} + \Vec{\gamma} \cdot \Vec{\partial_{1}}  \right) G_{\psi} 
   \\
   = & -\frac{1}{\rho_{1}} \left(\delta(\varphi_{1}-\varphi_{2})\mp \gamma^{\varphi}\delta(\varphi_{1}+\varphi_{2})\right)\delta(\rho_1-\rho_2)\delta^{(d-2)}\left(\vec{y}_{1}-\vec{y}_{2}\right)
    =-\frac{\delta^{(d)}(x_{1}-x_{2})}{\rho_{1}}\, ,
\fe 
where in the last step we have dropped $\delta(\varphi_{1}+\varphi_{2})$ since $0<\varphi<\pi$.

To calculate the one-point function of $\langle \bar{\psi}{\psi} (x) \rangle$, we evaluate the trace of \eqref{greenfunct}, expanding it around $\xi=0$, and keeping only the $\xi$ independent term
\begin{equation}
    \langle \bar{\psi}{\psi} (x) \rangle =  \mp \dfrac{ 2\pi c_d }{\theta}\sum_{n }\frac{  \sin \left( \frac{ 2\pi n}{\theta} \varphi  \right) \Gamma\left( \frac{3 - d}{2} \right)  \Gamma\left( \frac{d - 1}{2} + \frac{\pi n}{\theta} \right) }{ (4\pi)^{\frac{d}{2}} \sqrt{ \pi}  \Gamma\left( \frac{3 - d}{2} + \frac{\pi n}{\theta} \right) \rho^{d-1} }\, ,
    \label{psibarpsisum}
\end{equation}
where $c_d=2^{[\frac{d}{2}]}$, and 
the overall $-$ sign corresponds to either $++$ with $n \in \mathbb{Z}_{\geq 0}$ or $+-$ with $n \in \mathbb{Z}_{\geq 0}+\frac{1}{2}$. Similarly, the overall $+$ sign corresponds to either $--$ with $n \in \mathbb{Z}_{\geq 0}$ or $-+$ with $n \in \mathbb{Z}_{\geq 0}+\frac{1}{2}$. Note that there is no contribution of $n=0$ modes to the one-point function.

The $+-/-+$ case with $\theta=\pi$ describes free fermions on the hyperbolic space $\mH^d$. Indeed, boundary conditions \eqref{BCfermionswedge} in this case can be written in a compact form as $\left(\Vec{n}\cdot\Vec{\gamma}\right)G_{\psi}=\pm G_{\psi}$, where $\Vec{n}$ denotes the normal vector at the boundary of $\mH^d$. They correspond to the two conformal boundary conditions of \cite{Giombi:2021cnr}.
As a consistency check, we show below that the wedge one-point function $ \langle \bar{\psi}{\psi} (x) \rangle$ in this special case coincides with the one-point function with $+/-$ boundary conditions for the fermions in the hyperbolic space up to a Weyl factor. Indeed, the sum in \eqref{psibarpsisum} for this case can be performed explicitly,
\ie 
     &\langle \bar{\psi}{\psi} (x) \rangle_{\theta=\pi} =   \mp \ 2\sum_{n \in \rm \mathbb{Z}_{\geq 0} +1/2} \frac{  \sin \left( 2 n \varphi  \right)  c_d \Gamma\left( \frac{3 - d}{2} \right)  \Gamma\left( \frac{d - 1}{2} + n \right) }{ (4\pi)^{\frac{d}{2}} \sqrt{ \pi}  \Gamma\left( \frac{3 - d}{2} + n \right) \rho^{d-1} } 
     \\
     &=\mp i\dfrac{c_d \Gamma\left( \frac{3 - d}{2} \right)}{(4\pi)^{\frac{d}{2}} \sqrt{ \pi} \rho^{d-1}}\dfrac{\Gamma\left( \frac{ d}{2} \right)}{\Gamma\left( \frac{4 - d}{2} \right)}e^{-i \varphi}\left[{}_2 F_1 \left( 1,\frac{d}{2},2-\frac{d}{2},e^{-2i \varphi}\right)-e^{2 i \varphi} {}_2 F_1 \left(1,\frac{d}{2},2-\frac{d}{2},e^{2i \varphi} \right)\right] \, .
     \label{densityhalfinte}
\fe 
Now, using the hypergeometric identity which transforms $ {}_2 F_1(a,b,c,1/z)$ into linear combinations of $ {}_2 F_1(\Tilde{a},\Tilde{b},\Tilde{c},z)$ \cite{Abramowitz}, we find
\begin{gather}
    {}_2 F_1 \left( 1,\frac{d}{2},2-\frac{d}{2},\frac{1}{z}\right)-z {}_2 F_1 \left(1,\frac{d}{2},2-\frac{d}{2},z\right)
    =2^{d-1}\sqrt{\pi}\left(1-z\right)^{-d+1}\left(-z\right)^{d/2}\dfrac{\Gamma\left( \frac{4 - d}{2} \right)}{\Gamma\left( \frac{3 - d}{2} \right)}\, ,
    \label{hyperrel}
\end{gather}
with $z=e^{2i \varphi} $. Inserting \eqref{hyperrel} into \eqref{densityhalfinte}, we arrive at
\begin{gather}
    \langle \bar{\psi}{\psi} (x) \rangle_{\theta=\pi}=\mp\dfrac{ c_d\Gamma\left( \frac{ d}{2} \right)}{(4\pi)^{\frac{d}{2}}  \left(\rho \sin{\varphi}\right)^{d-1}}\, ,
    \label{psibarpsiTheta=pi+-/-+}
\end{gather}
which coincides with the result in \cite{Giombi:2021cnr} up to the Weyl factor $ \left(\rho \sin{\varphi}\right)^{d-1}$ as expected.

In the case of $++/--$ boundary conditions on the wedge and $\theta=\pi$, we can also perform the sum over $n$ explicitly,
\ie 
    &    \langle \bar{\psi}{\psi} (x) \rangle_{\theta=\pi}  
        =\mp \ 2\sum_{n \in \rm \mathbb{Z}_{\geq 0} } \frac{  \sin \left( 2 n \varphi  \right) c_d\Gamma\left( \frac{3 - d}{2} \right)  \Gamma\left( \frac{d - 1}{2} + n \right) }{ (4\pi)^{\frac{d}{2}} \sqrt{ \pi}  \Gamma\left( \frac{3 - d}{2} + n \right) \rho^{d-1} }
        =\mp i\dfrac{c_d\Gamma\left( \frac{3 - d}{2} \right)}{(4\pi)^{\frac{d}{2}} \sqrt{ \pi} \rho^{d-1}}\dfrac{\Gamma\left( \frac{ d+1}{2} \right)}{\Gamma\left( \frac{5 - d}{2} \right)}e^{-2i \varphi} 
        \\
        &\times \left[{}_2 F_1 \left( 1,\frac{d+1}{2},3-\frac{d+1}{2},e^{-2i \varphi}\right)-e^{4 i \varphi} {}_2 F_1 \left(1,\frac{d+1}{2},3-\frac{d+1}{2},e^{2i \varphi} \right)\right] \, .
    \label{psibarpsiTheta=pi++/--}
\fe 
In this case the $\langle \bar{\psi}{\psi} (x) \rangle_{\theta=\pi}$ no longer has the simple behavior associated with a homogeneous conformal boundary as in the previous case. Instead the corner of the wedge persists as a nontrivial codimension-two boundary-changing defect operator at $\theta=\pi$ and the above describes the one-point function of $\bar\psi\psi$ in the presence of this boundary changing operator.

We now come back to the general case and evaluate the one-point function $\langle \bar{\psi}{\psi} (x) \rangle$ at general $\theta$.
By separating finite and divergent pieces in \eqref{psibarpsisum} for $2<d<3$ and applying zeta regularization for the divergent piece, we obtain the following,
\ie 
 \langle \bar{\psi}{\psi} (x) \rangle =&  \mp \dfrac{2\pi c_d }{\theta}\sum_{n }\frac{  \sin \left( \frac{ 2\pi n}{\theta} \varphi  \right)   \Gamma\left( \frac{3 - d}{2} \right)   }{ (4\pi)^{\frac{d}{2}} \sqrt{ \pi}  \rho^{d-1} }\left[\frac{    \Gamma\left( \frac{d - 1}{2} + \frac{\pi n}{\theta} \right) }{ \Gamma\left( \frac{3 - d}{2} + \frac{\pi n}{\theta} \right) }-\left(\frac{\theta}{n \pi}\right)^{2-d}\right]   \\
 &\mp \dfrac{2\pi}{\theta}\sum_{n } c_d\frac{  \sin \left( \frac{ 2\pi n}{\theta} \varphi  \right)   \Gamma\left( \frac{3 - d}{2} \right)   }{ (4\pi)^{\frac{d}{2}} \sqrt{ \pi}  \rho^{d-1} }\left(\frac{\theta}{n \pi}\right)^{2-d}\,.
    \label{psibarpsiregularizaed}
\fe 
The first line above is finite and vanishes in $d=2$ case, while the sum in the second line can be performed explicitly. For example, in the case of $+-/-+$ boundary conditions, it can be represented through the Lerch transcendent function. In $d=2+\epsilon$, we obtain
\ie 
\langle \bar{\psi}{\psi} (x) \rangle=    
\begin{cases}
  \mp\dfrac{1}{2\theta \rho}\cot{\frac{\pi \varphi}{\theta}}+\mathcal{O}(\epsilon) &  ++/-- \,,
\\
\mp\dfrac{1}{2\theta \rho\sin{\frac{\pi \varphi}{\theta}}}+\mathcal{O}(\epsilon) &  +-/-+ \,. 
\end{cases}
    \label{psibarpsid=2}
\fe 
The special cases of $\theta=\pi$  agree with \eqref{psibarpsiTheta=pi+-/-+} and \eqref{psibarpsiTheta=pi++/--} as expected.

\subsubsection{Interacting Wedge in $\eps$ Expansion and Free Energy}
\label{sec:GNepsWedge}
We are now ready to determine the Casimir energy of fusion for conformal boundaries in the Gross-Neveu model. As before, the wedge free energy receives contributions from the free fermions and the interaction term. Using the hyperbolic space picture, we may write it as, 
\ie 
    &F(\theta) =  F_{0}(\theta)+I(\theta)  
    \\
    &=-\frac{ N \textrm{Vol} (\mH^{d - 1}) }{(4 \pi)^{\frac{d - 1}{2}} \Gamma \left( \frac{d - 1}{2} \right) }   \sum_{n}   \int_0^{\infty} d \lambda \frac{ | \Gamma \left( \frac{d - 1}{2} + i \lambda \right)|^2 \log \left( \lambda^2 + \frac{n^2 \pi^2 }{\theta^2}  \right) }{|\Gamma \left( \frac{1}{2} + i \lambda \right)|^2 }
    -  \frac{g_{*}}{2}  \int d^{d}x \langle\left( \bar{\Psi}_{i}\Psi^{i}\right)^2 \rangle\, .
    \label{freeenergyGN}
\fe 
The sum in the above equation is over $n \in \mathbb{Z}_{\geq 0}$ for $++/--$ boundary conditions and  over $n \in \mathbb{Z}_{\geq 0}+\frac{1}{2}$ for $+-/-+$ boundary conditions. In the first term of the second line above, we have used that the degeneracy of $N_f$ free massive fermions on $\mH^{d-1}$ is given by \cite{Camporesi:1992tm,Bytsenko:1996rr, Sato:2021eqo},
\begin{gather}
    \mu(\lambda)=\frac{ N \textrm{Vol} (\mH^{d - 1}) }{(4 \pi)^{\frac{d - 1}{2}} \Gamma \left( \frac{d - 1}{2} \right) }      \frac{ | \Gamma \left( \frac{d - 1}{2} + i \lambda \right)|^2 }{|\Gamma \left( \frac{1}{2} + i \lambda \right)|^2 }\, .
\end{gather}
Indeed, working on $S^1_\theta \times \mH^{d-1}$, we can decompose the fermion field $\psi$ in the wedge on $S^1_\theta$ as: 
\begin{equation}
\begin{split}
    \psi(x) &=  \frac{1}{\sqrt{2\theta}} \frac{1}{\rho^{\frac{d-1}{2}}}\sum_{n \in \mathbb{Z}_{\geq 0} + \frac{1}{2} } \hat{P}_{\pm}\left(n,\varphi\right)\chi_n (\rho, \vec{y}) \hspace{1cm}  +-/-+\\ 
    \psi(x) &= \frac{1}{2\sqrt{\theta}}\frac{1}{\rho^{\frac{d-1}{2}}}\hat{P}_{\pm}\left(0,\varphi\right)\chi_0 (\rho, \vec{y}) +   \frac{1}{\sqrt{2\theta}}\frac{1}{\rho^{\frac{d-1}{2}}} \sum_{n = 1}^{\infty} \hat{P}_{\pm}\left(n,\varphi\right)\chi_n (\rho, \vec{y})  \hspace{1cm}  ++/-- \\
\end{split}
\label{KKreductionGN}
\end{equation}
After performing a Kaluza-Klein reduction on $S^1_\theta$ and using that $\bar \psi= \psi^{\dagger}\gamma^{i}$ and $\left(\gamma^{\varphi}\right)^{\dagger}=\gamma^{\varphi}$, we obtain the following action for the massive fermion modes on $\mH^{d - 1}$ for the $+-/-+$ cases,
\begin{equation}
\begin{split}
     S &=  -\sum_{n \in \mathbb{Z}_{\geq 0} + \frac{1}{2} } \int \frac{d \rho d \vec{y}}{\rho^{d-1}} \bar \chi_n\left[\rho \left(\gamma^{\rho} \partial_{\rho} + \Vec{\gamma} \cdot \Vec{\partial} \right) - \frac{d-2}{2} \gamma^{\rho} + \frac{n\pi}{\theta} \right]\chi_n\,,
\end{split}
\end{equation}
while for $++/--$ we have,
\begin{equation}
\begin{split}
     S &=    -\int \frac{d \rho d \vec{y}}{\rho^{d-1}} \bar \chi_0 \left[\rho \left(\gamma^{\rho} \partial_{\rho} + \Vec{\gamma} \cdot \Vec{\partial} \right) - \frac{d-2}{2} \gamma^{\rho}  \right]\left[\dfrac{1 \pm \gamma^{\varphi}}{2}\right]\chi_0 \\
    &- \sum_{n=1}^{+\infty} \int \frac{d \rho d \vec{y}}{\rho^{d-1}} \bar \chi_n\left[\rho \left(\gamma^{\rho} \partial_{\rho} + \Vec{\gamma} \cdot \Vec{\partial} \right) - \frac{d-2}{2} \gamma^{\rho} + \frac{n\pi}{\theta} \right]\chi_n\,,
\end{split}
\end{equation} 
which includes the fermion zero mode. The formula for the free energy \eqref{freeenergyGN} then follows as before for the bosonic case.

To proceed with the evaluation of \eqref{freeenergyGN}, we first consider the free part $F_0$, which we determine in the small $\theta$ expansion and the result is given in \eqref{freeenerfreefermfinal}. 
We then study the interaction piece $I(\theta)$ that eventually produces the final result summarized in \eqref{eq:casimir_fermions_res}.

Applying the same trick as we have done in the scalar case, we differentiate $F_0(\theta)$ with respect to $\theta$ to obtain,
\begin{gather}
    \dfrac{\partial F_{0}(\theta)}{\partial \theta}=\frac{ N\textrm{Vol} (\mH^{d - 1}) }{(4 \pi)^{\frac{d - 1}{2}} \Gamma \left( \frac{d - 1}{2} \right) }   \sideset{}{'}{\sum}_{n}   \int \limits_{-\infty}^{+\infty} d \lambda \frac{ | \Gamma \left( \frac{d - 1}{2} + i \lambda \right)|^2 }{|\Gamma \left( \frac{1}{2} + i \lambda \right)|^2 } \dfrac{1}{\left( \lambda^2 + \frac{n^2 \pi^2 }{\theta^2}  \right) }\dfrac{n^2\pi^2}{\theta^{3}}\, ,
\end{gather}
where $\sideset{}{'}{\sum}_{n}$ means that  the $n=0$ mode is excluded and we will treat the contribution from the zero mode separately below. By closing the contour in upper half $\lambda$ plane, we have
\begin{gather}
     \dfrac{\partial F_{0}(\theta)}{\partial \theta}=\frac{N \textrm{Vol} (\mH^{d - 1}) }{(4 \pi)^{\frac{d - 1}{2}} \Gamma \left( \frac{d - 1}{2} \right) }   \sideset{}{'}{\sum}_{n}\dfrac{n^2\pi^2}{\theta^{3}}\bigg[\dfrac{\Gamma\left(\frac{d-1}{2}-\frac{n\pi}{\theta}\right)\Gamma\left(\frac{d-1}{2}+\frac{n\pi}{\theta}\right)}{\frac{n\pi}{\theta}}\cos{\frac{n\pi^2}{\theta}}\nonumber\\
     -\cot{\left(\frac{\left(d-1\right)\pi}{2}\right)}\dfrac{\Gamma\left(\frac{d-1}{2}-\frac{n\pi}{\theta}\right)\Gamma\left(\frac{d-1}{2}+\frac{n\pi}{\theta}\right)}{ \frac{n\pi}{\theta}}\sin{\frac{n\pi^2}{\theta}}\bigg] \, ,
\end{gather}
where the first term in the square brackets comes from the residues at $\lambda=i \frac{n\pi}{\theta}$ and the second term is the sum over residues at $\lambda=i\left(\frac{d-1}{2}+s\right)$ with $s=0,1,...$. The above can be further simplified to,
\begin{gather}
     \dfrac{\partial F_{0}(\theta)}{\partial \theta}=\frac{ N\textrm{Vol} (\mH^{d - 1})  \Gamma \left( \frac{3 - d}{2} \right) }{(4 \pi)^{\frac{d - 1}{2}}} \sideset{}{'}{\sum}_{n}\dfrac{n\pi}{\theta^{2}}\dfrac{\Gamma \left( \frac{d - 1}{2}+\frac{n\pi}{\theta} \right)}{\Gamma \left( \frac{3 - d}{2} + \frac{n\pi}{\theta} \right)}\, .
\end{gather}
Expanding in small $\theta$ with $\lim \limits_{z \to \infty}\frac{\Gamma(z+\alpha)}{\Gamma(z+\beta)}=z^{\alpha-\beta}$, and then integrating over $\theta$, we get
\ie 
    F_{0}(\theta)
    = \frac{ N\textrm{Vol} (\mH^{d - 1}) }{(4 \pi)^{\frac{d - 1}{2}}} \Gamma \left( \frac{1 - d}{2} \right)\dfrac{1}{2}\sideset{}{'}{\sum}_{n}\dfrac{\left(n\pi\right)^{d-1}}{\theta^{d-1}}\, .
\fe 
which can be evaluated using zeta regularization, 
\ie 
\label{freeenerfreefermfinal}
     &F_{0}^{++}(\theta)=F_{0}^{--}(\theta)=N\textrm{Vol} (\mH^{d - 1})\frac{   \pi^{-\frac{d}{2}} }{2^{d}\theta^{d-1}} \Gamma \left( \frac{ d}{2} \right)\zeta(d)\,, 
     \\
     &
    F_{0}^{+-}(\theta)=F_{0}^{-+}(\theta)=-N\textrm{Vol} (\mH^{d - 1})\frac{   \pi^{-\frac{d}{2}} }{2^{d}\theta^{d-1}} \Gamma \left( \frac{ d}{2} \right)\left(1-2^{1-d}\right)\zeta(d)\,.
\fe 
The above is consistent with the results in \cite{DePaola:1999im} where the slab geometry was considered.

For the case of $++/--$ boundary conditions we should be more careful, because we also need to take into account the contribution from the $n=0$ mode. As in the case of scalars, we need to incorporate the first correction to the free energy from the zero mode otherwise the wedge free energy will have IR divergences. By applying the same logic as in Section~\ref{sec:ONinterface} (see around \eqref{energycorrecONn=0}), we find that the correction to the zero mode energy in this case,
\begin{gather}
E^{(1)}_{0}= \frac{g_{*}}{2}\left(N_f-\frac{1}{2}\right)\int d^{d}x\bar{\chi}_{0}\chi_{0} \langle \bar{\psi}{\psi} (x) \rangle = \mathcal{O}(\epsilon^{2})\,,
\label{zeroenergycorrectGN++/--}
\end{gather}
where $\chi_{0}=\chi_{0}(\rho,\vec{y})$ is the zero mode and we have used \eqref{psibarpsiregularizaed}. 
Since we are only interested in the wedge free energy to leading orders in $\epsilon$, the $\theta$ dependent part of \eqref{freeenerfreefermfinal} does not receive additional contribution from the zero mode to this order. This is in contrast to the scalar case where the zero modes correct the Casimir energy to order $\epsilon^{3/2}$. Thus, as discussed around \eqref{GreenfuncHd-1zeromodes}, the two choices of boundary conditions on $\mH^{d-1}$ for  $n=0$ in the case of $++/--$  produce the same free energy in the leading order in $\epsilon$. 

To determine the contribution of the interaction term $I(\theta)$, at leading order in $\epsilon$, we  use  the one-point function computed in the previous subsection,
\begin{gather}
     \frac{g_{*}}{2}  \int d^{d}x \langle \bar{\psi}{\psi} (x)^2 \rangle= \frac{g_{*}}{2}N_f\left(N_f-\frac{1}{2}\right)  \int d^{d}x \langle \bar{\psi}{\psi} (x) \rangle^2 \, ,
     \label{intertermintegral}
\end{gather}
at $d=2$ given in \eqref{psibarpsid=2} since $g_{*} \sim \epsilon$. To regularize the divergent integral  over  $\varphi$, we could use the dimensional regularization similar to the scalar case (see around \eqref{dimreg1ptfunct}).
Alternatively we may also regularize the divergence in \eqref{intertermintegral} as follows: we start from the sum representation of the one-point function \eqref{psibarpsisum} and perform the integral over $\varphi$ first,
\ie 
     \int d^{d}x \langle \bar{\psi}{\psi} (x) \rangle^2
     = & \int dV_{\mH^{d-1}}\rho^{2-d} \frac{\theta}{2}\left(c_d\frac{2\pi    \Gamma\left( \frac{3 - d}{2} \right)  }{ \theta(4\pi)^{\frac{d}{2}} \sqrt{\pi} }\right)^{2}\sideset{}{'}{\sum}_{n}\left[ \left( \frac{ \Gamma\left( \frac{d - 1}{2} + \frac{\pi n}{\theta} \right) }{\Gamma\left( \frac{3 - d}{2} + \frac{\pi n}{\theta} \right)}\right)^2-\left(\dfrac{n \pi}{\theta}\right)^{2d-4}\right] 
     \\&
     +\int dV_{\mH^{d-1}}\rho^{2-d} \frac{\theta}{2}\left(c_d\frac{2\pi    \Gamma\left( \frac{3 - d}{2} \right)  }{ \theta(4\pi)^{\frac{d}{2}} \sqrt{\pi} }\right)^{2}\sideset{}{'}{\sum}_{n}\left(\dfrac{n \pi}{\theta}\right)^{2d-4}\,.
     \label{intertermsum}
\fe 
Now the sum in the first line above is convergent for $2<d<3$ and can be performed numerically, while the sum in the second line can be performed analytically with zeta regularization. Restricting to the $d=2$ case, where the sum in the first line of \eqref{intertermsum} vanishes, we find,
\ie 
    - \frac{g_{*}}{2}N_f\left(N_f-\frac{1}{2}\right) \int d^{2}x \langle \bar{\psi}{\psi} (x) \rangle^2=\begin{cases}
0 & +-/-+\,, \\
 \frac{\textrm{Vol}(H^{1})}{\theta}\frac{\pi}{16}\frac{N_f(2N_f-1)}{(N_f-1)}\epsilon &    ++/--\,.
 \end{cases} 
\label{freeenergcontrfrominterction}
\fe 

Thus, expanding the free fermion contribution \eqref{freeenerfreefermfinal} around $d=2+\epsilon$ and combining with \eqref{freeenergcontrfrominterction}, we obtain the Casimir energy below,
 \ie 
    \cE^{+-/-+}_{GN} =& -\dfrac{N_f \pi}{24}+\dfrac{\epsilon \pi}{48}N_f\left(\gamma+\log{\pi}-2\frac{\zeta'(2)}{\zeta(2)}\right)+\mathcal{O}(\epsilon^{2})\,, \\
    \cE^{++/--}_{GN} = &\dfrac{N_f \pi}{12}-\dfrac{\epsilon \pi}{24}N_f\left(\gamma+\log{4\pi}-2\frac{\zeta'(2)}{\zeta(2)}-\dfrac{3}{2}\dfrac{(2N_f-1)}{(N_f-1)}\right)+\mathcal{O}(\epsilon^{2})\,.
    \label{eq:casimir_fermions_res}
\fe 

\subsection{Numerical Treatment of the Large $N$ Model} \label{sec:NumericalLargeN}

In this section we will consider the critical $O(N)$ model in the wedge using lattice regularization. We will tackle this problem numerically directly at large $N$ (in the strict $N \rightarrow \infty$ limit), which should provide insights on the analytical treatment of the problem at the physical dimension $d=3$. Such an approach also serves as a complementary counterpart to what was discussed in the previous subsection. 

Let us briefly review the critical large $N$ vector models in flat space (see \cite{moshe2003quantum} for a more detailed review). We start with the following action as before
\begin{gather}
    S[\phi^I] = \int d^d x \left[\frac12 \left(\partial_\mu \phi^I\right)^2 + m_0^2 (\phi^I)^2 + \frac{\lambda_0}{4 N} \Lambda^{4-d} \left(\phi^I \phi^I\right)^2 \right]\,,
\end{gather}
where $m_0,\lambda_0$ are bare mass and coupling constants, which should be fine-tuned to bring the theory to criticality. We introduce the Hubbard-Stratonovich (HS) field $\sigma$ and integrate over the fields $\phi^I$ to obtain the following effective action for $\sigma$,
\begin{gather}
    S[\sigma] = N \left[\frac12 \tr \log\left[-\Delta+\sigma\right]+ \int d^d x \left(\mu_0 \sigma + \Lambda^{d-4} \frac{\sigma^2}{4\lambda_0}  \right) \right]\,, \quad \mu_0 = \frac{m_0}{\lambda_0}\,,
\end{gather}
which can be evaluated by saddle-point function in the large $N$ limit. The saddle-point equation for $\sigma$, also known as the gap equation, reads
\begin{gather}
    \left[\frac{1}{-\Delta+\sigma}\right]_{\mathcal{R}}(x,x) = \mu_0 + \Lambda^{d-4}\frac{\sigma}{4\lambda_0}\,,
    \label{gapeqn}
\end{gather}
after choosing a regularization scheme $\cR$. The notation $[..](x, x)$ above represents the trace over the matrix operator $[..]$. To reach criticality, we solve for $\mu_0=\mu_{\rm crit}$ by requiring that $\sigma=0$ in the flat space time. One can check that if $2<d<4$ the second term on the RHS of \eqref{gapeqn} is irrelevant in the IR and can be neglected. Thus to the leading order in the large $N$ limit, it boils down to solving the following equation,
\begin{gather}
    \left[\frac{1}{-\Delta+\sigma}\right]_{\mathcal{R}}(x,x) = \mu_{\rm crit}\,.
\end{gather}

Now we can apply this technique to our setup which involves the vector model on a wedge geometry. We work with the physical dimension $d=3$. Note that the metric and the HS field $\sigma$ on such a background take the following form
\begin{gather}
    ds^2 = d y^2 + d\rho^2 + \rho^2 d\varphi^2\,, \quad \varphi \in [0,\theta]\,,\quad
    \sigma(\rho,\varphi) = \frac{1}{\rho^2}\sigma(\varphi)\,,
    \label{wedgemetric}
\end{gather}
where $\sigma(\varphi)$ is some unknown function and to be determined by minimizing the effective action (free energy) for $\sigma$. Once $\sigma(\varphi)$ is fixed, then in principle we can determine any correlator in this geometry to leading non-trivial order in the large $N$ limit. Below, we will calculate the wedge free energy by taking the coincident point limit of the two-point function as we have done in the previous subsections using $\ep$ expansion. We will encounter some divergences in the process which we will regulate by working on a lattice. We will then minimize this free energy to determine the profile of $\sigma$ and hence get the free energy at the critical point. 

The scalar propagator of $\phi^I$ satisfies the following equation, 
\begin{gather*}
    \left(-\partial_y^2 -\partial_\rho^2 - \frac{1}{\rho}\partial_\rho + \frac{-\partial_\varphi^2 + \sigma(\varphi)}{\rho^2}\right)G(y,\rho,\varphi,y',\rho',\varphi') =\frac{1}{\rho}\delta(\rho-\rho') \delta(\varphi-\varphi')\delta(y-y')\,.
\end{gather*}
Note that the angular part is completely decoupled and we can find the eigenfunctions $\xi_n$ and the corresponding eigenvalues $E_n^2$ 
\begin{gather}
    \left(-\partial_\varphi^2 + \sigma(\varphi)\right) \xi_n(\varphi) = E_n^2 \xi_n(\varphi)\,, \quad \sum_n \xi^*_n(\varphi) \xi_n(\varphi') = \delta(\varphi-\varphi')\,.
    \label{eigenprobd3}
\end{gather}
Then using these eigenfunctions, we can expand the propagator into the following sum,
\ie 
G(y,\rho,\varphi,y',\rho',\varphi') = \sum_n \xi^*_n(\varphi) \xi_n(\varphi') \frac{1}{\sqrt{\rho \rho'}} G_n(\rho,\rho',y,y')\,, 
    \fe 
where $G_n$ is the propagator for a massive scalar field with mass $\sqrt{E_n^2 - 1/4}$ on $\mH^2$ which satisfies
\ie 
\left(-\rho^2(\partial_{y}^2 + \partial_\rho^2) + \left(E_n^2-1/4\right)\right)G_n(\rho,\rho',y,y') = \rho^2 \delta(\rho-\rho') \delta(y-y')\,.
\fe 
The explicit solutions are given by Legendre $Q$-functions \cite{POLYAKOV2008199}
\begin{gather}
   G_n(r,r',y,y') \equiv   G_n(Z) = Q_{-\frac12+E_n}(Z)\,,\quad Z \equiv 1+\frac{(\rho-\rho')^2 + (y-y')^2}{2 \rho \rho'}\,.
\end{gather}
Expanded near $Z=1$, the above becomes \cite{bateman1953higher}
\begin{gather}
    Q_{-\frac12+E_n}(Z) = \frac12 \log 2 - \gamma - \psi\left(E_n+\frac12\right) - \frac12 \log(Z-1) +\mathcal{O}(Z-1)\, ,
\end{gather}
where $\psi(x)$ is the digamma function. We will adopt the $\rm\overline{MS}$ scheme to regularize the propagator and obtain at coincident points,
\begin{gather}
   \rho  G(\rho,\varphi) =  -\sum_n \xi_n^*(\varphi) \xi_n(\varphi)\psi\left(E_n+\frac12\right)\,.
\end{gather}
This sum is still divergent, which we will handle with a lattice regularization. We first consider the case when there is no wedge (i.e. $\theta=2\pi$ in the metric \eqref{wedgemetric} with periodic boundary conditions) to benchmark the numerics. We approximate this circle by a lattice of $\cN$ sites located at the angles $\varphi_i= \frac{2\pi i}{\cN}$. In this situation, we can approximate the Schr\"{o}dinger operator in \eqref{eigenprobd3} by the following finite dimensional matrix $\cN \times \cN$,
\ie 
    H = -\left[\partial_\varphi^2\right]_{\cN\times \cN} + [\sigma(\varphi)]_{\cN\times \cN}\,, 
    \fe 
    with \ie 
    \left[\partial_\varphi^2\right]_{\cN\times \cN} = \frac{\cN^2}{(2\pi)^2} \begin{pmatrix}
    2 & -1 & 0 & \ldots & -1\\
    -1 & 2 &-1 & \ldots & 0\\
    0 & -1 & 2 & \ldots & 0\\
    \ldots & \ldots & \ldots & \ldots & \ldots \\
    -1 & 0 & 0 & \ldots & 2
    \end{pmatrix}\,, \quad [\sigma(\varphi)]_{\cN\times \cN} = \operatorname{diag}\left[\sigma\left(\frac{2\pi i}{\cN}\right)\right]^\cN_{i=1}\,.
\fe 
Such a matrix can be diagonalized using Matlab to find energies $\hat{E}_n$ and eigenfunctions $\hat{\xi}_n(\varphi_i)$, which are normalized as $\sum\limits^\cN_{n=1} \hat{\xi}_n(\varphi_i) \hat{\xi}^*_n(\varphi_j) = \cN \delta_{ij} $. The regularized propagator at coincident points then follows from
\begin{gather}
   \rho  G(\rho,\varphi_i) = - \sum^\cN_{n=1} \hat{\xi}_{n}^*(\varphi_i) \hat{\xi}_n(\varphi_i) \psi\left(E_n+\frac12\right)\,.
\end{gather}
In the large $N$ vector models, this regularized one-point function by the virtue of the equations of motions is equal to the parameter $\mu$
\begin{gather}
   -\sum^\cN_{n=1} \hat{\xi}_{n}^*(\varphi_i) \hat{\xi}_n(\varphi_i) \psi\left(E_n+\frac12\right) = \mu\,,
\end{gather}
and by setting $\mu$ to $\mu_{\rm crit}$ we tune the large $N$ vector model to criticality. In the absence of wedge, this simply corresponds to setting $\sigma(\varphi) \equiv 0$. In this situation we know the corresponding eigenfunctions and energies,
\begin{gather}
    \hat{\xi}_n(\varphi_i) = e^{\frac{2\pi i n}{\cN}}, \quad E_n = \frac{\cN^2}{\pi^2} \sin^2 \frac{\pi n}{\cN}\, , \quad \mu_{\rm crit} = -\sum^\cN_{n=1}  \psi\left(\frac{\cN^2}{\pi^2} \sin^2 \frac{\pi n}{\cN}+\frac12\right)\,.
\end{gather}
Now we will consider that in this lattice realization of the $S^1$ we chop a sub-lattice (a segment) of size $k$ and impose on the two ends the Dirichlet (ordinary)  boundary conditions.\footnote{The Neumann (special) boundary conditions could be implemented in a similar fashion, but the proposed numerical scheme becomes very unstable. This is because the special fixed point is not IR-stable in the boundary RG sense and can flow to ordinary fixed point as was discussed in \cite{Giombi:2020rmc}.} This would correspond to the study of the wedge CFT with angle $\theta=\frac{2\pi k}{\cN}$. Then we can apply the same techniques as before but instead of considering a Hamiltonian matrix of size $\cN\times \cN$, we consider one of size $k\times k$. Again, we need to solve the following spectral problem for new eigenvalues $\bar{E}_n$ and eigenfunctions $\bar{\xi}_n(\phi_i)$,
\begin{gather}
H_{k\times k} \bar{\xi}_{n} = \bar{E}_n \bar{\xi}_n, \quad \sum^k_{n=1}\bar{\xi}^*_{n}(\phi_i)\bar{\xi}_{n}(\phi_j) = \cN \delta_{ij}\,, \notag\\ 
-\sum^k_{n=1} \bar{\xi}_{n}^*(\varphi_i) \bar{\xi}_n(\varphi_i) \psi\left(\bar{E}_n+\frac12\right) = \mu_{\rm crit}=\sum^\cN_{n=1}  \psi\left(\frac{\cN^2}{\pi^2} \sin^2 \frac{\pi n}{\cN}+\frac12\right)\,.
\end{gather}
This equation could be solved using gradient descent by noticing that
\begin{gather}
    \frac{\partial F}{\partial \sigma} = -\sum^k_{n=1} \bar{\xi}_{n}^*(\varphi_i) \bar{\xi}_n(\varphi_i) \psi\left(\bar{E}_n+\frac12\right)  - \mu_{\rm crit}\,.
\end{gather}
In this way we can find numerically the $\sigma(\varphi)$ configuration for any opening angle $\theta$ by performing the above procedure for the corresponding value of $k$. For large enough $\cN$, we can approximate any opening angle in this way.  If we know the energy levels $E_n$, we can extract the free energy as below,
\begin{gather}
    \frac{F}{A} =\sum^k_{n=1}\left[-\frac{\bar{E}_n}{4\pi} \log\Gamma\left(\frac12+\bar{E}_n\right) +\frac{\psi^{(-2)}}{4\pi}\left(\frac12+\bar{E}_n\right) \right]  - \frac{\mu_{\rm crit}}{8\pi} \int d\varphi \sigma(\varphi) - f_0 \theta\,,
\end{gather}
where $f_0$ is a cosmological constant counterterm for the free energy. Let us emphasize that in the above formula, there is always an contribution coming from free energy of BCFT on the boundaries of the wedge (extensive in the boundary directions). These contributions must be explicitly subtracted to get the Casimir energy that captures the interactions between the boundaries.

\begin{figure}
    \centering
    \includegraphics[scale=0.9]{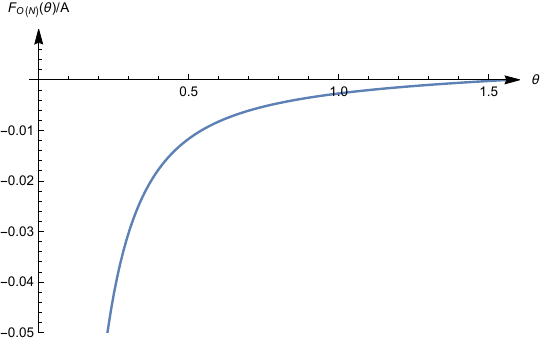}%
    \includegraphics[scale=0.8]{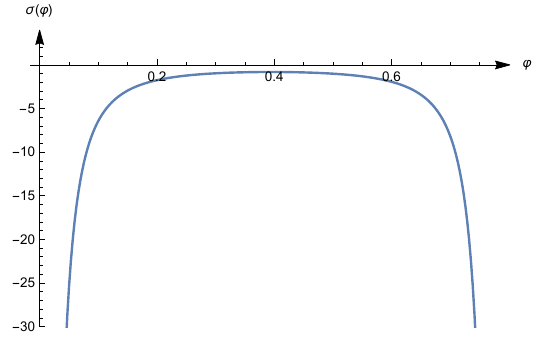}
    \caption{On the left panel the dependence of the Casimir energy on the angle of the wedge is displayed. On the right panel the typical profile of the HS field $\sigma$ for $\theta=\frac{\pi}{2}$.}
    \label{fig:FreeandHS}
\end{figure}

By implementing the above numerical procedure, we find that the typical profile of the HS field $\sigma$ is of the form presented in the right panel of the Figure~\ref{fig:FreeandHS}. Near the boundary of the wedge, we should recover the usual BCFT result. Indeed one can check that for small $\varphi$ the $\sigma$ field behaves as $\sigma \approx -\frac{0.25}{\theta^2}$, that is in agreement with the results of \cite{Giombi:2022vnz}.\footnote{For BCFT with ordinary boundary conditions, it was shown in \cite{Giombi:2022vnz} that at large $N$, the saddle point value of $\sigma$ is $(d - 2)(d - 4)/4$ which is $-1/4$ in $d = 3$.} In the left panel of the Figure~\ref{fig:FreeandHS} we present the numerically computed Casimir energy as a function of the opening angle of the wedge $\theta$. For small $\theta$, the dependence becomes power like $F(\theta) = -\frac{(2.45 \pm 0.05)\times 10^{-3}}{\theta^2}$. The $1/\theta^2$ dependence is again as expected at small $\theta$ from general considerations as discussed in Section~\ref{sec:Casimirandwedge}.

\section{Discussion}
\label{sec:discuss}

In this paper, we pursue a general investigation on the fusion of extended defect operators in conformal field theory of dimension $d>2$, which generalizes the more familiar operator product expansion for point operators. We define a fusion product for a pair of parallel defects after subtracting divergences at small transverse separation and explain general properties of the corresponding fusion algebra for defects. The aforementioned divergences are themselves physical observables that keep track of the interactions between the pair of defect. In particular, the leading divergence captures the Casimir energy due to the defect insertion. Furthermore we explain the relation between parallel defect fusion and a wedge-like defect configuration in CFT. The latter together with the utilization of hyperbolic spaces provides an efficient tool to extract the Casimir energy between interfaces (boundaries). 
To supplement the general discussion, we 
have determined these defect fusion data (e.g. fusion algebra and Casimir energy) for a variety of concrete examples including line defects and interfaces in the scalar $O(N)$ CFT and the fermionic Gross-Neveu(-Yukawa) CFT.

The ideas described here can be used to study the fusion of more general defects in CFT. One obvious target is the monodromy defect for global symmetries, such as
the $O(N)$ monodromy defect studied in \cite{Giombi:2021uae}. In the free theory of a single complex scalar $\Phi$, the $U(1)$ monodromy defect $\cD_{\vartheta}$ is defined by demanding that the scalar $\Phi$ picks up a phase as we go around the defect (using the same coordinates as in \eqref{cylindricalcoords} with the defect at $\rho=0$),
\begin{equation}
    \Phi (\vec{y},\rho, \varphi) = e^{i\vartheta} \Phi (\vec{y},\rho, \varphi + 2 \pi)\,, \quad \vartheta \sim \vartheta + 2 \pi\,.
\end{equation}
Intuitively, the fusion product for such a monodromy defect is $\cD_{\vartheta_1} \circ \cD_{\vartheta_2}  = \cD_{\vartheta_1 + \vartheta_2}$. However the Casimir energy is much less obvious. Another interesting example in this class of CFTs is the magnetic line defect in the $O(N)^3$ tensor model studied in \cite{Popov:2022nfq}. In this case, there are a large zoo of conformal line defects corresponding to different, inequivalent choices of the source term $J_{abc}$ under the action of $O(N)^3$, and because of that we expect the structure of defect fusion in the tensor models to be a lot more intricate.  
For the fusion of codimension-one defects, it would be interesting to consider cases with interactions localized on the boundary \cite{Giombi:2019enr, Behan:2020nsf, Behan:2021tcn}, which can define a nontrivial boundary condition even when the bulk theory is free. 
Another interesting phenomena to study will be the fusion of defects of different co-dimensions, for instance, the fusion of line defects with boundaries. Apart from exploring examples for defect fusion, it would also be important to understand general constraints on the fusion algebra and the Casimir energy from basic principles such as unitarity and locality. 

We have used the wedge geometry to analyze the fusion of conformal boundaries (also factorized interfaces) and in particular to extract the Casimir energy in Section~\ref{sec:ONinterface}. However, studying CFT on a wedge geometry is an interesting problem in its own right, which has received some recent attention \cite{Bissi:2022bgu, Antunes:2021qpy}. In particular, as was pointed out in \cite{Antunes:2021qpy}, one can write a crossing equation by demanding that the one-point function of bulk operators, when expanded in terms of operators on two different boundaries agree with each other. It will be interesting to exploit this crossing equation in explicit interacting examples such as the large $N$ critical $O(N)$ model or Gross-Neveu model. We expect the hyperbolic space method we have employed here to calculate the wedge free energy will be useful in this context. There is also potentially one additional channel where the bulk operator is expanded into operators at the edge which could lead to one more bootstrap equation for the wedge setup. It is worth exploring if this approach provides additional constraints on the wedge CFT data.

Finally, most of our calculations here in the explicit models utilized the $\epsilon$ expansion. It will be useful to develop complimentary approaches such as large $N$ expansion which may give us direct access to defect fusion in the strongly coupled CFT at $d=3$. It is possible to study this numerically directly in $d = 3$ as we have done in Section~\ref{sec:NumericalLargeN} for the fusion of ordinary boundary conditions in $d=3$ critical $O(N)$ model. But it will be desirable to develop an analytic approach at large $N$ in arbitrary dimensions which can serve as a cross-check of our results in $d = 4 - \epsilon$ which apply to arbitrary $N$.

 \section*{Acknowledgements}

The work of YW was
supported in part by the NSF grant PHY-2210420 and by the Simons Junior Faculty Fellows program.
F.K.P.
is currently a Simons Junior Fellow at New York University and supported
by a grant 855325FP from the Simons Foundation.
\appendix
\section{Fixed point for line defect in GNY to order $\epsilon$}\label{FixedPointNextOrder}
In this Appendix, we carry out the renormalization of the defect coupling $h_b$ to the
next order in bulk coupling constants $g^2_1, g_2$, by evaluating the one-point function of the bulk operator $s_0$. From this computation, we will also extract the scaling dimension of the defect operator $\hat s_0$. Note that this renormalization was previously studied in \cite{Barrat:2023ivo} and also \cite{Pannell:2023pwz} for more general setups. Here we work out the details explicitly for our case and fix a mistake in \cite{Barrat:2023ivo}.
We also provide consistency checks using large $N$ results.

Exploiting translation symmetry along the defect, we fix the position of the bulk operator to $X_{0}=(0,\eta)$. The points on the defect are denoted as $\hat{X}_{a}=(\tau_a,\bold{0})$, and we introduce the shorthand $X_{a \hat{b}}=\left(\tau_a-\tau_b,\eta_a\right)$. Throughout, we work within the minimal subtraction (MS) scheme.

When $g_{1,b}=g_{2,b}=0$, the scalar propagator and the fermion propagator (for a
fixed index $i$) are given by \cite{Fei:2015oha}
\begin{align}
\label{freeGreenfunctions}
G^\phi_d (X_1, X_2)= \langle s_{0}(X_1)s_0(X_2)\rangle_{0}=  \frac{C_\phi}{|X_{12}|^{d-2}}\,, \quad C_\phi\equiv \frac{\Gamma(\frac{d}{2}-1)}{4\pi^{\frac{d}{2}}}~,
\\
G^{\Psi}_d (X_1, X_2)= \langle \Psi_{0,i}(X_1)\bar{\Psi}_{0}^{i}(X_2)\rangle_{0}=  -C_\Psi\frac{\gamma^{\mu}(X_{12})_{\mu}}{|X_{12}|^{d}}\,, \quad C_\Psi\equiv \frac{\Gamma(\frac{d}{2})}{2\pi^{\frac{d}{2}}}~.
\end{align} 
The $\beta$-functions for the GNY model in $d=4-\epsilon$, up to the two-loop order,
are \cite{Fei:2016sgs}
\begin{gather}
\beta_{g^2_1}=-\epsilon g^2_1+\left(N+6\right)\frac{g^4_1}{(4\pi)^2}+\frac{1}{(4 \pi )^4}\left(-\frac{3}{2}  (4 N+3)g_1^6-4 g_1^4 g_2 +\frac{g^2_1 g_2 ^2}{6}\right)\,,\nonumber\\
    \beta_{g_2}=-  \epsilon g_2+\frac{1}{(4 \pi )^2}\left(3 g_{2} ^2+2 g_1^2 g_2  N-12 g_1^4 N\right)+\frac{1}{(4 \pi )^4}\left(96Ng_1^6 +7N g_1^4 g_2  -3N g^2_1 g_2 ^2 -\frac{17g_2^3}{3}\right)\,,
    \label{betafunctions}
\end{gather}
from which we can deduce the renormalization of $g^2_{1,b},g_{2,b}$ to the two-loop order:
\begin{gather}
 g^2_{1,b}=\mu^\epsilon \left(g^2_1+\frac{\left(N+6\right) g^4_1}{(4\pi)^2\epsilon}+\frac{1}{2(4 \pi )^4 \epsilon }\left(-\frac{3}{2}(4 N+3)g_1^6  -4g_1^4 g_2 +\frac{1}{6}g^2_1 g_2 ^2 \right)+\frac{(N+6)^2 g_1^6 }{(4 \pi )^4 \epsilon^2 }\right)\,, \nonumber \\
    g_{2,b}=\mu^{\epsilon}\Bigg(g_2+\frac{3g^2_2+2N g_2 g^2_1-12 N g^4_1}{(4\pi)^2\epsilon}+\frac{1}{2(4\pi)^4 \epsilon}\left(96 N g_1^6  +7Ng_1^4 g_2 -3Ng^2_1 g_2 ^2 - \frac{17g^3_2}{3}\right)\nonumber\\
    +\frac{1}{(4\pi)^4 \epsilon^2}\left(-24 N(N+3) g_1^6  +3N(N-10)g_1^4 g_2 +9Ng^2_1 g_2 ^2  +9g^3_2\right)\Bigg)\,.
    \label{barecouplings}
\end{gather}
There is an IR fixed point at:
\begin{gather}
   \frac{ g^2_{1\star}}{(4\pi)^2}=\frac{\epsilon}{N+6}+\frac{(N+66)\sqrt{N^2+132 N+36}-N^2+516N+882}{108(N+6)^3}\epsilon^2+\mathcal{O}(\epsilon^3)\,,\nonumber\\
   \frac{ g_{2\star}}{(4\pi)^2}=\frac{-N+6+\sqrt{N^2+132N+36}}{6(N+6)}\epsilon+\frac{1}{54 (N+6)^3 \sqrt{N^2+132N+36}}\nonumber\\
   \times\bigg(3N^4+155N^3+2745N^2-2538N+7344\nonumber\\
   -(3N^3-43N^2-1545N-1224)\sqrt{N^2+132N+36}\bigg)\epsilon^2+\mathcal{O}(\epsilon^3)\,.
   \label{fixedpoint}
\end{gather}

We introduce a line defect in the GNY model by adding  to $S_{\rm GNY}$ the following defect action,
\begin{align}\label{SDdef}
    S_D \equiv  h_b\int_{\mathbb{R}} d \tau\, s_0(\tau,\bold{0})\,, 
\end{align}

The one-point function of the bare operator $s_0$ to the first order in bulk coupling constants $g^2_1,g_2$ was computed in \cite{Giombi:2022vnz}. Their results can be summarized as
\begin{align}
\left\langle \phi_0(X_0)\right\rangle_{1-\text{order}} = A_0+B_{g^2_1}+B_{g_2}\,,
\end{align}
where
\begin{align}
       A_0&= -\frac{1}{\eta ^{d-3}}\frac{h_b \Gamma \left(\frac{d-3}{2}\right)}{4 \pi ^{\frac{d-1}{2}} }\,, \qquad
       B_{g_1}=\frac{1}{\eta ^{2 d-7}}\Gamma \left(\frac{d}{2}-1\right)^2 \Gamma \left(d-\frac{7}{2}\right)\frac{g^2_{1,b}N h_b }{ 64 \pi ^{d-\frac{1}{2}} (4-d) \Gamma (d-2)}\,,\\
       B_{g_2}&=\frac{1}{\eta ^{3 d-11}}\left(\frac{\Gamma \left(\frac{d-3}{2}\right)}{4 \pi ^{\frac{d-1}{2}}}\right)^3 \frac{g_{2,b}h^3_{b}}{12 (3 d-11) (4-d) }\,.
\end{align}
At next order in the coupling constants $g^2_1,g_2$, there are nine diagrams as shown in Figure~\ref{phinextorder}. We evaluate them systematically using the Mellin-Barnes method (see for example Appendices in \cite{Diatlyk:2024ngd}).

Diagrams $\mathcal{I}^{}_{0,2,1}$, $\mathcal{I}^{}_{0,2,3}$, $\mathcal{I}^{}_{0,2,5}$ were computed in \cite{PhysRevB.90.035131} when discussing scalar $O(N)$ model, and the result for $N=1$ can be written as
\begin{figure}[h!]
\centering
\begin{tikzpicture} 
\draw[densely dashed] [line width = 0.5mm] (-1.2, -1) to (1.2,-1);
\draw  [densely dashed] (0,1) node[vertex]{}   (0,0.3)node[vertex]{} to[out=-45,in=45]  (0,-0.3)node[vertex]{}  (0,-1)node[vertex]{} to[out=90,in=90]  (0,-0.3) to[out=135,in=-135]  (0,0.3) to[out=90,in=90]  (0,1);
\draw [densely dashed] (0,0.3)node[vertex]{} to[out=90,in=90]  (0,-0.3);
\node at (0.1,-1.5) {$\mathcal{I}^{}_{0,2,1}$};
\draw  (0,0.3)node[vertex,red]{} ;
\draw  (0,-0.3)node[vertex,red]{} ;
\end{tikzpicture}
\qquad
\begin{tikzpicture} 
\draw[densely dashed] [line width = 0.5mm] (-1.2, -1) to (1.2,-1);
\draw [densely dashed] (-0.5,1) to  (-0.5,-1);
\draw [densely dashed] (-0.5,0) to  (0.5,0);
\draw [densely dashed] (-0.5,0) to [out=45,in=135]  (0.5,0);
\draw [densely dashed] (0.5,0) to  (0,-1);
\draw [densely dashed] (0.5,0) to  (0.9,-1);
\node at (0.1,-1.5) {$\mathcal{I}^{}_{0,2,3}$};
\draw  (-0.5,0)node[vertex,red]{} ;
\draw  (0.5,0)node[vertex,red]{} ;
\draw  (-0.5,1)node[vertex,black]{} ;
\draw  (0,-1)node[vertex,black]{} ;
\draw  (0.9,-1)node[vertex,black]{} ;
\draw  (-0.5,-1)node[vertex,black]{} ;
\end{tikzpicture}
\qquad
\begin{tikzpicture} 
\draw[densely dashed] [line width = 0.5mm] (-1.2, -1) to (1.2,-1);
\draw  [densely dashed] (0,1) node[vertex]{}   (0,0.3)node[vertex]{}  (0,-0.3)node[vertex]{}  (0,-1)node[vertex]{} to[out=90,in=90]  (0,-0.3)   (0,0.3) to[out=90,in=90]  (0,1);
\draw [densely dashed] (0,0.3)node[vertex]{} to[out=90,in=90]  (0,-0.3);
\draw [densely dashed] (0,0.3) to  (-0.6,-1);
\draw [densely dashed] (0,0.3) to  (0.6,-1);
\draw [densely dashed] (0,-0.3) to  (0.3,-1);
\draw [densely dashed] (0,-0.3) to  (-0.3,-1);
\node at (0.1,-1.5) {$\mathcal{I}^{}_{0,2,5}$};
\draw  (0,0.3)node[vertex,red]{} ;
\draw  (0,-0.3)node[vertex,red]{} ;
\draw  (-0.3,-1)node[vertex,black]{} ;
\draw  (-0.6,-1)node[vertex,black]{} ;
\draw  (0.3,-1)node[vertex,black]{} ;
\draw  (0.6,-1)node[vertex,black]{} ;
\end{tikzpicture}
\qquad
\begin{tikzpicture} 
\draw[densely dashed] [line width = 0.5mm] (-1.2, -1) to (1.2,-1);

\draw [rotate=90, shift={(0.4,0.6)}] 
    (-0.25,-0.6) node[]{} 
    to [out=45,in=135] (0.25,-0.6) node[]{} 
    [out=-135,in=-45] 
    to (-0.25,-0.6); 

\draw [densely dashed] (0,1) to  (0,0.6);
\draw [densely dashed] (0,0.2) to  (0,-1);
\draw [densely dashed] (0,-0.3) to  (0.3,-1);
\draw [densely dashed] (0,-0.3) to  (-0.3,-1);
\node at (0.1,-1.5) {$\mathcal{I}^{(1)}_{2,1,3}$};
\draw  (0,1)node[vertex,black]{} ;
\draw (0,0.6) node[rectangle, fill=blue, inner sep=0pt, minimum width=4pt, minimum height=4pt]{} ;
\draw (0,0.2) node[rectangle, fill=blue, inner sep=0pt, minimum width=4pt, minimum height=4pt]{} ;
\draw  (0,-0.3)node[vertex,red]{} ;
\draw  (-0.3,-1)node[vertex,black]{} ;
\draw  (0.3,-1)node[vertex,black]{} ;
\draw  (0,-1)node[vertex,black]{} ;
\end{tikzpicture}
\qquad
\begin{tikzpicture} 
\draw[densely dashed] [line width = 0.5mm] (-1.2, -1) to (1.2,-1);
\draw  [densely dashed] (0,1) node[vertex]{}   (0,0.3)  (0,-0.3)  (0,-1) to[out=90,in=90]  (0,-0.3)   (0,0.3) to[out=90,in=90]  (0,1);
\draw [densely dashed] (0,0.3) to[out=90,in=90]  (0,-0.3);
\draw [densely dashed] (0,0.3) to  (-0.6,-1);
\draw [densely dashed] (0,0.3) to  (0.25,-0.2);
\draw [densely dashed] (0.6,-1) to  (0.42,-0.6);

\draw [rotate=-65, shift={(0.5,0.73)}] 
    (-0.25,-0.6) node[]{} 
    to [out=45,in=135] (0.25,-0.6) node[]{} 
    [out=-135,in=-45] 
    to (-0.25,-0.6); 

\draw (0.42,-0.6) node[rectangle, fill=blue, inner sep=0pt, minimum width=4pt, minimum height=4pt,rotate =-65]{} ;
\draw (0.25,-0.2) node[rectangle, fill=blue, inner sep=0pt, minimum width=4pt, minimum height=4pt,rotate =-65]{} ;
\node at (0.1,-1.5) {$\mathcal{I}^{(2)}_{2,1,3}$};
\draw  (0,0.3)node[vertex,red]{} ;
\draw  (-0.6,-1)node[vertex,black]{} ;
\draw  (0,-1)node[vertex,black]{} ;

\draw  (0.6,-1)node[vertex,black]{} ;
\end{tikzpicture}
\qquad
\begin{tikzpicture} [rotate=90,scale=0.6]
\draw [densely dashed][line width = 0.5mm] (-1, 2) to (-1,-2);
\draw[color=black,densely dashed] (-1,0) to (-0.4,0); 
\draw[color=black,densely dashed] (0.4,0) to (1,0);
\draw[color=black,densely dashed] (1.8,0) to (2.4,0); 
\draw(-1,0)node[vertex]{} ;
\draw (2.4,0)node[vertex]{} ;
\draw (-0.4, 0) to[out=-45,in=-135]    (0.4,0) ;
\draw (-0.4, 0) to[out=45,in=135]    (0.4,0) ;
\draw [densely dashed] (0.4, 0) to    (1,0) ;
\node[rectangle, fill=blue, inner sep=0pt, minimum width=4pt, minimum height=4pt] at (-0.4, 0) {};
\node[rectangle, fill=blue, inner sep=0pt, minimum width=4pt, minimum height=4pt] at (0.4, 0) {};
\draw (1, 0) to[out=-45,in=-135]    (1.8,0) ;
\draw (1, 0) to[out=45,in=135]    (1.8,0) ;
\node[rectangle, fill=blue, inner sep=0pt, minimum width=4pt, minimum height=4pt] at (1, 0) {};
\node[rectangle, fill=blue, inner sep=0pt, minimum width=4pt, minimum height=4pt] at (1.8, 0) {};
\node at  (-1.75,0) {$\mathcal{I}^{(1)}_{4,0,1}$};
\end{tikzpicture} 
\qquad
\begin{tikzpicture} [rotate=270]
\draw [densely dashed][line width = 0.5mm] (1, 1) to (1,-1);
\draw[densely dashed] (1,0) to (0.6,0);
\draw[densely dashed] (-1,0) to (-0.4,0);
\draw (-1,0)node[vertex]{} ;
\draw (1,0)node[vertex]{} ;
\draw (-0.4, 0) to    (0.4,0) ;
\draw (-0.4, 0) to    (-0.4,0.55) ;
\draw (-0.4, 0.55) to    (0.4,0.55) ;
\draw (0.4, 0) to   (0.4,0.55) ;
\draw [densely dashed] (-0.4, 0.55) to[out=90,in=90]    (0.4,0.55) ;
\node[rectangle, fill=blue, inner sep=0pt, minimum width=4pt, minimum height=4pt] at (-0.4, 0) {};
\node[rectangle, fill=blue, inner sep=0pt, minimum width=4pt, minimum height=4pt] at (0.4, 0) {};

\node[rectangle, fill=blue, inner sep=0pt, minimum width=4pt, minimum height=4pt] at (-0.4, 0.55) {};
\node[rectangle, fill=blue, inner sep=0pt, minimum width=4pt, minimum height=4pt] at (0.4, 0.55) {};
\node at (1.5,0) {$\mathcal{I}^{(2)}_{4,0,1}$};
\end{tikzpicture} 
\qquad
\begin{tikzpicture} [rotate=270]
\draw [densely dashed][line width = 0.5mm] (1, 1) to (1,-1);
\draw[densely dashed] (1,0) to (0.6,0);
\draw[densely dashed] (-1,0) to (-0.4,0);
\draw (-1,0)node[vertex]{} ;
\draw (1,0)node[vertex]{} ;

\draw (-0.4, 0) to    (-0.1,0.20625) ;
\draw    (0.1,0.34375) to (0.4,0.55);
\draw (-0.4, 0) to    (-0.4,0.55) ;
\draw (-0.4, 0.55) to    (0.4,0) ;
\draw (0.4, 0) to   (0.4,0.55) ;
\draw [densely dashed] (-0.4, 0.55) to[out=90,in=90]    (0.4,0.55) ;
\node[rectangle, fill=blue, inner sep=0pt, minimum width=4pt, minimum height=4pt] at (-0.4, 0) {};
\node[rectangle, fill=blue, inner sep=0pt, minimum width=4pt, minimum height=4pt] at (0.4, 0) {};

\node[rectangle, fill=blue, inner sep=0pt, minimum width=4pt, minimum height=4pt] at (-0.4, 0.55) {};
\node[rectangle, fill=blue, inner sep=0pt, minimum width=4pt, minimum height=4pt] at (0.4, 0.55) {};
\node at (1.5,0) {$\mathcal{I}^{(3)}_{4,0,1}$};
\end{tikzpicture} 
\qquad
\begin{tikzpicture} [rotate=270]
\draw [densely dashed][line width = 0.5mm] (1, 1) to (1,-1);
\begin{scope}[shift={(0,-0.4)}]
\draw[densely dashed] (1,0) to (0.6,0);
\draw[densely dashed] (-1,0) to (-0.4,0);
\draw (-1,0)node[vertex]{} ;
\draw (1,0)node[vertex]{} ;
\draw [densely dashed] (0.4, 0.55) to    (1,0.55) ;
\draw [densely dashed] (-0.4, 0.55) to    (1,1.1) ;
\draw (-0.4, 0) to    (0.4,0) ;
\draw (-0.4, 0) to    (-0.4,0.55) ;
\draw (-0.4, 0.55) to    (0.4,0.55) ;
\draw (0.4, 0) to   (0.4,0.55) ;
\draw (1,1.1)node[vertex]{} ;
\draw (1,0.55)node[vertex]{} ;
\node[rectangle, fill=blue, inner sep=0pt, minimum width=4pt, minimum height=4pt] at (-0.4, 0) {};
\node[rectangle, fill=blue, inner sep=0pt, minimum width=4pt, minimum height=4pt] at (0.4, 0) {};

\node[rectangle, fill=blue, inner sep=0pt, minimum width=4pt, minimum height=4pt] at (-0.4, 0.55) {};
\node[rectangle, fill=blue, inner sep=0pt, minimum width=4pt, minimum height=4pt] at (0.4, 0.55) {};
\end{scope}
\node at (1.5,0) {$\mathcal{I}^{(4)}_{4,0,1}$};
\end{tikzpicture} 
\caption{Diagrams contributing to the one-point function  $\langle s_{0}(X_0)\rangle$ in GNY model up to second order in bulk coupling constants $g^2_{1,b}$ and $g_{2,b}$. Blue vertices represent insertions of the Yukawa coupling $g_{1,b}$ , while red vertices denote scalar self-interaction vertices associated with the coupling $g_{2,b}$.}
\label{phinextorder}
\end{figure}

\begin{gather}
     \mathcal{I}^{}_{0,2,1}=\frac{-h_b g^2_{2,b}}{\eta^{3d-11}}\left(\frac{\sqrt{\pi } \Gamma \left(\frac{3 d-11}{2}\right)}{\Gamma \left(\frac{3 d-10}{2}\right)}\right)\frac{(d-4)^2  \Gamma \left(\frac{d-4}{2}\right)^3}{3\times 2^{12} \pi^{\frac{3d}{2}} (d-3) (3 d-10) (3 d-8)}\,,\nonumber\\
     \mathcal{I}^{}_{0,2,3}=-\frac{g^2_{2,b} h^3_b}{\eta ^{4 d-15}}\frac{\csc (\pi  d) \csc \left(\frac{3 \pi  d}{2}\right) \Gamma \left(\frac{d-3}{2}\right)}{ 2^{d+8} \pi ^{2 d-5} (4 d-15) \Gamma \left(\frac{14-3 d}{2}\right) \Gamma \left(\frac{9}{2}-d\right) \Gamma (d-2)}\,,\nonumber\\
     \mathcal{I}^{}_{0,2,5}=-\frac{g^2_{2,b}h^5_b}{\eta^{5d-19}}\frac{\Gamma \left(\frac{d-3}{2}\right)^5 }{3 \times 2^{15} \pi ^{\frac{5 (d-1)}{2}} (d-4)^2 (3 d-11) (5 d-19) }\,,
\end{gather}
Next two diagrams are easy and are given by
\begin{gather}
\mathcal{I}^{(1)}_{2,1,3}=\frac{ g^2_{1,b}g_{2,b }N h^3_b}{3!}C^5_{\phi}C^2_{\psi}\int \frac{d^{d} X_1d^{d} X_2 d^{d} X_3}{|X_{01}|^{d-2}|X_{12}|^{2d-2}|X_{23}|^{d-2}}\left(\int \limits_{-\infty}^{+\infty} \frac{d\tau}{\left(\tau^2+\eta^2_3\right)^{\frac{d-2}{2}}}\right)^3\nonumber\\
    =\frac{ g^2_{1,b}g_{2,b }N h^3_b}{\eta^{4d-15}}\frac{\csc (\pi  d) \Gamma \left(\frac{4-d}{2}\right) \Gamma \left(\frac{d-3}{2}\right) \Gamma \left(\frac{4d-15}{2}\right)}{2^{2 d+7} \pi ^{\frac{1}{2} (4 d-7)} \Gamma \left(\frac{14-3 d}{2}\right) \Gamma \left(\frac{3 d-7}{2}\right)}\,,
\end{gather}
as well as
\begin{gather}
\mathcal{I}^{(2)}_{2,1,3}=\frac{ g^2_{1,b}g_{2,b }N h^3_0}{2}C^5_{\phi}C^2_{\psi}\int \frac{d^{d} X_1d^{d} X_2 d^{d} X_3 d \tau_4 d \tau_5 d \tau_6}{|X_{03}|^{d-2}|X_{3\hat{4}}|^{d-2}|X_{3\hat{5}}|^{d-2}|X_{13}|^{d-2}|X_{2\hat{6}}|^{d-2}|X_{12}|^{2d-2}}\nonumber\\
    =\frac{g^2_{1,b}g_{2,b }N h^3_0}{\eta^{4d-15}}\frac{ \Gamma (2 d-8)}{3(4 \pi )^{2 d-3} (4-d) (d-3) (4 d-15)}\,.
\end{gather}

Calculating diagrams $\mathcal{I}^{(1)}_{4,0,1}-\mathcal{I}^{(3)}_{4,0,1}$ is equivalent to evaluating the integrated bulk two-point function $\int_{\mathbb R} d \tau_2 \langle s_0(X_1)s_0(X_2)\rangle$ without the line defect with $X_{2}=(\tau_2,0)$. Divergent part of 
$\langle s_0(X_1)s_0(X_2)\rangle$ can be deduced from bulk renormalization. 
The integral over $\tau_2$ takes the following simple form
\begin{gather}
    \int_{\mathbb{R}}  \frac{d \tau_{2}}{\left(\tau^2_{2}+\eta^{2}_{1}\right)^{\frac{3d-10}{2}}}=\frac{\sqrt{\pi } \Gamma \left(\frac{3 d-11}{2} \right)}{\Gamma \left(\frac{3 d-10}{2} \right)}\frac{1}{\eta^{3d-11}_{1}}\,.
\end{gather}
By denoting term of order $g^4_{1,b}$ in bulk two point function as $\langle s_0(X_1)s_0(X_2)\rangle_{g^4_{1,b}}$, we find
\begin{gather}
\mathcal{I}^{(1)}_{4,0,1}+\mathcal{I}^{(2)}_{4,0,1}+\mathcal{I}^{(3)}_{4,0,1}=-h_b\int_{\mathbb R} d \tau_2 \langle s_0(X_1)s_0(X_2)\rangle_{g^4_{1,0}}\nonumber\\
=\frac{g^4_{1,b}N h_b}{\eta^{3d-11}}\left(-\frac{N+3}{2^{10} \pi ^5 \epsilon ^2}-\frac{29+8 N+6 (N+3) \log \left(4 \pi  e^{\gamma_{E} }\right)}{2^{12} \pi ^5 \epsilon }\right)+\mathcal{O}(\epsilon^0)\,.
\end{gather}
Finally, the last diagram is slightly more difficult to compute. However, we can determine its divergent part without evaluating the full expression.\footnote{We also evaluated this diagram explicitly and  confirm this result.} This diagram is linear in $N$ and should have at most quadratic divergence in $\epsilon$:
\begin{gather}
\mathcal{I}^{(4)}_{4,0,1}=\frac{g^{4}_{1,b}N h^3_b}{\eta^{4d-15}}\left(\frac{a_2}{\epsilon^2}\frac{2}{3(4\pi)^5}+\frac{1}{\epsilon}\frac{a_1+60-\pi ^2+24 \log \left(4 \pi  e^{\gamma_E }\right)}{18(4\pi)^5}\right)+\mathcal{O}(\epsilon^0)\,,
    \label{I^(4)divergentpart}
\end{gather}
where $a_1,a_2$ are constants to be determined. We can set $a_2=1$ by requiring that renormalized one-point function of $s$ should not have $\frac{g^{4}_{1}N h^3}{\epsilon^2}$ and $\frac{g^{4}_{1}N h^3\log{\eta}}{\epsilon}$ terms. We will further show $a_1=0$ from large $N$ result of fixed point in \cite{Giombi:2022vnz}.

Altogether, up to the second order in bulk coupling constants $g^2_1,g_2$, the one-point function of the renormalized bulk operator $s \equiv Z_{\phi}^{-\frac{1}{2}} s_0$ reads
\begin{gather}
\label{dd}
    \left\langle s(0,\eta)\right\rangle_{2-\text{order}} = Z_{\phi}^{-\frac{1}{2}}\bigg(A_0+B_{g^2_1}+B_{g_2}+ \mathcal{I}^{}_{0,2,1}+\mathcal{I}^{}_{0,2,3}+\mathcal{I}^{}_{0,2,5}+\mathcal{I}^{(1)}_{2,1,3}+\mathcal{I}^{(2)}_{2,1,3}\\
+ \mathcal{I}^{(1)}_{4,0,1}+\mathcal{I}^{(2)}_{4,0,1}+\mathcal{I}^{(3)}_{4,0,1}+\mathcal{I}^{(4)}_{4,0,1}\bigg)\,, \nonumber
\end{gather}
where  the wavefunction renormalization $Z_{\phi}$ is fully determined by the bulk theory (see for example \cite{Fei:2016sgs}):
\begin{gather}\label{Zphi}
    Z_{\phi}=1-\frac{g^2_{1}N}{(4\pi)^2 \epsilon}-\frac{g_{2}^{2}}{(4\pi)^{4}}\frac{1}{12 \epsilon}+\frac{g^4_{1}N}{(4\pi)^4}\left(-\frac{3}{\epsilon^2}+\frac{5}{4\epsilon}\right)\,.
\end{gather}

\subsection{Fixed point and the scaling dimension of $\hat\phi$}
By requiring the one-point function $\left\langle s\right\rangle$ to be finite in the $\epsilon\to 0$ limit after making the substitution \eqref{barecouplings}, we find that $a_2=1$ and the  renormalization of the defect coupling to be 
\begin{gather}
    h_b = \mu^{\frac{\epsilon}{2}}\bigg[h+\frac{g^2_1 N h }{2 (4 \pi )^2 \epsilon }+\frac{g_2 h^3  }{12 (4 \pi )^2 \epsilon }+\frac{g_{2}^2}{\epsilon (4 \pi )^4}\left(\frac{3 h }{72 }-\frac{9 h^3 }{108 }-\frac{h^5 }{48 
    }\right)+\frac{g_{2}^2}{\epsilon ^2 (4 \pi )^4}\left(\frac{9 h^3 }{108 }+\frac{h^5 }{96 }\right)\nonumber\\
    +\frac{g_1^2 g_2 h^3 N }{(4 \pi )^4  }\left(\frac{1}{8\epsilon^2}-\frac{1}{12\epsilon}\right)+\frac{g_1^4 N h  }{(4 \pi )^4  }\left(\frac{3(N+4)}{8\epsilon^2}-\frac{5}{8\epsilon}\right)
    +\frac{g_1^4 N  h^3  }{(4 \pi )^4  }\left(-\frac{1}{3\epsilon^2}+\frac{a_1+6-\pi ^2}{18\epsilon}\right)\bigg]\,,
\end{gather}
where in the first line terms which contains only powers of $g_2$ coincide with the result on ${\rm O}(N)$ model in \cite{Cuomo:2021kfm} (see equations (3.15) -- (3.17)).
Imposing $\mu\partial_\mu h_b=0$ and using $\beta$-functions from \eqref{betafunctions}, we obtain the beta function of $h$,\footnote{
Note that this corrects a mistake in (B.11) of \cite{Barrat:2023ivo}.}
\begin{gather}
\beta_h = -\frac{\epsilon}{2} h+\frac{g_2}{(4\pi) ^2}\frac{ h^3  }{6 }+\frac{g^2_1 N h}{2(4\pi)^2}+\frac{g^2_2}{(4\pi) ^4}\left(\frac{h}{12}-\frac{h^3}{4}-\frac{h^5}{12}\right)\nonumber\\
-\frac{g^2_1  g_2 Nh^3 }{4 (4 \pi )^4}+\frac{g^4_1N}{(4\pi)^4}\left(-\frac{5h}{4}+\frac{a_1+6-\pi ^2}{6}h^3  \right)\,.
\end{gather}

We are looking for the solution in the form:
\begin{gather}
    h^2_{\star}=\frac{108}{6-N+\sqrt{N^2+132 N+36}}+H \epsilon\,.
\end{gather}
By plugging the bulk fixed point $g_{1\star},g_{2\star}$ from \eqref{fixedpoint} into $\beta_h = 0$, we identify the defect fixed point to the $\epsilon$ order as a function of $a_1$:
\begin{gather}
    H=\frac{31 N^3+8739 N^2+180954 N+31536}{2 (N+6) \left(-N^3-126 N^2+756 N+216+\left(N^2+60 N+36\right) \sqrt{N^2+132 N+36}\right)}\nonumber\\
    +\frac{\sqrt{N^2+132 N+36} \left(-31 N^2+\left(648 \pi ^2-648 a_1-3453\right) N+5256\right)}{2 (N+6) \left(-N^3-126 N^2+756 N+216+\left(N^2+60 N+36\right) \sqrt{N^2+132 N+36}\right)}\,.
    \label{Hgena1}
\end{gather}
Its large $N$ expansion has the following simple form:
\begin{gather}
    h^2_{\star}=\frac{3}{2}+\frac{5+\pi ^2-a_1}{8} \epsilon +\mathcal{O}\left(\frac{1}{N}\right)\,.
\end{gather}

In \cite{Giombi:2022vnz} the fixed point $h_\star$ at large $N$ was computed. In order to use their result we need to relate $\sigma$ field at large $N$ to the field $s$ in GNY in $d=4-\epsilon$:
\begin{gather}
    h^2_\star=g^2_{1,\star} N \left(h^{large N}_{\star,there}\right)^2=g^2_{1,\star} N \left(\frac{2^{3-2 d} (d-3) \csc \left(\frac{\pi  d}{2}\right) \Gamma (d)}{(d-2)^2 d \Gamma \left(\frac{d-1}{2}\right)^2\left((d-3) H_{\frac{d}{2}-2}+(3-d) H_{d-4}-1\right)}\right)\bigg|_{d=4-\epsilon}\nonumber\\
=\frac{3}{2}+\frac{5+\pi ^2}{8} \epsilon+\mathcal{O}\left(\frac{1}{N},\epsilon^2\right)\,.
\end{gather}
From this we see that we need to set $a_1=0$ in \eqref{Hgena1} which then completely determines $h_\star$ at finite $N$ to order $\ep$.


Let us provide some further consistency checks on the fixed point coupling by comparing to large $N$ DCFT data.
The scaling dimension of the  defect operator $\hat{s}$ at the fixed point is,
\begin{gather}
\Delta_{\hat s} = 1+\left.\frac{\partial\beta_h}{\partial h}\right|_{h_\star,g_{1\star},g_{2\star}} = 1+\frac{6}{N+6}\epsilon\nonumber\\
-\frac{\epsilon^2}{18(N+6)^3}\left(52 N^2+915 N+5868+(11 N+6) \sqrt{N^2+132N+36}\right)\,.
\label{scalingHATphi^2epsilon}
\end{gather}
At large $N$ the leading behaviour is \begin{gather}
    \Delta_{\hat s}=1+\frac{1}{N}\left(6 \epsilon-\frac{7}{2}\epsilon^2\right)\,,
\end{gather}
which coincide with large $N$ result in  \cite{Giombi:2022vnz} (see eq. (3.15)):
\begin{gather}
    \Delta(\hat{\sigma})=1-\frac{1}{N}\frac{2^{d+2} (d-1) \sin \left(\frac{\pi  d}{2}\right) \Gamma \left(\frac{d-1}{2}\right)}{\pi ^{3/2} d (d-2) \Gamma \left(\frac{d}{2}-1\right)}\bigg|_{d=4-\epsilon}=1+\frac{1}{N}\left(6 \epsilon-\frac{7}{2}\epsilon^2\right)\,.
\end{gather}
Additionally, we  have also matched the one-point function of $s$ with that from the large $N$ analysis in \cite{Giombi:2022vnz} to leading order in $\ep$ (i.e. focusing on the first three  terms in \eqref{dd}).
 
\bibliographystyle{JHEP}
\bibliography{fusion}

@article{Pannell:2023pwz,
    author = "Pannell, William H. and Stergiou, Andreas",
    title = "{Line defect RG flows in the {\ensuremath{\varepsilon}} expansion}",
    eprint = "2302.14069",
    archivePrefix = "arXiv",
    primaryClass = "hep-th",
    doi = "10.1007/JHEP06(2023)186",
    journal = "JHEP",
    volume = "06",
    pages = "186",
    year = "2023"
}

@article{Kapustin:2007wm,
    author = "Kapustin, Anton and Saulina, Natalia",
    title = "{The Algebra of Wilson-'t Hooft operators}",
    eprint = "0710.2097",
    archivePrefix = "arXiv",
    primaryClass = "hep-th",
    doi = "10.1016/j.nuclphysb.2009.02.004",
    journal = "Nucl. Phys. B",
    volume = "814",
    pages = "327--365",
    year = "2009"
}

@article{Drukker:2012de,
    author = "Drukker, Nadav",
    title = "{Integrable Wilson loops}",
    eprint = "1203.1617",
    archivePrefix = "arXiv",
    primaryClass = "hep-th",
    doi = "10.1007/JHEP10(2013)135",
    journal = "JHEP",
    volume = "10",
    pages = "135",
    year = "2013"
}

@article{Drukker:2022beq,
    author = "Drukker, Nadav and Tr\'epanier, Maxime",
    title = "{Ironing out the crease}",
    eprint = "2204.12627",
    archivePrefix = "arXiv",
    primaryClass = "hep-th",
    doi = "10.1007/JHEP08(2022)193",
    journal = "JHEP",
    volume = "08",
    pages = "193",
    year = "2022"
}

@article{Drukker:2011za,
    author = "Drukker, Nadav and Forini, Valentina",
    title = "{Generalized quark-antiquark potential at weak and strong coupling}",
    eprint = "1105.5144",
    archivePrefix = "arXiv",
    primaryClass = "hep-th",
    reportNumber = "IMPERIAL-TP-2011-ND-02, NSF-KITP-11-073, AEI-2011-027",
    doi = "10.1007/JHEP06(2011)131",
    journal = "JHEP",
    volume = "06",
    pages = "131",
    year = "2011"
}

@article{Correa:2012at,
    author = "Correa, Diego and Henn, Johannes and Maldacena, Juan and Sever, Amit",
    title = "{An exact formula for the radiation of a moving quark in N=4 super Yang Mills}",
    eprint = "1202.4455",
    archivePrefix = "arXiv",
    primaryClass = "hep-th",
    doi = "10.1007/JHEP06(2012)048",
    journal = "JHEP",
    volume = "06",
    pages = "048",
    year = "2012"
}

@article{Polyakov:1972ay,
    author = "Polyakov, A. M.",
    title = "{Effect of strong interactions on vacuum polarization}",
    journal = "Zh. Eksp. Teor. Fiz.",
    volume = "63",
    pages = "24--34",
    year = "1972"
}

@article{Bachas:2004sy,
    author = "Bachas, Constantin and Gaberdiel, Matthias",
    title = "{Loop operators and the Kondo problem}",
    eprint = "hep-th/0411067",
    archivePrefix = "arXiv",
    reportNumber = "LPTENS-04-46",
    doi = "10.1088/1126-6708/2004/11/065",
    journal = "JHEP",
    volume = "11",
    pages = "065",
    year = "2004"
}

@article{Bachas:2013ora,
    author = "Bachas, C. and Brunner, I. and Roggenkamp, D.",
    title = "{Fusion of Critical Defect Lines in the 2D Ising Model}",
    eprint = "1303.3616",
    archivePrefix = "arXiv",
    primaryClass = "cond-mat.stat-mech",
    reportNumber = "LMU-ASC-13-13, LPTENS-13-06",
    doi = "10.1088/1742-5468/2013/08/P08008",
    journal = "J. Stat. Mech.",
    volume = "1308",
    pages = "P08008",
    year = "2013"
}

@article{affleck1991kondo,
  title={The Kondo effect, conformal field theory and fusion rules},
  author={Affleck, Ian and Ludwig, Andreas WW},
  journal={Nuclear Physics B},
  volume={352},
  number={3},
  pages={849--862},
  year={1991},
  publisher={Elsevier}
}

@article{Gadde:2016fbj,
    author = "Gadde, Abhijit",
    title = "{Conformal constraints on defects}",
    eprint = "1602.06354",
    archivePrefix = "arXiv",
    primaryClass = "hep-th",
    doi = "10.1007/JHEP01(2020)038",
    journal = "JHEP",
    volume = "01",
    pages = "038",
    year = "2020"
}

@article{affleck1990current,
  title={A current algebra approach to the Kondo effect},
  author={Affleck, Ian},
  journal={Nuclear Physics B},
  volume={336},
  number={3},
  pages={517--532},
  year={1990},
  publisher={Elsevier}
}

@article{affleck1991critical,
  title={Critical theory of overscreened Kondo fixed points},
  author={Affleck, Ian and Ludwig, Andreas WW},
  journal={Nuclear Physics B},
  volume={360},
  number={2-3},
  pages={641--696},
  year={1991},
  publisher={Elsevier}
}

@article{Konechny:2015qla,
    author = "Konechny, Anatoly",
    title = "{Fusion of conformal interfaces and bulk induced boundary RG flows}",
    eprint = "1509.07787",
    archivePrefix = "arXiv",
    primaryClass = "hep-th",
    doi = "10.1007/JHEP12(2015)114",
    journal = "JHEP",
    volume = "12",
    pages = "114",
    year = "2015"
}

@article{mitchell2012universal,
  title={Universal low-temperature crossover in two-channel Kondo models},
  author={Mitchell, Andrew K and Sela, Eran},
  journal={Physical Review B},
  volume={85},
  number={23},
  pages={235127},
  year={2012},
  publisher={APS}
}

@article{lopes2020anyons,
  title={Anyons in multichannel Kondo systems},
  author={Lopes, Pedro LS and Affleck, Ian and Sela, Eran},
  journal={Physical Review B},
  volume={101},
  number={8},
  pages={085141},
  year={2020},
  publisher={APS}
}

@article{gabay2022multi,
  title={Multi-impurity chiral Kondo model: Correlation functions and anyon fusion rules},
  author={Gabay, Dor and Han, Cheolhee and Lopes, Pedro LS and Affleck, Ian and Sela, Eran},
  journal={Physical Review B},
  volume={105},
  number={3},
  pages={035151},
  year={2022},
  publisher={APS}
}

@article{OBannon:2015cqy,
    author = "O'Bannon, Andy and Papadimitriou, Ioannis and Probst, Jonas",
    title = "{A Holographic Two-Impurity Kondo Model}",
    eprint = "1510.08123",
    archivePrefix = "arXiv",
    primaryClass = "hep-th",
    reportNumber = "OUTP-15-28P, SISSA-49-2015-FISI",
    doi = "10.1007/JHEP01(2016)103",
    journal = "JHEP",
    volume = "01",
    pages = "103",
    year = "2016"
}

@article{Affleck:1991yq,
    author = "Affleck, Ian and Ludwig, Andreas W. W.",
    title = "{Exact critical theory of the two impurity Kondo model}",
    reportNumber = "UBCTP-91-026",
    doi = "10.1103/PhysRevLett.68.1046",
    journal = "Phys. Rev. Lett.",
    volume = "68",
    pages = "1046--1049",
    year = "1992"
}

@article{affleck1995conformal,
  title={Conformal-field-theory approach to the two-impurity Kondo problem: comparison with numerical renormalization-group results},
  author={Affleck, Ian and Ludwig, Andreas WW and Jones, Barbara A},
  journal={Physical Review B},
  volume={52},
  number={13},
  pages={9528},
  year={1995},
  publisher={APS}
}

@article{gan1995mapping,
  title={Mapping the critical point of the two-impurity Kondo model to a two-channel problem},
  author={Gan, Junwu},
  journal={Physical review letters},
  volume={74},
  number={13},
  pages={2583},
  year={1995},
  publisher={APS}
}

@article{gan1995solution,
  title={Solution of the two-impurity Kondo model: Critical point, Fermi-liquid phase, and crossover},
  author={Gan, Junwu},
  journal={Physical Review B},
  volume={51},
  number={13},
  pages={8287},
  year={1995},
  publisher={APS}
}

@article{georges1995solution,
  title={Solution of the two-impurity, two-channel Kondo model},
  author={Georges, Antoine and Sengupta, Anirvan M},
  journal={Physical review letters},
  volume={74},
  number={14},
  pages={2808},
  year={1995},
  publisher={APS}
}

@article{jones1989critical,
  title={Critical point in the solution of the two magnetic impurity problem},
  author={Jones, BA and Varma, CM},
  journal={Physical Review B},
  volume={40},
  number={1},
  pages={324},
  year={1989},
  publisher={APS}
}

@article{McGreevy:2022oyu,
    author = "McGreevy, John",
    title = "{Generalized Symmetries in Condensed Matter}",
    eprint = "2204.03045",
    archivePrefix = "arXiv",
    primaryClass = "cond-mat.str-el",
    doi = "10.1146/annurev-conmatphys-040721-021029",
    month = "4",
    year = "2022"
}

@inproceedings{Cordova:2022ruw,
    author = "Cordova, Clay and Dumitrescu, Thomas T. and Intriligator, Kenneth and Shao, Shu-Heng",
    title = "{Snowmass White Paper: Generalized Symmetries in Quantum Field Theory and Beyond}",
    booktitle = "{Snowmass 2021}",
    eprint = "2205.09545",
    archivePrefix = "arXiv",
    primaryClass = "hep-th",
    month = "5",
    year = "2022"
}

@article{Schafer-Nameki:2023jdn,
    author = "Schafer-Nameki, Sakura",
    title = "{ICTP Lectures on (Non-)Invertible Generalized Symmetries}",
    eprint = "2305.18296",
    archivePrefix = "arXiv",
    primaryClass = "hep-th",
    month = "5",
    year = "2023"
}

@article{Brennan:2023mmt,
    author = "Brennan, T. Daniel and Hong, Sungwoo",
    title = "{Introduction to Generalized Global Symmetries in QFT and Particle Physics}",
    eprint = "2306.00912",
    archivePrefix = "arXiv",
    primaryClass = "hep-ph",
    month = "6",
    year = "2023"
}

@article{Bhardwaj:2023kri,
    author = "Bhardwaj, Lakshya and Bottini, Lea E. and Fraser-Taliente, Ludovic and Gladden, Liam and Gould, Dewi S. W. and Platschorre, Arthur and Tillim, Hannah",
    title = "{Lectures on Generalized Symmetries}",
    eprint = "2307.07547",
    archivePrefix = "arXiv",
    primaryClass = "hep-th",
    month = "7",
    year = "2023"
}

@article{Shao:2023gho,
    author = "Shao, Shu-Heng",
    title = "{What's Done Cannot Be Undone: TASI Lectures on Non-Invertible Symmetry}",
    eprint = "2308.00747",
    archivePrefix = "arXiv",
    primaryClass = "hep-th",
    reportNumber = "YITP-SB-2023-19",
    month = "8",
    year = "2023"
}

@article{Copetti:2023mcq,
    author = "Copetti, Christian and Del Zotto, Michele and Ohmori, Kantaro and Wang, Yifan",
    title = "{Higher Structure of Chiral Symmetry}",
    eprint = "2305.18282",
    archivePrefix = "arXiv",
    primaryClass = "hep-th",
    month = "5",
    year = "2023"
}

@article{Allais:2014fqa,
    author = "Allais, Andrea and Sachdev, Subir",
    title = "{Spectral function of a localized fermion coupled to the Wilson-Fisher conformal field theory}",
    eprint = "1406.3022",
    archivePrefix = "arXiv",
    primaryClass = "cond-mat.str-el",
    doi = "10.1103/PhysRevB.90.035131",
    journal = "Phys. Rev. B",
    volume = "90",
    number = "3",
    pages = "035131",
    year = "2014"
}

@article{Kapustin:2010if,
    author = "Kapustin, Anton and Saulina, Natalia",
    editor = "Sati, Hisham and Schreiber, Urs",
    title = "{Surface operators in 3d Topological Field Theory and 2d Rational Conformal Field Theory}",
    eprint = "1012.0911",
    archivePrefix = "arXiv",
    primaryClass = "hep-th",
    reportNumber = "PI-STRINGS-195",
    pages = "175--198",
    month = "12",
    year = "2010"
}

@article{Choi:2022zal,
    author = "Choi, Yichul and Cordova, Clay and Hsin, Po-Shen and Lam, Ho Tat and Shao, Shu-Heng",
    title = "{Non-invertible Condensation, Duality, and Triality Defects in 3+1 Dimensions}",
    eprint = "2204.09025",
    archivePrefix = "arXiv",
    primaryClass = "hep-th",
    reportNumber = "YITP-SB-2022-16, MIT/CTP-5423, YITP-SB-2022-16, MIT/CTP-5423",
    doi = "10.1007/s00220-023-04727-4",
    journal = "Commun. Math. Phys.",
    volume = "402",
    number = "1",
    pages = "489--542",
    year = "2023"
}

@article{Roumpedakis:2022aik,
    author = "Roumpedakis, Konstantinos and Seifnashri, Sahand and Shao, Shu-Heng",
    title = "{Higher Gauging and Non-invertible Condensation Defects}",
    eprint = "2204.02407",
    archivePrefix = "arXiv",
    primaryClass = "hep-th",
    reportNumber = "YITP-SB-2022-14",
    doi = "10.1007/s00220-023-04706-9",
    journal = "Commun. Math. Phys.",
    volume = "401",
    number = "3",
    pages = "3043--3107",
    year = "2023"
}

@article{Bachas:2007td,
    author = "Bachas, C. and Brunner, I.",
    title = "{Fusion of conformal interfaces}",
    eprint = "0712.0076",
    archivePrefix = "arXiv",
    primaryClass = "hep-th",
    doi = "10.1088/1126-6708/2008/02/085",
    journal = "JHEP",
    volume = "02",
    pages = "085",
    year = "2008"
}

@book{bateman1953higher,
  title={Higher transcendental functions [volumes i-iii]},
  author={Bateman, Harry},
  volume={1},
  year={1953},
  publisher={McGRAW-HILL book company}
}

@article{POLYAKOV2008199,
title = {De Sitter space and eternity},
journal = {Nuclear Physics B},
volume = {797},
number = {1},
pages = {199-217},
year = {2008},
issn = {0550-3213},
doi = {https://doi.org/10.1016/j.nuclphysb.2008.01.002},
url = {https://www.sciencedirect.com/science/article/pii/S0550321308000369},
author = {A.M. Polyakov}
}

@article{moshe2003quantum,
  title={Quantum field theory in the large N limit: A Review},
  author={Moshe, Moshe and Zinn-Justin, Jean},
  journal={Physics Reports},
  volume={385},
  number={3-6},
  pages={69--228},
  year={2003},
  publisher={Elsevier}
}

@inproceedings{Hartman:2022zik,
    author = "Hartman, Thomas and Mazac, Dalimil and Simmons-Duffin, David and Zhiboedov, Alexander",
    title = "{Snowmass White Paper: The Analytic Conformal Bootstrap}",
    booktitle = "{2022 Snowmass Summer Study}",
    eprint = "2202.11012",
    archivePrefix = "arXiv",
    primaryClass = "hep-th",
    month = "2",
    year = "2022"
}

@inproceedings{Poland:2022qrs,
    author = "Poland, David and Simmons-Duffin, David",
    title = "{Snowmass White Paper: The Numerical Conformal Bootstrap}",
    booktitle = "{2022 Snowmass Summer Study}",
    eprint = "2203.08117",
    archivePrefix = "arXiv",
    primaryClass = "hep-th",
    reportNumber = "CALT-TH 2022-013",
    month = "3",
    year = "2022"
}

@article{Poland:2018epd,
    author = "Poland, David and Rychkov, Slava and Vichi, Alessandro",
    title = "{The Conformal Bootstrap: Theory, Numerical Techniques, and Applications}",
    eprint = "1805.04405",
    archivePrefix = "arXiv",
    primaryClass = "hep-th",
    doi = "10.1103/RevModPhys.91.015002",
    journal = "Rev. Mod. Phys.",
    volume = "91",
    pages = "015002",
    year = "2019"
}

@article{Fei:2016sgs,
    author = "Fei, Lin and Giombi, Simone and Klebanov, Igor R. and Tarnopolsky, Grigory",
    title = "{Yukawa CFTs and Emergent Supersymmetry}",
    eprint = "1607.05316",
    archivePrefix = "arXiv",
    primaryClass = "hep-th",
    reportNumber = "PUPT-2504",
    doi = "10.1093/ptep/ptw120",
    journal = "PTEP",
    volume = "2016",
    number = "12",
    pages = "12C105",
    year = "2016"
}

@article{Liendo:2012hy,
    author = "Liendo, Pedro and Rastelli, Leonardo and van Rees, Balt C.",
    title = "{The Bootstrap Program for Boundary CFT$_d$}",
    eprint = "1210.4258",
    archivePrefix = "arXiv",
    primaryClass = "hep-th",
    reportNumber = "YITP-SB-12-37",
    doi = "10.1007/JHEP07(2013)113",
    journal = "JHEP",
    volume = "07",
    pages = "113",
    year = "2013"
}

@article{Behan:2021tcn,
    author = "Behan, Connor and Di Pietro, Lorenzo and Lauria, Edoardo and van Rees, Balt C.",
    title = "{Bootstrapping boundary-localized interactions II. Minimal models at the boundary}",
    eprint = "2111.04747",
    archivePrefix = "arXiv",
    primaryClass = "hep-th",
    doi = "10.1007/JHEP03(2022)146",
    journal = "JHEP",
    volume = "03",
    pages = "146",
    year = "2022"
}

@article{Behan:2020nsf,
    author = "Behan, Connor and Di Pietro, Lorenzo and Lauria, Edoardo and Van Rees, Balt C.",
    title = "{Bootstrapping boundary-localized interactions}",
    eprint = "2009.03336",
    archivePrefix = "arXiv",
    primaryClass = "hep-th",
    reportNumber = "CPHT-RR112.122020",
    doi = "10.1007/JHEP12(2020)182",
    journal = "JHEP",
    volume = "12",
    pages = "182",
    year = "2020"
}

@article{Metlitski:2020cqy,
    author = "Metlitski, Max A.",
    title = "{Boundary criticality of the O(N) model in d = 3 critically revisited}",
    eprint = "2009.05119",
    archivePrefix = "arXiv",
    primaryClass = "cond-mat.str-el",
    month = "9",
    year = "2020"
}

@article{Rodriguez-Gomez:2022gbz,
    author = "Rodriguez-Gomez, Diego",
    title = "{A Scaling Limit for Line and Surface Defects}",
    eprint = "2202.03471",
    archivePrefix = "arXiv",
    primaryClass = "hep-th",
    month = "2",
    year = "2022"
}

@article{Affleck:1995ge,
    author = "Affleck, Ian",
    editor = "Nowak, Maciej A. and Wegrzyn, Pawel",
    title = "{Conformal field theory approach to the Kondo effect}",
    eprint = "cond-mat/9512099",
    archivePrefix = "arXiv",
    journal = "Acta Phys. Polon. B",
    volume = "26",
    pages = "1869--1932",
    year = "1995"
}

@article{Gaiotto:2014kfa,
    author = "Gaiotto, Davide and Kapustin, Anton and Seiberg, Nathan and Willett, Brian",
    title = "{Generalized Global Symmetries}",
    eprint = "1412.5148",
    archivePrefix = "arXiv",
    primaryClass = "hep-th",
    doi = "10.1007/JHEP02(2015)172",
    journal = "JHEP",
    volume = "02",
    pages = "172",
    year = "2015"
}

@article{Kondo:1964nea,
    author = "Kondo, J.",
    title = "{Resistance Minimum in Dilute Magnetic Alloys}",
    doi = "10.1143/PTP.32.37",
    journal = "Prog. Theor. Phys.",
    volume = "32",
    number = "1",
    pages = "37--49",
    year = "1964"
}

@article{Giombi:2020rmc,
    author = "Giombi, Simone and Khanchandani, Himanshu",
    title = "{CFT in AdS and boundary RG flows}",
    eprint = "2007.04955",
    archivePrefix = "arXiv",
    primaryClass = "hep-th",
    doi = "10.1007/JHEP11(2020)118",
    journal = "JHEP",
    volume = "11",
    pages = "118",
    year = "2020"
}

@article{Giombi:2021uae,
    author = "Giombi, Simone and Helfenberger, Elizabeth and Ji, Ziming and Khanchandani, Himanshu",
    title = "{Monodromy defects from hyperbolic space}",
    eprint = "2102.11815",
    archivePrefix = "arXiv",
    primaryClass = "hep-th",
    doi = "10.1007/JHEP02(2022)041",
    journal = "JHEP",
    volume = "02",
    pages = "041",
    year = "2022"
}

@article{Gliozzi:2015qsa,
    author = "Gliozzi, Ferdinando and Liendo, Pedro and Meineri, Marco and Rago, Antonio",
    title = "{Boundary and Interface CFTs from the Conformal Bootstrap}",
    eprint = "1502.07217",
    archivePrefix = "arXiv",
    primaryClass = "hep-th",
    reportNumber = "HU-EP-15-08, HU-EP-15/08",
    doi = "10.1007/JHEP05(2015)036",
    journal = "JHEP",
    volume = "05",
    pages = "036",
    year = "2015",
    note = "[Erratum: JHEP 12, 093 (2021)]"
}

@article{Cuomo:2021kfm,
    author = "Cuomo, Gabriel and Komargodski, Zohar and Mezei, M\'ark",
    title = "{Localized magnetic field in the O(N) model}",
    eprint = "2112.10634",
    archivePrefix = "arXiv",
    primaryClass = "hep-th",
    doi = "10.1007/JHEP02(2022)134",
    journal = "JHEP",
    volume = "02",
    pages = "134",
    year = "2022"
}

@article{Billo:2016cpy,
    author = "Bill\`o, Marco and Gon\c{c}alves, Vasco and Lauria, Edoardo and Meineri, Marco",
    title = "{Defects in conformal field theory}",
    eprint = "1601.02883",
    archivePrefix = "arXiv",
    primaryClass = "hep-th",
    doi = "10.1007/JHEP04(2016)091",
    journal = "JHEP",
    volume = "04",
    pages = "091",
    year = "2016"
}

@article{McAvity:1995zd,
      author         = "McAvity, D. M. and Osborn, H.",
      title          = "{Conformal field theories near a boundary in general
                        dimensions}",
      journal        = "Nucl. Phys.",
      volume         = "B455",
      year           = "1995",
      pages          = "522-576",
      doi            = "10.1016/0550-3213(95)00476-9",
      eprint         = "cond-mat/9505127",
      archivePrefix  = "arXiv",
      primaryClass   = "cond-mat",
      reportNumber   = "DAMTP-95-1, UBC-TP-95-002",
      SLACcitation   = "%%CITATION = COND-MAT/9505127;%%"
}

@article{Bissi:2022bgu,
    author = {Bissi, Agnese and Dey, Parijat and Sisti, Jacopo and S\"oderberg, Alexander},
    title = "{Interacting conformal scalar in a wedge}",
    eprint = "2206.06326",
    archivePrefix = "arXiv",
    primaryClass = "hep-th",
    reportNumber = "UUITP-55/21",
    month = "6",
    year = "2022"
}

@article{Kaviraj:2018tfd,
      author         = "Kaviraj, Apratim and Paulos, Miguel F.",
      title          = "{The Functional Bootstrap for Boundary CFT}",
      year           = "2018",
      eprint         = "1812.04034",
      archivePrefix  = "arXiv",
      primaryClass   = "hep-th",
      SLACcitation   = "%%CITATION = ARXIV:1812.04034;%%"
}

@article{PhysRevA.46.1886,
  title = {Free energy and specific heat of critical films and surfaces},
  author = {Krech, M. and Dietrich, S.},
  journal = {Phys. Rev. A},
  volume = {46},
  issue = {4},
  pages = {1886--1921},
  numpages = {0},
  year = {1992},
  month = {Aug},
  publisher = {American Physical Society},
  doi = {10.1103/PhysRevA.46.1886},
  url = {https://link.aps.org/doi/10.1103/PhysRevA.46.1886}
}

@article{Diehl:2011sy,
    author = "Diehl, H. W. and Schmidt, Felix M.",
    title = "{Critical Casimir effect in films for generic non-symmetry-breaking boundary conditions}",
    eprint = "1110.1241",
    archivePrefix = "arXiv",
    primaryClass = "cond-mat.stat-mech",
    doi = "10.1088/1367-2630/13/12/123025",
    journal = "New J. Phys.",
    volume = "13",
    pages = "123025",
    year = "2011"
}

@article{pjm/1102613160,
author = {A. Erd{\'e}lyi and F. G. Tricomi},
title = {{The asymptotic expansion of a ratio of gamma functions.}},
volume = {1},
journal = {Pacific Journal of Mathematics},
number = {1},
publisher = {Pacific Journal of Mathematics, A Non-profit Corporation},
pages = {133 -- 142},
year = {1951},
}

@article{McAvity:1993ue,
      author         = "McAvity, D. M. and Osborn, H.",
      title          = "{Energy momentum tensor in conformal field theories near
                        a boundary}",
      journal        = "Nucl. Phys.",
      volume         = "B406",
      year           = "1993",
      pages          = "655-680",
      doi            = "10.1016/0550-3213(93)90005-A",
      eprint         = "hep-th/9302068",
      archivePrefix  = "arXiv",
      primaryClass   = "hep-th",
      reportNumber   = "DAMTP-93-01",
      SLACcitation   = "%%CITATION = HEP-TH/9302068;%%"
}

@article{Bissi:2018mcq,
      author         = "Bissi, Agnese and Hansen, Tobias and Söderberg,
                        Alexander",
      title          = "{Analytic Bootstrap for Boundary CFT}",
      journal        = "JHEP",
      volume         = "01",
      year           = "2019",
      pages          = "010",
      doi            = "10.1007/JHEP01(2019)010",
      eprint         = "1808.08155",
      archivePrefix  = "arXiv",
      primaryClass   = "hep-th",
      SLACcitation   = "%%CITATION = ARXIV:1808.08155;%%"
}

@article{10.1143/PTP.70.1226,
    author = {Ohno, Kaoru and Okabe, Yutaka},
    title = "{The 1/n Expansion for the n-Vector Model in the Semi-Infinite Space}",
    journal = {Progress of Theoretical Physics},
    volume = {70},
    number = {5},
    pages = {1226-1239},
    year = {1983},
    month = {11},
    issn = {0033-068X},
    doi = {10.1143/PTP.70.1226},
    url = {https://doi.org/10.1143/PTP.70.1226},
    eprint = {https://academic.oup.com/ptp/article-pdf/70/5/1226/5465952/70-5-1226.pdf},
}

@incollection{Die86a,
Address = {London},
Author = {H. W. Diehl},
Booktitle = {Phase Transitions and Critical Phenomena},
Editor = {C. Domb and J. L. Lebowitz},
Pages = {75--267},
Publisher = {Academic},
Title = {Field-theoretical Approach to Critical Behaviour at Surfaces},
Volume = {10},
Year = {1986}}

@article{Die97,
Author = {H. W. Diehl},
Journal = {International Journal of Modern Physics B},
Number = {30},
Pages = {3503--3523},
Title = {The Theory of Boundary Critical Phenomena},
Volume = {11},
Year = {1997}}

@book{cardy_1996, place={Cambridge}, series={Cambridge Lecture Notes in Physics}, title={Scaling and Renormalization in Statistical Physics}, DOI={10.1017/CBO9781316036440}, publisher={Cambridge University Press}, author={Cardy, John}, year={1996}, collection={Cambridge Lecture Notes in Physics}}

@article{PhysRevLett.38.1046,
  title = {Surface Critical Exponents in Terms of Bulk Exponents},
  author = {Bray, A. J. and Moore, M. A.},
  journal = {Phys. Rev. Lett.},
  volume = {38},
  issue = {19},
  pages = {1046--1048},
  numpages = {0},
  year = {1977},
  month = {May},
  publisher = {American Physical Society},
  doi = {10.1103/PhysRevLett.38.1046},
  url = {https://link.aps.org/doi/10.1103/PhysRevLett.38.1046}
}

@article{Carmi:2018qzm,
      author         = "Carmi, Dean and Di Pietro, Lorenzo and Komatsu, Shota",
      title          = "{A Study of Quantum Field Theories in AdS at Finite
                        Coupling}",
      journal        = "JHEP",
      volume         = "01",
      year           = "2019",
      pages          = "200",
      doi            = "10.1007/JHEP01(2019)200",
      eprint         = "1810.04185",
      archivePrefix  = "arXiv",
      primaryClass   = "hep-th",
      SLACcitation   = "%%CITATION = ARXIV:1810.04185;%%"
}

@article{Diehl:1981zz,
      author         = "Diehl, H. W. and Dietrich, S.",
      title          = "{Field-theoretical approach to multicritical behavior
                        near free surfaces}",
      journal        = "Phys. Rev.",
      volume         = "B24",
      year           = "1981",
      pages          = "2878-2880",
      doi            = "10.1103/PhysRevB.24.2878",
      SLACcitation   = "%%CITATION = PHRVA,B24,2878;%%"
}

@article{Diehl:2006mz,
    author = "Diehl, H. W. and Gruneberg, D. and Shpot, M. A.",
    title = "{Fluctuation-induced forces in periodic slabs: Breakdown of epsilon expansion at the bulk critical point and revised field theory}",
    eprint = "cond-mat/0605293",
    archivePrefix = "arXiv",
    doi = "10.1209/epl/i2006-10090-0",
    journal = "EPL",
    volume = "75",
    pages = "241--247",
    year = "2006"
}

@article{Gruneberg:2007av,
    author = "Gruneberg, Daniel and Diehl, H. W.",
    title = "{Thermodynamic Casimir effects involving interacting field theories with zero modes}",
    eprint = "0710.4436",
    archivePrefix = "arXiv",
    primaryClass = "cond-mat.stat-mech",
    doi = "10.1103/PhysRevB.77.115409",
    journal = "Phys. Rev. B",
    volume = "77",
    pages = "115409",
    year = "2008"
}

@article{JLCardy_1983,
doi = {10.1088/0305-4470/16/15/026},
url = {https://dx.doi.org/10.1088/0305-4470/16/15/026},
year = {1983},
month = {oct},
publisher = {},
volume = {16},
number = {15},
pages = {3617},
author = {J L Cardy},
title = {Critical behaviour at an edge},
journal = {Journal of Physics A: Mathematical and General}
}

@article{Antunes:2021qpy,
    author = "Antunes, Ant\'onio",
    title = "{Conformal bootstrap near the edge}",
    eprint = "2103.03132",
    archivePrefix = "arXiv",
    primaryClass = "hep-th",
    doi = "10.1007/JHEP10(2021)057",
    journal = "JHEP",
    volume = "10",
    pages = "057",
    year = "2021"
}

@article{Goldberger:2001tn,
    author = "Goldberger, Walter D. and Wise, Mark B.",
    title = "{Renormalization group flows for brane couplings}",
    eprint = "hep-th/0104170",
    archivePrefix = "arXiv",
    reportNumber = "CALT-68-2327",
    doi = "10.1103/PhysRevD.65.025011",
    journal = "Phys. Rev. D",
    volume = "65",
    pages = "025011",
    year = "2002"
}

@article{Michel:2014lva,
    author = "Michel, Ben and Mintun, Eric and Polchinski, Joseph and Puhm, Andrea and Saad, Philip",
    title = "{Remarks on brane and antibrane dynamics}",
    eprint = "1412.5702",
    archivePrefix = "arXiv",
    primaryClass = "hep-th",
    doi = "10.1007/JHEP09(2015)021",
    journal = "JHEP",
    volume = "09",
    pages = "021",
    year = "2015"
}

@book{arnol2013mathematical,
  title={Mathematical methods of classical mechanics},
  author={Arnol'd, Vladimir Igorevich},
  volume={60},
  year={2013},
  publisher={Springer Science \& Business Media}
}

@article{Fredenhagen:2006dn,
    author = "Fredenhagen, Stefan and Gaberdiel, Matthias R. and Keller, Christoph A.",
    title = "{Bulk induced boundary perturbations}",
    eprint = "hep-th/0609034",
    archivePrefix = "arXiv",
    doi = "10.1088/1751-8113/40/1/F03",
    journal = "J. Phys. A",
    volume = "40",
    pages = "F17",
    year = "2007"
}

@article{hybridfusion,
    author = "Diatlyk, Oleksandr and Khanchandani, Himanshu and Popov, Fedor and Wang, Yifan",
    title = "to appear",
    eprint = "",
    archivePrefix = "",
    primaryClass = "",
    doi = "",
    journal = "",
    year = ""
}

@article{Bartsch:2022ytj,
    author = "Bartsch, Thomas and Bullimore, Mathew and Ferrari, Andrea E. V. and Pearson, Jamie",
    title = "{Non-invertible Symmetries and Higher Representation Theory II}",
    eprint = "2212.07393",
    archivePrefix = "arXiv",
    primaryClass = "hep-th",
    month = "12",
    year = "2022"
}

@article{Bhardwaj:2022yxj,
    author = "Bhardwaj, Lakshya and Bottini, Lea E. and Schafer-Nameki, Sakura and Tiwari, Apoorv",
    title = "{Non-invertible higher-categorical symmetries}",
    eprint = "2204.06564",
    archivePrefix = "arXiv",
    primaryClass = "hep-th",
    doi = "10.21468/SciPostPhys.14.1.007",
    journal = "SciPost Phys.",
    volume = "14",
    number = "1",
    pages = "007",
    year = "2023"
}

@article{Giombi:2021cnr,
    author = "Giombi, Simone and Helfenberger, Elizabeth and Khanchandani, Himanshu",
    title = "{Fermions in AdS and Gross-Neveu BCFT}",
    eprint = "2110.04268",
    archivePrefix = "arXiv",
    primaryClass = "hep-th",
    doi = "10.1007/JHEP07(2022)018",
    journal = "JHEP",
    volume = "07",
    pages = "018",
    year = "2022"
}

@article{Metlitski:2009iyg,
    author = "Metlitski, Max A. and Fuertes, Carlos A. and Sachdev, Subir",
    title = "{Entanglement Entropy in the O(N) model}",
    eprint = "0904.4477",
    archivePrefix = "arXiv",
    primaryClass = "cond-mat.stat-mech",
    doi = "10.1103/PhysRevB.80.115122",
    journal = "Phys. Rev. B",
    volume = "80",
    number = "11",
    pages = "115122",
    year = "2009"
}

@article{Johnson:1975zp,
    author = "Johnson, K.",
    title = "{The M.I.T. Bag Model}",
    reportNumber = "MIT-CTP-494",
    journal = "Acta Phys. Polon. B",
    volume = "6",
    pages = "865",
    year = "1975"
}

@article{DePaola:1999im,
    author = "De Paola, R. D. M. and Rodrigues, R. B. and Svaiter, N. F.",
    title = "{Casimir energy of massless fermions in the slab bag}",
    eprint = "hep-th/9905039",
    archivePrefix = "arXiv",
    reportNumber = "CBPF-NF-020-99",
    doi = "10.1142/S0217732399002431",
    journal = "Mod. Phys. Lett. A",
    volume = "14",
    pages = "2353--2362",
    year = "1999"
}

@book{Milonni:1994xx,
    author = "Milonni, P. W.",
    title = "{The Quantum vacuum: An Introduction to quantum electrodynamics}",
    year = "1994"
}

@article{Giombi:2022vnz,
    author = "Giombi, Simone and Helfenberger, Elizabeth and Khanchandani, Himanshu",
    title = "{Line defects in fermionic CFTs}",
    eprint = "2211.11073",
    archivePrefix = "arXiv",
    primaryClass = "hep-th",
    reportNumber = "PUPT-2644",
    doi = "10.1007/JHEP08(2023)224",
    journal = "JHEP",
    volume = "08",
    pages = "224",
    year = "2023"
}

@article{Diehl:2014bpa,
    author = {Diehl, H. W. and Gr\"uneberg, Daniel and Hasenbusch, Martin and Hucht, Alfred and Rutkevich, Sergei B. and Schmidt, Felix M.},
    title = "{Large-$n$ approach to thermodynamic Casimir effects in slabs with free surfaces}",
    eprint = "1402.3510",
    archivePrefix = "arXiv",
    primaryClass = "cond-mat.stat-mech",
    doi = "10.1103/PhysRevE.89.062123",
    journal = "Phys. Rev. E",
    volume = "89",
    number = "6",
    pages = "062123",
    year = "2014"
}

@Book{Abramowitz,
 author    = "Milton {Abramowitz} and Irene A. {Stegun}",
 title     = "Handbook of Mathematical Functions with
              Formulas, Graphs, and Mathematical Tables, 9th printing.",
 year      =  1972,
 address   = "New York City",

}

@article{Soderberg:2021kne,
    author = {S\"oderberg, Alexander},
    title = "{Fusion of conformal defects in four dimensions}",
    eprint = "2102.00718",
    archivePrefix = "arXiv",
    primaryClass = "hep-th",
    reportNumber = "UUITP-07/21",
    doi = "10.1007/JHEP04(2021)087",
    journal = "JHEP",
    volume = "04",
    pages = "087",
    year = "2021"
}

@article{SoderbergRousu:2023zyj,
    author = {S\"oderberg Rousu, Alexander},
    title = "{Fusion of conformal defects in interacting theories}",
    eprint = "2304.10239",
    archivePrefix = "arXiv",
    primaryClass = "hep-th",
    reportNumber = "UUITP-09/23",
    doi = "10.1007/JHEP10(2023)183",
    journal = "JHEP",
    volume = "10",
    pages = "183",
    year = "2023"
}

@article{Correa:2012hh,
    author = "Correa, Diego and Maldacena, Juan and Sever, Amit",
    title = "{The quark anti-quark potential and the cusp anomalous dimension from a TBA equation}",
    eprint = "1203.1913",
    archivePrefix = "arXiv",
    primaryClass = "hep-th",
    doi = "10.1007/JHEP08(2012)134",
    journal = "JHEP",
    volume = "08",
    pages = "134",
    year = "2012"
}

@article{Popov:2022nfq,
    author = "Popov, Fedor K. and Wang, Yifan",
    title = "{Non-perturbative defects in tensor models from melonic trees}",
    eprint = "2206.14206",
    archivePrefix = "arXiv",
    primaryClass = "hep-th",
    doi = "10.1007/JHEP11(2022)057",
    journal = "JHEP",
    volume = "11",
    pages = "057",
    year = "2022"
}

@article{Sato:2021eqo,
    author = "Sato, Yoshiki",
    title = "{Free energy and defect $C$-theorem in free fermion}",
    eprint = "2102.11468",
    archivePrefix = "arXiv",
    primaryClass = "hep-th",
    doi = "10.1007/JHEP05(2021)202",
    journal = "JHEP",
    volume = "05",
    pages = "202",
    year = "2021"
}

@article{Camporesi:1992tm,
    author = "Camporesi, R.",
    title = "{The Spinor heat kernel in maximally symmetric spaces}",
    doi = "10.1007/BF02100862",
    journal = "Commun. Math. Phys.",
    volume = "148",
    pages = "283--308",
    year = "1992"
}

@article{Bytsenko:1996rr,
    author = "Bytsenko, Andre A. and Cognola, Guido and Zerbini, Sergio",
    title = "{Quantum fields in hyperbolic space-times with finite spatial volume}",
    eprint = "hep-th/9605209",
    archivePrefix = "arXiv",
    reportNumber = "UTF-377",
    doi = "10.1088/0264-9381/14/3/008",
    journal = "Class. Quant. Grav.",
    volume = "14",
    pages = "615--627",
    year = "1997"
}

@article{Camporesi:1994ga,
    author = "Camporesi, R. and Higuchi, A.",
    title = "{Spectral functions and zeta functions in hyperbolic spaces}",
    doi = "10.1063/1.530850",
    journal = "J. Math. Phys.",
    volume = "35",
    pages = "4217--4246",
    year = "1994"
}

@article{Krishnan:2023cff,
    author = "Krishnan, Abijith and Metlitski, Max A.",
    title = "{A plane defect in the 3d O(N) model}",
    eprint = "2301.05728",
    archivePrefix = "arXiv",
    primaryClass = "cond-mat.str-el",
    doi = "10.21468/SciPostPhys.15.3.090",
    journal = "SciPost Phys.",
    volume = "15",
    number = "3",
    pages = "090",
    year = "2023"
}

@article{Giombi:2023dqs,
    author = "Giombi, Simone and Liu, Bowei",
    title = "{Notes on a surface defect in the O(N) model}",
    eprint = "2305.11402",
    archivePrefix = "arXiv",
    primaryClass = "hep-th",
    reportNumber = "PUPT-2643",
    doi = "10.1007/JHEP12(2023)004",
    journal = "JHEP",
    volume = "12",
    pages = "004",
    year = "2023"
}

@article{Raviv-Moshe:2023yvq,
    author = "Raviv-Moshe, Avia and Zhong, Siwei",
    title = "{Phases of surface defects in Scalar Field Theories}",
    eprint = "2305.11370",
    archivePrefix = "arXiv",
    primaryClass = "hep-th",
    doi = "10.1007/JHEP08(2023)143",
    journal = "JHEP",
    volume = "08",
    pages = "143",
    year = "2023"
}

@article{Trepanier:2023tvb,
    author = "Tr\'epanier, Maxime",
    title = "{Surface defects in the O(N) model}",
    eprint = "2305.10486",
    archivePrefix = "arXiv",
    primaryClass = "hep-th",
    doi = "10.1007/JHEP09(2023)074",
    journal = "JHEP",
    volume = "09",
    pages = "074",
    year = "2023"
}

@article{Giombi:2019enr,
    author = "Giombi, Simone and Khanchandani, Himanshu",
    title = "{$O(N)$ models with boundary interactions and their long range generalizations}",
    eprint = "1912.08169",
    archivePrefix = "arXiv",
    primaryClass = "hep-th",
    reportNumber = "PUPT-2606",
    doi = "10.1007/JHEP08(2020)010",
    journal = "JHEP",
    volume = "08",
    number = "08",
    pages = "010",
    year = "2020"
}

@article{PhysRevD.10.3235,
  title = {Dynamical symmetry breaking in asymptotically free field theories},
  author = {Gross, David J. and Neveu, Andr\'e},
  journal = {Phys. Rev. D},
  volume = {10},
  issue = {10},
  pages = {3235--3253},
  numpages = {0},
  year = {1974},
  month = {Nov},
  publisher = {American Physical Society},
  doi = {10.1103/PhysRevD.10.3235},
  url = {https://link.aps.org/doi/10.1103/PhysRevD.10.3235}
}

@article{CARDY1988377,
title = {Finite-size dependence of the free energy in two-dimensional critical systems},
journal = {Nuclear Physics B},
volume = {300},
pages = {377-392},
year = {1988},
issn = {0550-3213},
doi = {https://doi.org/10.1016/0550-3213(88)90604-9},
url = {https://www.sciencedirect.com/science/article/pii/0550321388906049},
author = {John L. Cardy and Ingo Peschel}
}

@article{Affleck_1994,
doi = {10.1088/0305-4470/27/16/007},
url = {https://dx.doi.org/10.1088/0305-4470/27/16/007},
year = {1994},
month = {aug},
publisher = {},
volume = {27},
number = {16},
pages = {5375},
author = {I Affleck and  A W W Ludwig},
title = {The Fermi edge singularity and boundary condition changing operators},
journal = {Journal of Physics A: Mathematical and General}
}

@article{Affleck:1996mm,
    author = "Affleck, Ian",
    editor = "Froehlich, J. and Rittenberg, V. and Schwimmer, A.",
    title = "{Boundary condition changing operators in conformal field theory and condensed matter physics}",
    eprint = "hep-th/9611064",
    archivePrefix = "arXiv",
    doi = "10.1016/S0920-5632(97)00411-8",
    journal = "Nucl. Phys. B Proc. Suppl.",
    volume = "58",
    pages = "35--41",
    year = "1997"
}

@article{PhysRevB.90.035131,
  title = {Spectral function of a localized fermion coupled to the Wilson-Fisher conformal field theory},
  author = {Allais, Andrea and Sachdev, Subir},
  journal = {Phys. Rev. B},
  volume = {90},
  issue = {3},
  pages = {035131},
  numpages = {11},
  year = {2014},
  month = {Jul},
  publisher = {American Physical Society},
  doi = {10.1103/PhysRevB.90.035131},
  url = {https://link.aps.org/doi/10.1103/PhysRevB.90.035131}
}

@article{Fei:2015oha,
    author = "Fei, Lin and Giombi, Simone and Klebanov, Igor R. and Tarnopolsky, Grigory",
    title = "{Generalized $F$-Theorem and the $\epsilon$ Expansion}",
    eprint = "1507.01960",
    archivePrefix = "arXiv",
    primaryClass = "hep-th",
    reportNumber = "PUPT-2481",
    doi = "10.1007/JHEP12(2015)155",
    journal = "JHEP",
    volume = "12",
    pages = "155",
    year = "2015"
}

@article{Diatlyk:2024ngd,
    author = "Diatlyk, Oleksandr and Sun, Zimo and Wang, Yifan",
    title = "{Surprises in the Ordinary: $O(N)$ Invariant Surface Defect in the $\epsilon$-expansion}",
    eprint = "2411.16522",
    archivePrefix = "arXiv",
    primaryClass = "hep-th",
    reportNumber = "PUPT-2655",
    month = "11",
    year = "2024"
}

@article{Barrat:2023ivo,
    author = "Barrat, Julien and Liendo, Pedro and van Vliet, Philine",
    title = "{Line defect correlators in fermionic CFTs}",
    eprint = "2304.13588",
    archivePrefix = "arXiv",
    primaryClass = "hep-th",
    reportNumber = "HU-EP-23/10-RTG, DESY-23-055",
    doi = "10.1007/JHEP05(2025)146",
    journal = "JHEP",
    volume = "05",
    pages = "146",
    year = "2025"
}
\end{document}